\documentclass[12pt,a4paper]{article}

\usepackage[T1]{fontenc}
\usepackage{lmodern}
\usepackage{microtype}

\usepackage{geometry}
\usepackage{setspace}
\usepackage[main=english,ngerman]{babel}
\usepackage{csquotes}

\usepackage{amsmath, amssymb}
\usepackage{siunitx}

\usepackage{graphicx}
\usepackage{caption}
\usepackage{subcaption}

\usepackage{booktabs}
\usepackage{array}

\usepackage{float}

\usepackage{placeins}

\usepackage{tikz}

\renewcommand{\topfraction}{0.95}
\renewcommand{\bottomfraction}{0.8}
\renewcommand{\textfraction}{0.05}
\renewcommand{\floatpagefraction}{0.85}

\usepackage{hyperref}

\hypersetup{hidelinks}
\usepackage[nameinlink]{cleveref}
\Crefname{section}{Section}{Sections}
\crefname{section}{Section}{Sections}

\Crefname{subsection}{Section}{Sections}
\crefname{subsection}{Section}{Sections}

\Crefname{subsubsection}{Section}{Sections}
\crefname{subsubsection}{Section}{Sections}

\crefname{figure}{Fig.}{Figs.}
\Crefname{figure}{Fig.}{Figs.}

\crefname{table}{Table}{Tables}
\Crefname{table}{Table}{Tables}

\usepackage[
backend=biber,
style=apa, uniquename=false
]{biblatex}

\newcommand{\LOOFigureCaption}[1]{\textbf{Harmonic leave-one-out analysis after omitting $P_{#1}$ from the $f_6$ denominator.}\par\smallskip
	The sixfold numerator $P_6$ was retained, and the compositional statistic was recomputed as
	$f_6^{\neg #1}=P_6/\sum_{k=2,\,k\neq #1}^{11}P_k$.
	\textbf{(A)} Population mean $\pm$ SEM of $f_6^{\neg #1}$ as a function of relative radius for experimental cells, the matched circular control, and the matched sixfold control.
	\textbf{(B)} Population mean $\pm$ SEM of the corresponding cell-matched standardized deviations, calculated separately relative to the circular and sixfold reference distributions.
	\textbf{(C)} Survival functions of the per-cell $AUC(Z)$ summaries within the predefined local-field window ($0.1\lambda$--$0.4\lambda$); the dashed line marks $AUC(Z)=0$.
	\textbf{(D)} Paired cell-wise $AUC(Z)$ values relative to the circular and sixfold controls. Dots represent cells and lines connect the two reference comparisons for the same cell. Violin plots summarize the distributions. Black points mark the median; black error bars show percentile-bootstrap $95$\% confidence intervals of the median. Colored $p$-values are two-sided sign-flip permutation tests of the population mean $AUC(Z)$ relative to each reference; the gray $p$-value tests the paired difference between the two reference comparisons. The omission analyses are dependent sensitivity analyses of the $f_6$ normalization and are not interpreted as independent hypothesis tests.
}

\hypersetup{
	hidelinks,
	pdftitle={Grid-cell firing fields lack local sixfold symmetry},
	pdfauthor={Anian Kerscher}
}

\title{Grid-cell firing fields lack local sixfold symmetry}
\author{Anian Kerscher\\
	\footnotesize Faculty of Biology, Ludwig-Maximilians-Universität München, Munich, Germany\\
	\footnotesize Graduate School of Systemic Neurosciences (GSN), Ludwig-Maximilians-Universität München, Germany\\
	\footnotesize Bernstein Center for Computational Neuroscience Munich, Munich, Germany
}

\date{September 25, 2026}

\begin{document}

	\maketitle

	\begin{abstract}
		\noindent
		The firing fields of a grid cell form a hexagonal lattice, yet each field's intrinsic geometric structure remains unclear: Are individual grid fields radially symmetric or do they exhibit (weak) sixfold modulation inherited from the global lattice? To address this question, we quantified the within-field angular structure using harmonic analysis and per-cell matched simulations that preserved sampling statistics, field scale, and lattice geometry after correction for global elliptic deformation. We found that, across the investigated cells, the sixfold harmonic power fraction was consistent with values from matched locally circular reference fields. Experimental fields further showed substantially less sixfold modulation than matched simulations with an imposed local sixfold component, demonstrating sensitivity to sixfold structure at that level. This conclusion was supported by a complementary sixfold regression metric, harmonic leave-one-out sensitivity analyses, and robustness checks across radial-window choices, smoothing, and simulated background-noise levels. An observation-matched Burak--Fiete continuous-attractor network likewise showed no selective local sixfold enhancement despite its globally periodic grid structure, and this conclusion persisted when grid scale and relative field width were matched more closely to the experimental regime. These results suggest that the global hexagonal lattice 
organization need not be accompanied by detectable local sixfold angular modulation of individual firing fields, placing quantitative constraints on grid-field microstructure and mechanistic models of grid formation.
	\end{abstract}	
	
	\section{Introduction}
	
	Grid cells in the medial entorhinal cortex~(MEC) exhibit spatially periodic firing patterns with multiple discrete firing fields arranged as a hexagonal lattice across the environment \parencite{hafting2005microstructure}. Grid cells with similar spacing and orientation are organized into discrete modules along the dorsoventral axis of MEC \parencite{stensola2012entorhinal}. At the population level, this structured activity is widely interpreted as providing a metric representation of physical space that supports spatial navigation 
	\parencite{moser2014grid}.
	
	The global hexagonal arrangement of firing fields has been extensively characterized, including its spacing, orientation, modular structure, and sensitivity to the geometry of the animal's environment \parencite{hafting2005microstructure, stensola2012entorhinal, krupic2015grid, stensola2015shearing}. In contrast, the within-field angular structure of individual grid fields has not been systematically quantified.     
	In theoretical models, individual grid fields are
	usually represented as compact, radially symmetric activity bumps \parencite{wei2015principle, sanzeni2016complete}.
However, it remains unclear whether in reality single firing fields exhibit any sixfold angular modulation beyond small sixfold variations that reflect neighboring grid fields. 
It is also unknown whether such sixfold modulation would exhibit a consistent phase across cells.
	
	Importantly, global lattice symmetry and local field symmetry are logically distinct. A hexagonal lattice can be generated by repeating approximately radially symmetric fields, whereas an intrinsic local sixfold component would indicate that lattice symmetry is also expressed within the tuning profile of an individual field. If such local structure is systematically related to the global lattice, its sixfold phase may additionally show consistent alignment across cells.

	Local angular activity modulations have implications for both neural coding and biophysical mechanisms. A subtle angular structure could influence the fidelity with which spatial position is encoded at small scales by introducing anisotropy in local tuning gradients, thereby affecting the directional dependence of decoding accuracy \parencite{averbeck2006neural, mathis2012optimal}. In addition, the presence or absence of intrinsic sixfold modulation tests whether the local field structure of commonly studied continuous attractor network models is consistent with empirical observations \parencite{burakAccuratePathIntegration2009}. Distinguishing true radial symmetry from weak but systematic angular structure requires quantitative analysis, as such effects are unlikely to be reliably detectable by visual inspection alone. In fact, detecting weak angular structure at the level of single fields is technically challenging, as finite sampling, spatial smoothing, and environmental geometry can induce spurious harmonic structure in firing-rate maps.
	
	Environmental boundaries are known to distort the global grid symmetry and local lattice geometry \parencite{stensola2015shearing, krupic2015grid}. To isolate intrinsic field structure from boundary-induced deformations, the analysis focuses on well-isolated grid fields that are centrally located in large open-field environments, where such effects are minimized.
	
	We therefore asked whether individual grid fields exhibit a selective local sixfold angular modulation beyond that expected from matched locally circular fields.
To address this question, we developed a quantitative analysis pipeline that extracts angle-resolved firing profiles from isolated grid fields and compares their angular structure to matched simulated reference models. Using long recordings from grid cells within a single grid-cell module, we analyzed angular firing statistics across spatial scales and assessed the presence of sixfold angular modulation using both qualitative and quantitative measures.
	
	Our results show that, after correction for global elliptic deformation, local sixfold modulation in experimental firing fields is consistent with matched circular-field expectations and substantially weaker than in matched simulations with an imposed local sixfold component. These findings provide a quantitative constraint on the microstructure of grid fields and support a distinction between global hexagonal lattice organization and local firing-field shape.

	\section{Methods}
	
	The data analysis consisted of the following steps: (i) Spatial firing rate maps were constructed from spike and position data (\cref{methods-rate-map}) and used to compute spatial autocorrelograms. (ii) Grid scale and global elliptical lattice distortions were estimated from the first ring of autocorrelation peaks and reduced via ellipse-based normalization (\cref{methods-grid-scale-est}). (iii) Firing fields located in the arena's central region were identified in the normalized coordinate system (\cref{methods-sogc}). (iv) For each field, angle-resolved firing profiles were extracted across radial distances expressed relative to grid scale (\cref{methods-annulus-firing-rates}). (v) Local angular symmetry was quantified using harmonic regression (\cref{methods-harmonic-regression}). (vi) Results were compared to matched simulated grid-cell data with circular or weakly sixfold-modulated local field structure (\cref{methods-model-simulator}) to assess sensitivity and establish null expectations (\cref{methods-statistical-tests}).
	
	\subsection{Experimental data}
	
	The analysis used extracellular recordings of grid cells from the MEC obtained during open-field foraging experiments previously published by \textcite{gardnerToroidalTopologyPopulation2022}. Recordings were made using Neuropixels probes targeting MEC layers II/III in adult male Long-Evans rats. The present analysis used data from rat R, day 1, module 1 (the recorded module with the smallest grid scale), during a session in a square open field ($\SI{150}{\centi\meter} \times \SI{150}{\centi\meter}$) lasting approximately 140~minutes. The animal's trajectory covered nearly the entire arena, resulting in dense spatial sampling without large unsampled regions.
	
	The selected module comprises 166 grid cells with similar grid scale and lattice orientation but different spatial phases.
	The grid scale of this module allowed multiple firing fields per cell to be at least partially contained within the arena, reducing boundary truncation at central fields and supporting stable estimation of lattice geometry for subsequent radius-normalized analyses.
	The combination of long recording duration, dense spatial coverage, and multiple well-resolved firing fields per cell supports reliable estimation of the local angular field structure.
	
	\subsection{Rate-map construction}\label{methods-rate-map}
	
	Two-dimensional spatial firing rate maps were constructed from spike and position data. Spike times were mapped to spatial coordinates by temporal interpolation of the recorded trajectory, yielding two-dimensional spike locations within the arena.
	
	Spatial firing rate maps were computed as occupancy-corrected spike density estimates. Spike locations and positional samples were accumulated on a regular two-dimensional grid covering the arena ($201~\times~201$ bins). To reduce sampling noise while preserving spatial structure, spike and occupancy maps were smoothed using isotropic Gaussian kernels, consistent with quantitative evaluations of rate map accuracy \parencite{grievesPLOSEstimatingNeuronal2023}. Spike maps were smoothed with standard deviation $\sigma_s = \SI{5}{\centi\meter}$ and occupancy maps with $\sigma_o = \SI{4}{\centi\meter}$. The smoothing widths were constrained such that $\sigma_s > \sigma_o$ to account for the sparser sampling of spike events relative to occupancy measurements. A systematic bandwidth sweep was used to assess the stability and adequacy of the canonical setting and to define sensitivity analyses across alternative smoothing widths (Supplementary \cref{sec:supp-ratemap}).
	
	Gaussian smoothing near the arena boundary was corrected for finite spatial support by normalizing each smoothed map by the correspondingly smoothed binary support mask.
	Formally, omitting this boundary-support correction from the notation, the firing rate map $r(x, y)$ was computed as:
	\begin{equation}
		r(x, y) = \frac{(G_{\sigma_s} \ast S)(x, y)}{(G_{\sigma_o} \ast O)(x, y)},
	\end{equation}
where $S(x, y)$ denotes the spatial spike count map and $O(x, y)$ the occupancy map. Each map is convolved ($\ast$) with a Gaussian smoothing kernel $G_\sigma$ with standard deviation $\sigma$.
	
	To prevent artificially inflated firing rates in sparsely sampled regions, occupancy values within the valid spatial support that fell below the 1st percentile of the non-zero occupancy distribution were clipped to this percentile prior to division, stabilizing rate estimates without imposing hard spatial thresholds.
	
	Occupancy-corrected rate maps were normalized to unit sum prior to ACF-based geometric estimation and map-based center detection. Downstream annular profiles were recomputed directly from spikes and occupancy in the destretched coordinate system. When lattice destretching was applied (\cref{methods-grid-scale-est}), spike and trajectory coordinates were transformed and the rate map was recomputed in the transformed coordinate system. Valid spatial support was restricted to the affine image of the original square arena, generally a parallelogram within the rectangular computational array.
	
	\subsection{Grid-scale estimation and reduction of global distortions}\label{methods-grid-scale-est}
	
To define a common spatial reference frame across cells and enable radius-normalized analyses, the grid scale $\lambda$ was estimated from each cell's two-dimensional spatial autocorrelation function (ACF), which provides a robust representation of lattice
	geometry in grid cells \parencite{stensola2012entorhinal}. ACFs were computed as Pearson spatial correlations between the occupancy-corrected rate map and spatially shifted copies of itself. For each spatial displacement, the Pearson correlation was evaluated only over bins that belonged to the valid spatial support in both overlapping maps, with the mean and variance recomputed separately for that overlap. Spatial displacements containing fewer than 10 overlapping valid bins were not evaluated. The correlations were calculated using an FFT-accelerated formulation that is numerically equivalent to explicit shift-wise Pearson correlation.
	
	For maps reconstructed after affine destretching (see below), the valid support corresponds to the affine image of the original square arena, generally a parallelogram contained within the rectangular computational array. Pixels outside this transformed arena were therefore excluded from ACF estimation rather than treated as zero-valued observations.
	
	For a grid cell the ACF exhibits a central peak surrounded by six first-order peaks in an approximately symmetric arrangement. The locations of these six peaks were identified, and grid scale $\lambda$ was defined as their mean distance from the central peak.
	
	In many cells, the first-order peaks deviated from circular symmetry and instead formed an elliptical configuration, resulting in systematic variability among the six distance estimates. Such global anisotropy can bias radial normalization and may induce spurious lower-order angular components when angular firing profiles are computed in uncorrected Euclidean coordinates. To reduce this geometric nuisance, an ellipse was fitted to the six peak locations using a direct least-squares conic method \parencite{fitzgibbonDirectLeastsquaresFitting1996}. A linear transformation was then applied to map the fitted ellipse onto a circle, yielding a destretched coordinate system in which global anisotropy was reduced.
	
	Grid scale was re-estimated in the destretched coordinate system using the same peak-based procedure. Subsequent center-based spatial analyses were performed in this normalized (destretched) coordinate system. Radial distances were expressed relative to the destretched grid scale $\lambda$. Because $\lambda$ therefore directly determines the radial normalization, accurate cell-level scale estimation is important for the present analysis. The relatively small grid scale of the module analyzed here lies well below the finite-arena regime in which ACF-based scale estimates become increasingly biased as grid spacing approaches the dimensions of the environment \parencite{veselov2026latticescales}. Spatial coordinates otherwise retained the linearly transformed centimeter units defined by the affine mapping. Because the affine transform is estimated from the global ACF, matched control simulations verified that the transform generalizes across independent split-half spike samples and predominantly suppresses twofold angular structure while leaving absolute sixfold harmonic power essentially unchanged  (Supplementary \cref{sec:supp-ellipse}). Because experimental data and matched reference simulations (\cref{methods-model-simulator}) were processed through the same ACF-based affine-normalization pipeline, effects of this preprocessing were calibrated relative to identically processed reference distributions.
	
\subsection{Field-center detection and selection}\label{methods-sogc}
	
	To quantify the local angular structure within individual grid fields, accurate and robust field-center estimates are essential. All subsequent analyses were performed relative to these centers. This was done to minimize systematic center offsets that can bias angular estimates, particularly when angular structure is weak and confined to specific radial ranges. 
	
	Grid-cell rate maps can contain irregularities, such as partial fields near arena boundaries, spurious local maxima, and distortions arising from sampling variability. Simple peak-detection or threshold-based methods can therefore yield unstable or spurious field centers, particularly in noisy data. To obtain robust center estimates, candidate field centers were identified using a self-organized grid clustering (SOGC) procedure applied in the destretched coordinate system.
	
	SOGC is a mode-seeking algorithm related to mean-shift clustering. Candidate centers are initialized on a dense grid and iteratively updated by two interacting forces: (i) an attraction term that moves each candidate toward the kernel-weighted centroid of nearby spikes, equivalent to ascending the local spike-density landscape, and (ii) a weaker isotropic repulsive interaction between nearby candidate nodes that discourages crowding and provides spatial regularization at the scale of a firing field. Nearby nodes are merged during iteration, yielding a compact set of putative field centers. The repulsive interaction does not impose a lattice orientation; global lattice consistency is assessed separately by the subsequent step described next.
	
	To suppress isolated spurious maxima and assess consistency with the global grid structure, a second, lattice-constrained filtering step was applied. Using the grid spacing $\lambda$ obtained from the ACF (\cref{methods-grid-scale-est}), an anchor-free hexagonal lattice was fitted while holding its spacing fixed to $\lambda$ and optimizing lattice orientation and translation (phase).
	Candidate centers were retained only if they lay within a specified distance of a fitted lattice site. Retained centers remained at their empirically detected locations and were not moved onto the fitted lattice; the lattice fit was used only to reject inconsistent candidate centers.
	
	For each included cell (\cref{sec:inclusion-exclusion}), a single field was selected using a fixed rule: the eligible center closest to the arena center in the destretched coordinate system. Further algorithmic details, parameter choices, convergence criteria, and validation are provided in Supplementary~\cref{sec:supp-SOGC,sec:supp-lattice-fit}. 
	
	\subsection{Inclusion and exclusion criteria}\label{sec:inclusion-exclusion}
	
	Cells were included in the angular field analysis only if all preprocessing steps satisfied predefined quality-control criteria intended to exclude unstable or degenerate cases.
	
	Cells were excluded if the first-ring ACF peaks failed a hexagonality check (angular deviation exceeding $\ang{30}$ from the ideal $\ang{60}$ spacing), if the spread of first-ring peak distances exceeded $\SI{5}{\centi\meter}$ ($\sim 10\%$ of the module grid spacing) after affine normalization, or if destretching increased the peak-distance interval by more than $50$\% relative to the raw ACF. Cells were additionally excluded if the axis ratio of the ellipse fitted to the first-ring ACF peaks increased following affine normalization.
	To ensure stable rate-map and center estimation, cells were required to have fired at least $10{,}000$ spikes within the approximately 140-minute recording session, corresponding to approximately $1.18$~Hz over the valid trajectory support.
	
	Following center detection and lattice-constrained filtering, candidate centers within $0.3\lambda$ of the original square-arena boundary were excluded to reduce boundary truncation of angular profiles. Cells were required to retain at least three candidate centers after this step to ensure stable lattice phase estimation and center selection. Cells were further excluded if affine normalization changed the estimated mean first-ring grid scale by more than $\SI{20}{\centi\meter}$.
	
	Applying these criteria yielded $140$ cells for the angular symmetry analysis.
	
	\subsection{Angle-resolved firing profiles from annuli}\label{methods-annulus-firing-rates}
	
	Having identified reliable field centers, the angular structure of firing activity was then quantified within individual grid fields. Because grid fields do not exhibit sharply defined boundaries and can vary in size and shape, local symmetry cannot be assessed using a single radius or hard spatial cutoff.
	
	For each selected field, spikes and occupancy were expressed in polar coordinates $(\rho, \theta)$ relative to the estimated field center in the distortion-reduced coordinate system (\cref{methods-grid-scale-est}). Radial distances were normalized by the destretched grid scale $\lambda$ to allow a comparison across cells.
	
	Angle-resolved firing rates were computed within thin annuli spanning a specified range of relative radii. Within each annulus, the angle-dependent firing rate was estimated schematically as the ratio of spike counts to occupancy after separable radial and circular smoothing,
	\begin{equation}
		r(\rho, \theta) = \frac{(K_{\rho} \ast K_{\theta} \ast S)(\rho, \theta)}{(K_{\rho} \ast K_{\theta} \ast O)(\rho, \theta)},
	\end{equation}
where $S(\rho, \theta)$ denotes spike counts, $O(\rho, \theta)$ the corresponding occupancy, and $K_\rho$ and $K_\theta$ are radial and angular kernels defining the annulus width and angular resolution, respectively. A small regularization term was included to stabilize estimates in sparsely sampled regions (Supplementary \cref{sec:supp-annulus-rate-estimation}).
	
	Annuli were stepped outward from the field center in overlapping increments, yielding a two-dimensional representation of the firing rate as a function of radius and angle. This approach avoids assumptions about the shape of field boundaries and allows the angular modulation to be assessed locally across the extent of the field.
	
	To parameterize the field width for matched simulations (\cref{methods-param-matching}), each field's radial envelope was estimated by computing the annulus-averaged rate profile $y(\rho) = \langle r(\rho, \theta) \rangle_\theta$ from $0.1\lambda$ to $0.5 \lambda$ (approximately halfway to neighboring fields), and fitting a Gaussian envelope 
	$
G(\rho) = A \exp[-\rho^2 / (2 \sigma^2)] + b
	$
	with nonnegative parameters. Here, $A$ is the Gaussian amplitude above baseline, $b$ is the fitted baseline, and $\sigma$ is the Gaussian standard deviation used as the field-width estimate.
	
	The resulting radius-angle firing profiles form the basis for subsequent harmonic analysis of angular symmetry. Parameter choices governing radial resolution, angular resolution, and smoothing widths are provided in Supplementary \cref{sec:supp-annulus-rate-estimation}.
	
	Angular symmetry analyses were restricted a priori to radii between $0.1\lambda$ and $0.4\lambda$, corresponding to the local-field region while minimizing potential contamination from neighboring firing fields.
	
	\subsection{Angular autocovariance analysis}
	
	To visualize periodic angular structure independent of overall firing amplitude and offset, the circular autocovariance (ACV) of the angular firing profile $r(\theta)$ was computed within each radial annulus. For each annulus, $r(\theta)$ was scaled by its maximum value, mean-centered, and its circular ACV computed using FFT-based convolution. Angular offsets within $\pm \ang{30}$ of zero were excluded from visualization to suppress the dominant peak at zero offset and the elevated covariance at small angular offsets arising from local angular smoothness. For comparability across radii and cells, each ACV trace was normalized by its maximum absolute value prior to averaging across cells. Because the ACV amplitude depends on sampling variability and may reflect non-sixfold periodicities or lattice-scale structure, ACV was used as a qualitative diagnostic only. Formal quantification of angular symmetry was performed using harmonic regression (\cref{methods-harmonic-regression}).
	
	\subsection{Quantification of sixfold angular modulation}\label{methods-harmonic-regression}
	
	To detect and quantify sixfold angular structure in the extracted angle-resolved firing profiles, the magnitude of any sixfold periodic component was evaluated. For each radial annulus, the angular firing profile was first normalized by its circular mean, $\tilde{r}(\theta) = r(\theta)/\langle r(\theta) \rangle_\theta$, such that subsequent metrics reflect relative angular modulation independent of overall annulus gain. For each annulus, the resulting profile was modeled using linear harmonic regression:
	\begin{equation}
		\tilde{r}(\theta) = a_k \cos(k \theta) + b_k \sin(k \theta) + c + \epsilon,
	\end{equation}
where $k$ denotes angular frequency, $a_k$ and $b_k$ parameterize the amplitude and phase of the $k$-th harmonic, $c$ is a constant offset, and $\epsilon$ captures residual structure. This formulation is mathematically equivalent to projecting the angular profile onto the $k$-th Fourier basis, while explicitly accounting for a nonzero mean and missing angular samples. Because $\tilde{r}(\theta)$ is normalized by its circular mean, the fitted constant term $c$ is expected to be approximately 1 and was included primarily as a consistency check. Fits were obtained by least squares.
	
	Harmonic components were evaluated for angular frequencies $k = 2, \dots, 11$. Angular Hann smoothing attenuates higher angular frequencies more strongly. Accordingly, cross-harmonic spectral shape was interpreted only relative to matched references processed with the identical smoothing and estimation procedure, rather than against a flat theoretical spectrum. The first harmonic was excluded because it is particularly sensitive to residual field-center displacement, while frequencies $k\geq12$ have angular periods of $\ang{30}$ or less, approaching the $\ang{29}$ angular smoothing scale; $k=12$ also represents a higher harmonic rather than the fundamental sixfold component targeted here. For each frequency, harmonic power was defined as $P_k = A^2_k = a^2_k + b^2_k$. The primary analysis focused on the sixfold component ($k=6$), consistent with the global hexagonal lattice symmetry of grid-cell firing. To quantify the relative enhancement of the sixfold modulation, the sixfold power fraction was computed:
	\begin{equation}
		f_6 = \frac{P_6}{\sum_{k=2}^{11}P_k}.
	\end{equation}
$f_6$ is the fraction of the total fitted harmonic power across $k=2, \dots, 11$ at the same radius that is assigned to $k=6$. It is therefore a compositional measure: it can change either because $P_6$ changes or because one or more of the non-sixfold powers in the denominator change. It should not be interpreted as an absolute measure of sixfold modulation.
	
	To determine whether the primary $f_6$ comparison depended disproportionately on any single non-sixfold component, a harmonic leave-one-out analysis was performed. For each $j \neq 6$, $f_6$ was recomputed after omitting $P_j$ from the denominator:
	\begin{equation}
		f_6^{\neg j} = \frac{P_6}{\sum_{k=2, k \neq j}^{11}P_k}, \quad
		j \in \{2,3,4,5,7,8,9,10,11\}.
	\end{equation}
	These highly dependent variants were treated as a sensitivity analysis of the statistic.
	
	All harmonic metrics were computed independently for each annulus (radius). As a complementary sixfold-specific sensitivity measure, the coefficient of determination $R_6^2$ of the $k=6$ harmonic fit was additionally evaluated. This measure was used as a robustness analysis and did not replace the predefined primary statistic $f_6$.
	
	As an additional diagnostic, the sixfold phase was extracted as $\phi = \operatorname{atan2}(-b_6, a_6)$. The population phase alignment at each radius was quantified using a weighted circular mean,
	\begin{equation}
		V(\rho) = \left| \frac{\sum_i w_i(\rho) \exp(i\phi_i(\rho))}{\sum_i w_i(\rho)} \right|,
	\end{equation}
	with weights $w_i(\rho) = P_{6,i}(\rho)$. Weighting by $P_6$ reduces the influence of poorly determined phases at radii with negligible sixfold power. Uncertainty in $V(\rho)$ was estimated by bootstrap resampling of cells ($20{,}000$ resamples).
	
	All fits were performed independently for each cell and each analyzed radial annulus.
	
	\subsection{Experimental parameters for matched simulations}\label{methods-param-matching}
	
	To construct matched reference simulations that preserve sampling statistics and geometric nuisance structure, cell-specific parameters were extracted from the selected central field in the distortion-reduced coordinate system. Using the radial envelope procedure described in \cref{methods-annulus-firing-rates}, an estimate of field width $\hat{\sigma}$ was obtained for each cell, expressed both in absolute units (cm) and relative to the cell's grid scale $\lambda$. Only the fitted width $\hat{\sigma}$ was retained for the simulated local-field envelope. A separate uniform spatial-noise component was included in the simulations as described below.
	
	In addition, we retained each cell's spike count $N_i^{\mathrm{spk}}$, destretched grid scale $\lambda$, affine inverse transform $T^{-1}$, selected local-field location, the occupancy map in the distortion-reduced coordinate system, and the corresponding valid spatial support (bounding box) used for annulus extraction. These per-cell parameters were used to generate matched simulated spikes under different local symmetry priors (\cref{methods-model-simulator}).
	
	\subsection{Simulated firing-field models}\label{methods-model-simulator}
	
	Synthetic grid-cell spike data were generated under controlled local field symmetry priors while matching each experimental cell's sampling statistics and geometric transformations. Simulations were performed on a discretized square arena ($201 \times 201$ grid) in the distortion-reduced (\enquote{destretched}) coordinate system (Section~\ref{methods-grid-scale-est}), using the cell-specific transformed spatial support. Subsequently the simulations were mapped back to the original coordinate system so that the simulated data underwent the identical preprocessing and geometric normalization steps as the experimental data.
	The simulations were designed to preserve the global grid geometry, using the per-cell lattice scale and a common lattice orientation ($\ang{16}$), while allowing systematic manipulation of the local angular structure within individual fields.
	
	\subsubsection{Local field model}
	
	Each local firing field was defined as an isotropic Gaussian envelope centered at $(x_c, y_c)$, with $\rho = \sqrt{(x-x_c)^2 + (y-y_c)^2}$,
	\begin{equation}
		g_\mathrm{rad}(\rho) = \exp(-\frac{\rho^2}{2 \sigma^2}),
	\end{equation}
with $\sigma$ set to the per-cell estimate $\hat{\sigma}$. These fields instantiate the null model of locally isotropic grid fields (circular). The overall multiplicative amplitude is arbitrary under fixed-spike-count normalized sampling (see below) and was fixed to one; no additive baseline term was included.
	
	To generate a weak sixfold local prior with a fixed angular phase $\theta_\mathrm{loc}=\ang{0}$, the Gaussian was modulated by a sixfold von Mises term with unit mean,
	\begin{equation}
		\label{eq:I-hex}
		g_\mathrm{hex}(\rho, \theta) = g_\mathrm{rad}(\rho)[1 + \beta (\tilde{a}(\theta) - 1)]_+,
	\end{equation}
	
	\begin{equation}
		\tilde{a}(\theta) \propto \exp(\kappa_\mathrm{eff} \cos(6(\theta - \theta_\mathrm{loc}))),
	\end{equation}
where $[\cdot]_+$ denotes rectification at zero (clipping) to ensure nonnegative intensity. The angular modulation term was normalized to have unit mean over the discretized field grid, thereby centering the angular gain around unity. The term $\beta$ controls the strength of angular modulation relative to the isotropic Gaussian envelope. Circular controls used $\beta = 0$. Simulations with a sixfold prior used $\beta > 0$ (swept over a small set of values for robustness). In practice, this was implemented using a mixture of three von Mises terms with twofold periodicity separated by $\ang{120}$, each with concentration $\kappa=2.1$, which yields an overall sixfold modulation and is equivalent in periodicity to a single $\exp(\kappa \cos(6\cdot))$ von Mises term.
	
	\subsubsection{Hexagonal lattice tiling}
	
	Grid-cell-like intensity maps were generated by summing identical local fields placed on a hexagonal lattice with spacing $\lambda$ set to the per-cell estimate obtained from the experimental ACF and global rotation $\theta_\mathrm{lat}=\ang{16}$,
	\begin{equation}
		g_\mathrm{grid}(x, y) = \sum_j g_j(x,y),
	\end{equation}
without individual normalization prior to summation. Within a simulated cell, all lattice sites used the same cell-specific field-width parameter $\sigma$, and the Gaussian field envelopes were not explicitly radially truncated.	Consequently, contributions from neighboring modeled fields increased naturally
	with increasing $\sigma/\lambda$. For spike generation, the resulting spatial weights were subsequently combined with the experimental occupancy map and normalized to obtain a discrete spatial probability distribution $p(x, y)$. 
	
	\subsubsection{Spike sampling, occupancy matching, and geometric transforms}
	
	For each experimental cell, $R=100$ replicate simulated datasets were generated under each of the two priors (circular and sixfold). For each replicate, $N_i^{\mathrm{spk}}$ spike locations were requested, with $N_i^{\mathrm{spk}}$ matched to the experimental spike count of cell $i$, before inverse transformation and arena cropping. Of these, $10\%$ constituted a uniform spatial-background component defined over the original physical arena, while the remaining $90\%$ were sampled independently from the occupancy-weighted lattice distribution $p(x,y)$. Structured spike locations were subsequently mapped back to the original coordinate system using the per-cell inverse affine transform $T^{-1}$ and locations outside the valid arena were discarded. Consequently, the final retained spike count could be slightly smaller than the requested $N_i^{\mathrm{spk}}$; the uniform-background component, which was drawn directly in the physical arena, was retained by construction. All simulated datasets were processed through the same spatial-analysis pipeline and subjected to the same downstream geometric and field-selection quality-control criteria. For null-referenced comparisons, a cell was retained only if at least $K=75$ of the $100$ attempted replicates passed the complete pipeline in each reference condition. This yielded $N=137$ cells for population inference.
	
	\subsection{Continuous-attractor-network model}\label{CAN}
	
	As an additional mechanistic control, grid-cell activity generated by the continuous attractor network (CAN) model of \textcite{burakAccuratePathIntegration2009} was analyzed. The implementation followed the released periodic Burak--Fiete (BF) model on a $128\times128$ neural sheet with four interleaved preferred-direction populations. The canonical condition used the released recurrent parameter $\bar\alpha=1.05$, which is the relative-width parameter of the two Gaussian components in the recurrent kernel $W_0(r) = a \cdot \mathrm{exp}(-\bar\alpha \beta r^2) - \mathrm{exp}(-\beta r^2)$. Network formation comprised $1000$ updates using the released formation protocol, with linear recurrent coupling and a spatially tapered input envelope for the first $799$ updates followed by a uniform envelope for the remaining $201$. The resulting pattern was subsequently stabilized under periodic recurrent coupling and uniform input using three $250$~ms velocity drives at $0.8$~m/s in the fixed directions $0$, $\pi/5$, and $3\pi/10$. Behavioral trajectory integration then used periodic recurrent coupling and uniform input across the neural sheet.
	
	The network was driven with the valid portions of the same Gardner trajectory used for the experimental analysis. The canonical analysis used the complete valid trajectory, which consisted of two continuous segments separated by an approximately 2~min period without valid behavioral sampling. Velocity was integrated only within observed segments; no displacement was interpolated across the gap. At the beginning of the later observed segment, the stabilized attractor state was re-anchored to the globally defined BF spatial phase corresponding to the first observed position of that segment. Between consecutive trajectory samples, the network dynamics were integrated at the internal BF time step of $0.5$~ms. For each such trajectory interval, readout activity was defined as the mean activity of the corresponding unit across these internal network updates.
	
	\noindent
	A fixed bank of $200$ neural readouts was selected uniformly without replacement from the neural sheet. The first $137$ readouts were paired, in predefined order, with the $137$ cells of the final experimental population. This pairing served only to match observation depth and has no biological cell-to-cell interpretation. Each BF readout was assigned the number of experimental spikes observed for its paired cell over the same temporal support.
	
	We then generated exactly this number of model events along the observed trajectory. For each event, a trajectory interval was selected with probability proportional to its mean readout activity multiplied by its duration. The event time was sampled uniformly within the selected interval, and its position was obtained by linear interpolation between the corresponding trajectory endpoints. No additional spatial-background component was added. Because the total event count was fixed in this way, the absolute BF activity scale was not interpreted as an intrinsic firing rate; instead, the observation model tested whether the model's spatial structure persisted under experimentally matched finite sampling.

	The resulting event locations were processed with the same spatial analysis used for the experimental cells, including rate-map construction, Pearson spatial autocorrelation, affine normalization, grid-scale estimation, field-center selection, field-width estimation, and annulus-based angular profiles. For each BF readout, separate circular and imposed-sixfold reference populations were generated using that readout's own recovered event count, grid scale, field width, affine transform, selected center, occupancy, and valid spatial support. One hundred realizations were generated per reference condition with the same zero-background observation model, and reference events were passed directly to the same annulus estimator without re-estimating spatial geometry. Harmonic metrics and null-referenced inference were then computed using the same $f_6$ and cell-wise $AUC(Z)$ procedures as for the experimental analysis. Additionally, the $P_2$-omitted sensitivity statistic $f_6^{\neg2}$ was evaluated. Because all BF readouts within each condition originate from one deterministic network realization, the corresponding population statistics quantify consistency across spatial readout phases rather than replication across independent networks.
	
	The canonical model produced considerably narrower fields relative to grid spacing than the experimental population. One targeted field-width-matched sensitivity condition was therefore included by changing the recurrent parameter to $\bar\alpha=1.012$ and applying one global velocity-to-phase scaling factor ($1.03153$). This condition was selected using grid scale and radial field width only; no angular or harmonic statistic was used for parameter selection. Model-validity diagnostics showed substantial deterioration of spatial registration during longer integrations of this parameterization. Quantitative symmetry analysis was therefore restricted to the first $30$~min of the first continuous trajectory segment (Supplementary \cref{sec:supp-CAN-diagnostics}).
	
	\subsection{Statistical testing of sixfold angular modulation}\label{methods-statistical-tests}
	
	\subsubsection{Null hypothesis and reference distributions}
	
	The primary reference hypothesis was that local angular structure is adequately described by the matched circular-field model, with no selective sixfold modulation beyond that produced by sampling variability under a shared analysis pipeline. To obtain cell-specific simulation-based reference distributions, matched simulations were generated under a circular local prior ($\beta=0$ in Eq.~\ref{eq:I-hex}) and, for comparison and sensitivity calibration, under an imposed sixfold local prior ($\beta=0.2$ in Eq.~\ref{eq:I-hex}), as described in \cref{methods-model-simulator}.
	
	For each cell and radial position $\rho$, null distributions were obtained from the corresponding set of matched simulations, allowing a cell-wise comparison of the experimental angular metrics with their empirical null expectation.
	
	\subsubsection{Null-referenced effect sizes}
	
	Let $M_i(\rho)$ denote an angular metric (e.g. sixfold power fraction $f_6$) for cell $i$ at radius $\rho$. For each cell and radius, the null-standardized effect was defined as
	\begin{equation}
		Z_i(\rho) = \frac{M_i^{\mathrm{exp}}(\rho) - \mu_i^{\mathrm{null}}(\rho)}{\sigma_i^{\mathrm{null}}(\rho)},
	\end{equation}
	where $\mu_i^{\mathrm{null}}(\rho)$ and $\sigma_i^{\mathrm{null}}(\rho)$ denote the mean and standard deviation across matched null simulations for that cell. These curves quantify deviation from null expectation in units of the cell-specific null standard deviation and were used for descriptive visualization across radii. Population curves were obtained by averaging $Z_i(\rho)$ across cells. These radius-resolved population mean $\pm$ SEM curves were used for descriptive visualization only.
	
	\subsubsection{Population-level inference}
	
	For population inference each cell's null-standardized deviation curve $Z_i(\rho)$ was summarized within the predefined local window ($0.1\lambda$--$0.4\lambda$) by its area under the curve ($AUC$), 
	\begin{equation}
		A_i \equiv {AUC}[Z_i] = \int_{0.1}^{0.4} Z_i(\rho_{\mathrm{rel}})\,
		d\rho_{\mathrm{rel}} 	
	\end{equation}
	computed numerically using the trapezoidal rule.
	
	Under the null hypothesis, $A_i$ is expected to be centered near zero. A two-sided sign-flip permutation test across cells ($20{,}000$ permutations) assessed whether the population mean of $A_i$ differed from zero. This test preserves the magnitude of each cell's effect since it uses the mean as test statistic, and randomizes only the sign of each magnitude. The sign-flip test assumes sign-exchangeability of the cell-wise $A_i$ values under the null, i.e. that under the null their distribution is sufficiently symmetric about zero for random sign assignment to reproduce the relevant null distribution. 
	
	To directly compare the two reference conditions, we also tested whether $A_i^{\mathrm{Six}} - A_i^{\mathrm{Circ}}$ differed from zero using the same paired sign-flip procedure. For descriptive visualization of cell-level summary statistics, confidence intervals around median values were estimated using percentile bootstrap resampling across cells (20,000 resamples, with replacement). 
	
	To verify that inference was not materially limited by Monte Carlo estimation of the cell-specific reference moments, the primary analysis was repeated using the first $25$, $50$, and $75$ successful matched replicates per cell, and the resulting estimates were
	compared with the complete retained reference bank. Cell membership was held fixed to the final $N=137$ population for this convergence analysis.
	
	\subsection{Temporal firing-mode analyses}\label{methods-temporal-modes}
	
	To test whether local sixfold structure was only apparent during burst firing, the analysis was repeated using burst events.
	Except for the event definition and sampling modifications described below, burst-event profiles were analyzed using the same frozen annulus estimator, harmonic metrics, cell-wise reference standardization, $AUC$ summary, and population sign-flip inference as described in
	\cref{methods-annulus-firing-rates,methods-harmonic-regression,methods-statistical-tests}.
	The burst analysis started from the $140$ cells for which the all-spike
	preprocessing and local-field geometry had been successfully defined. Grid scale $\lambda$, the selected field center, affine transformation, occupancy map, valid spatial support, and fitted field width were inherited from the all-spike analysis and were not re-estimated from burst or tonic subsets.
	
	A burst was defined as a sequence (consecutive spike events) of at least two spikes with interspike intervals (ISI) of $\leq\SI{12}{\milli\second}$, with each burst represented by one spatial location at its onset. The reported profiles were evaluated at the same radial centers from $0.1\lambda$ to $0.4\lambda$ as in the canonical analysis (Supplementary \cref{sec:supp-temporal-modes}). Cells with at least $282$ independent burst-onset events within the complete annulus support entered the primary burst analysis.
	
	The circular and imposed-sixfold spatial model families were inherited from the matched all-spike analysis, but the previously simulated all-spike realizations were not reused. Instead, for every included cell, $100$ new circular and $100$ new sixfold reference event sets were sampled directly within the frozen local coordinate system. Each reference realization was conditioned to reproduce exactly the observed burst-event counts within $0.05\lambda$ radial matching strata, while the local polar angle remained unconstrained. Further implementation details of the radial conditioning are given in Supplementary \cref{sec:supp-temporal-modes}.
	
	For a direct burst--vs--tonic comparison, tonic spikes were defined as spikes not belonging to any detected burst. Within each $0.05\lambda$ radial matching stratum, burst onsets and tonic spikes were sampled without replacement to the same count, $M_j=\min(B_j,T_j)$. Thus, each paired realization had identical binned radial counts and total sample size in the two firing modes, while local polar angle remained unconstrained (Supplementary \cref{sec:supp-temporal-modes}). $25$ paired resamples were generated per cell. For each resample, $f_6$ was integrated over the predefined $0.1\lambda$--$0.4\lambda$ interval, the burst-minus-tonic $AUC(f_6)$ difference was calculated, and, in contrast to the canonical analysis, these differences were averaged within each cell. Population inference was then performed by the same two-sided cell-wise sign-flip procedure; resamples were not treated as independent observations. Thus, the generated tonic references were treated differently from the simulated data.
	
	Because representing bursts by independent onset events substantially reduced the available sampling, we complemented the within-cell firing-mode analysis with an all-spike analysis of temporal grid-cell phenotypes. Temporal classes were obtained by applying the temporal-autocorrelation/ToMATo classification procedure of \textcite{gardnerToroidalTopologyPopulation2022} to the corresponding day-1 recordings analyzed here. Because cluster identifiers are arbitrary, the three resulting day-1 clusters were assigned to the bursty, theta-modulated, and non-bursty phenotypes by one-to-one cosine matching of their mean temporal autocorrelograms to the reconstructed reference-class prototypes (Supplementary \cref{sec:supp-temporal-modes}).
	
	No spatial preprocessing, field analysis, or matched simulations were repeated for this comparison. The endpoint for each of the $137$ canonical cells was its already determined all-spike $AUC(Z)$ relative to the circular reference. Differences among the three temporal classes were tested using a one-way permutation ANOVA: the conventional $F$ statistic quantified the ratio of between-class to within-class variability, and significance was determined by permuting class labels across cells while preserving class sizes. $\eta^2$ was reported as an effect-size estimate, and $R_6^2$ provided a complementary endpoint.

	\section{Results}	
	
	The firing fields of grid cells in the rodent MEC are arranged in highly regular hexagonal lattices \parencite{hafting2005microstructure, stensola2012entorhinal}. 
	This study investigates whether individual firing fields show the same sixfold symmetry.
	Several coding and normative treatments idealize individual grid cells as periodic arrays of simple localized tuning profiles, including periodically repeated Gaussian or circular fields \parencite{mathis2012optimal, sanzeni2016complete, mosheiff2017efficient}.
	Radially symmetric individual fields would therefore support a common modeling assumption. More generally, our analysis provides quantitative constraints on mechanisms underlying grid-cell formation and coding.

	\subsection{Data quality, geometric normalization, and field selection}
	
	We analyzed spike data from a single grid-cell module recorded during long explorations of an open-field ($\SI{150}{\centi\meter} \times \SI{150}{\centi\meter}$) arena \parencite{gardnerToroidalTopologyPopulation2022}. Of $166$ cells in the selected module, $140$ met predefined preprocessing and geometric-stability criteria. Of these cells, $137$ retained sufficient matched simulation replicates under both the circular and the sixfold control models and were included in null-referenced population analyses. Across this final $137$-cell population, cells fired with a median session-averaged firing rate of $3.40$~Hz (IQR $2.27$--$4.61$~Hz; range $1.20$--$8.32$~Hz). Before unit-sum normalization, the occupancy-corrected rate maps had a median peak firing rate of $17.49$~Hz (IQR $11.39$--$23.98$~Hz; range $2.77$--$45.87$~Hz).
	
\begin{figure}[!t]
		\centering
		\includegraphics[width=0.9\textwidth]{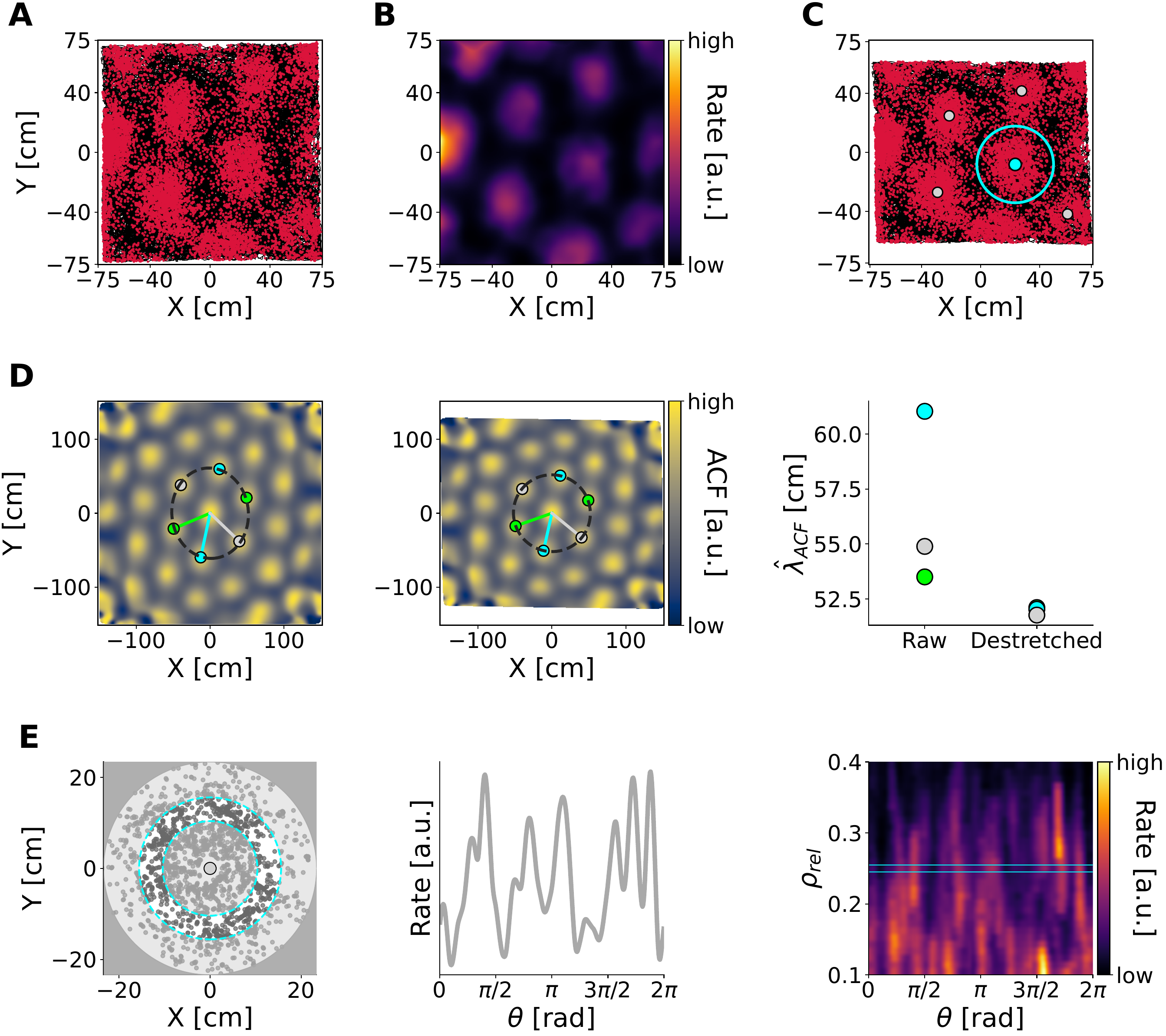}
		\caption[Example grid cell: Analysis pipeline]{\textbf{Firing fields of an example grid cell and workflow for analyzing the local angular activity profile.}\par\smallskip
			\textbf{(A)} Spike locations (red) overlaid on the animal trajectory in the arena (black). 
			\textbf{(B)} Occupancy-corrected, smoothed, and normalized firing-rate map. 
			\textbf{(C)} Distortion-reduced (affine-normalized) representation used for field-level analyses. The detected field centers are shown in light gray; the analyzed central field is highlighted in cyan. Circle: Radius $0.5\lambda$ around the analyzed field.
			\textbf{(D)} Grid-scale estimation and affine normalization from the spatial autocorrelogram (ACF). Left: Raw ACF with first-ring peak locations and fitted ellipse. Middle: ACF after ellipse-derived affine normalization. Right: First-ring peak distances before and after normalization, illustrating reduced anisotropy.
			\textbf{(E)} Annulus-based extraction of angle-resolved firing. Left: Spikes near the analyzed field with an example annulus indicated (dashed lines). Middle: Angular firing-rate profile within the example annulus. Right: Angular profiles across relative radii ($0.1\lambda$--$0.4\lambda$); the highlighted trace corresponds to the example annulus.
		}
		\label{fig:fig1}
	\end{figure}

	Spike distributions and rate maps exhibited the expected hexagonal lattice structure (\cref{fig:fig1}A-B). To prevent local angular measurements from being confounded by global elliptical lattice distortion, we estimated the grid scale $\lambda$ from the first ring of spatial ACF peaks and applied affine normalization to reduce the elliptical deformation (\cref{fig:fig1}D). Destretching reduced the variability in first-ring peak distances and decreased systematic axis-ratio biases (Supplementary \cref{sec:supp-ellipse}), yielding a distortion-reduced coordinate system for subsequent radial analyses.
	
	Field centers were identified using a self-organized grid clustering (SOGC) procedure followed by lattice-constrained filtering. To minimize boundary effects and field truncation, we restricted angular analyses to well-isolated central fields (\cref{fig:fig1}C). Radial distances were expressed relative to grid scale ($\rho_{\mathrm{rel}} = \rho/\lambda$) to enable comparisons across cells with different absolute spacings.
	
	Annulus-based angular firing profiles were extracted from overlapping circular rings (\cref{fig:fig1}E). The resulting radius-angle representations showed smooth radial envelopes but did not reveal an obvious sixfold pattern. Angular profiles varied with radius and sampling density, and no sharp field boundary was apparent (Supplementary \cref{fig:supp-gauss-field_width}), motivating window-based summary statistics rather than single-radius inference.
		
	The population-averaged radial envelope exhibited a smooth decay from the field center and a rise beyond $\sim0.5 \lambda$, consistent with increasing contributions from neighboring lattice fields (Supplementary \cref{fig:supp-gauss-field_width}). Gaussian-plus-baseline fits to per-cell radial profiles within $0.1\lambda$--$0.5\lambda$ were used to estimate field width $\hat{\sigma}$ for matched simulation construction. The local-field analysis window ($0.1\lambda$--$0.4\lambda$) was chosen a priori to remain within the local-field region and to avoid both the central sampling instability and the between-field rise.
	
	\FloatBarrier
	
	\subsection{Matched simulations reproduce sampling and geometry}
	
	To establish a reference for local angular structure under matched sampling and lattice geometry, we constructed per-cell null simulations (\cref{fig:fig2}A). These simulations closely matched spike count and preserved occupancy sampling, grid spacing, estimated field width, geometric transformations and central-field location (up to rare near-center substitutions). Two local-field priors were used: (i) circular ($\beta = 0$), and (ii) sixfold ($\beta = 0.2$), unless otherwise stated.
	
	\begin{figure}[tbp]
		\centering
		\includegraphics[width=0.9\textwidth]{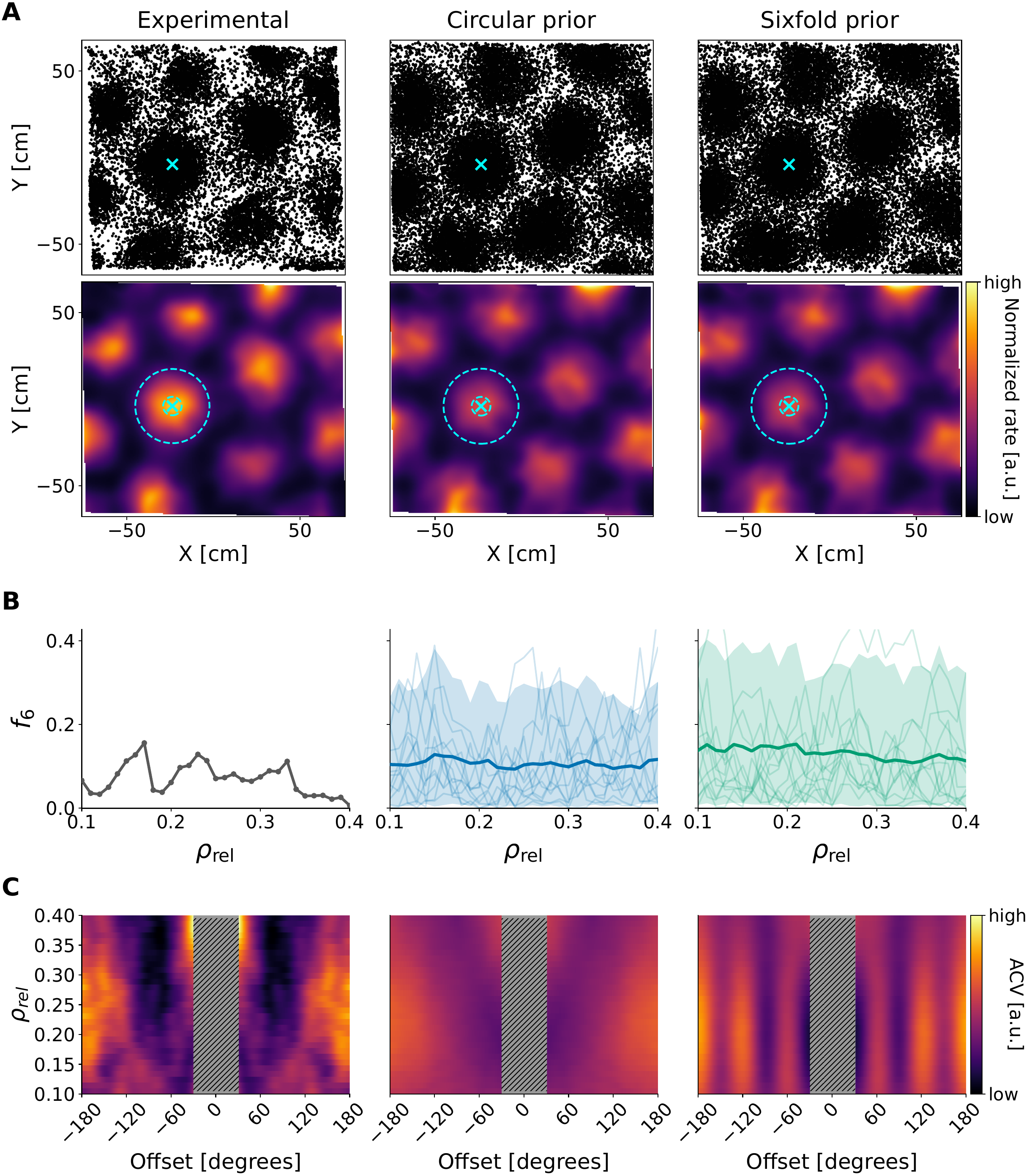}
		\caption[Example grid cell: matched simulations]{\textbf{Example cell, matched simulations, and population angular structure.}\par\smallskip
			\textbf{(A)} Spike locations in distortion-reduced coordinates (left) and matched simulations under a circular local-field prior (middle) and a weak sixfold-modulation prior (right). Cross: Center of the analyzed firing field. Bottom row: Corresponding occupancy-corrected rate maps; dashed circles mark the inner and outer boundaries of the predefined local-field window ($0.1\lambda$--$0.4\lambda$). Rate maps share a common color scale within the row.
			\textbf{(B)} Radial profiles of the sixfold harmonic power fraction $f_6$ as a function of relative radius for the example experimental cell and its matched reference simulations. Each radius summarizes the harmonic	decomposition of the corresponding annular angular profile introduced in \cref{fig:fig1}E. The experimental curve is the single observed profile. For each reference condition, the thick line shows the mean across passing matched realizations, the shaded region denotes the pointwise 5th--95th percentile range, and a deterministic subset of individual realizations is shown as faint curves. These traces illustrate the substantial radius-to-radius variability present in individual matched fields; inferential comparisons therefore use the full matched distribution rather than isolated profile maxima.
}
		\label{fig:fig2}
	\end{figure}
	
	\begin{figure}[tbp]
		\ContinuedFloat
		\caption[]{\textbf{Continued.}
			
			\textbf{(C)}
			Cell-balanced population mean angular autocovariance (ACV) across the final
			matched population ($N=137$) within the local-field window
			($0.1\lambda$--$0.4\lambda$). For simulated branches, realizations were
			first averaged within each cell and then averaged across cells, giving each cell equal weight. Each annular ACV was normalized by its maximum absolute value before population
			averaging; the heat maps therefore visualize the consistency of angular ACV structure across cells rather than absolute ACV magnitude. The gray hatched region
			denotes the excluded $\pm\ang{30}$ neighborhood around zero offset. A common
			color scale is used across conditions.
		}
	\end{figure}
	
	For an example cell, destretched simulated rate maps closely matched the destretched experimental map in scale and spatial extent (\cref{fig:fig2}A), while the weak imposed local sixfold prior was not readily identifiable by visual inspection of the full rate map alone. Radial profiles of the sixfold harmonic power fraction $f_6$, which summarizes the harmonic decomposition of the annular profiles introduced in \cref{fig:fig1}E, are shown for the recorded cell and the matched simulations (\cref{fig:fig2}B). The experimental profile represents a single realization, whereas the simulation curves summarize variability across replicated datasets. In the example cell, the sixfold-prior distribution was elevated most clearly over part of the inner-to-intermediate local-field range, whereas individual matched realizations showed substantial radius-to-radius variability. This illustrates why inference used cell-specific distributions over many matched realizations rather than individual simulated fields.
	
	At the population level, mean angular ACV profiles within the local-field window ($0.1\lambda$--$0.4\lambda$) provided a qualitative view of the angular structure (\cref{fig:fig2}C). Because each annular ACV was normalized before cell-balanced averaging, these maps emphasize the consistency of angular ACV structure across cells rather than absolute ACV magnitude. All three conditions exhibited anisotropic structure, including a prominent feature near $\ang{180}$, consistent with substantial low-order, particularly twofold, angular structure. In addition to this shared low-order structure, the sixfold-prior simulations displayed clear sixfold periodicity, with recurrent peaks near multiples of $\ang{60}$ across most radii in the window. Such consistent sixfold modulation was not visually apparent in the experimental data or in the circular-prior simulations. Because ACV magnitude reflects contributions from both sixfold and lower-order angular components, formal inference was based on per-cell harmonic regression metrics evaluated against matched null distributions. Extending the ACV analysis to $0.8\lambda$ showed pronounced $\ang{60}$ periodicity in all three conditions at larger radii, where neighboring fields increasingly reflect the global hexagonal lattice, while the imposed-sixfold condition showed this periodicity already within the local-field range (Supplementary \cref{fig:supp-acv-full}). This provides a qualitative positive control for recovery of sixfold periodicity and illustrates the rationale for the predefined window.
	
	Parameters were closely recovered in the matched simulations across cells (Supplementary \cref{fig:supp-model-matching}). Across the final matched population ($N=137$), the fitted experimental relative field width was $\hat{\sigma}/\lambda = 0.188 \pm 0.032$ (mean $\pm$ SD; median $= 0.185$, IQR $= 0.165$--$0.203$). The corresponding recovered widths were $0.180 \pm 0.021$ in both the circular and sixfold-prior simulation branches. The recovered grid scale was slightly larger than the matched experimental value in both simulation branches, with median absolute recovery error between recovered and matched experimental grid scale of $\SI{0.697}{\centi\meter}$ for the circular control and $\SI{0.707}{\centi\meter}$ for the sixfold-prior simulations. Relative to the
	corresponding experimental input value, the recovered local-field width differed
	by a median of $1.77\%$ and $1.85\%$, respectively, with compression at the
	broad end of the observed width range. The experimental fields contained $3340\pm1615$ spikes (mean $\pm$ SD) within the analyzed $0.1\lambda$--$0.4\lambda$ window, compared to $3214\pm1462$ and $3216\pm1463$ spikes in the matched circular and sixfold-prior simulations, respectively. The origin and influence of the broad-field recovery compression and the small grid-scale recovery offset were examined in supplementary diagnostics (Supplementary \cref{sec:supp-radius-excursion}).

	\FloatBarrier
	
\subsection{Sixfold harmonic modulation is consistent with matched circular-field expectations}
	
\begin{figure}[tbp]
		\centering
		\includegraphics[width=0.9\textwidth]{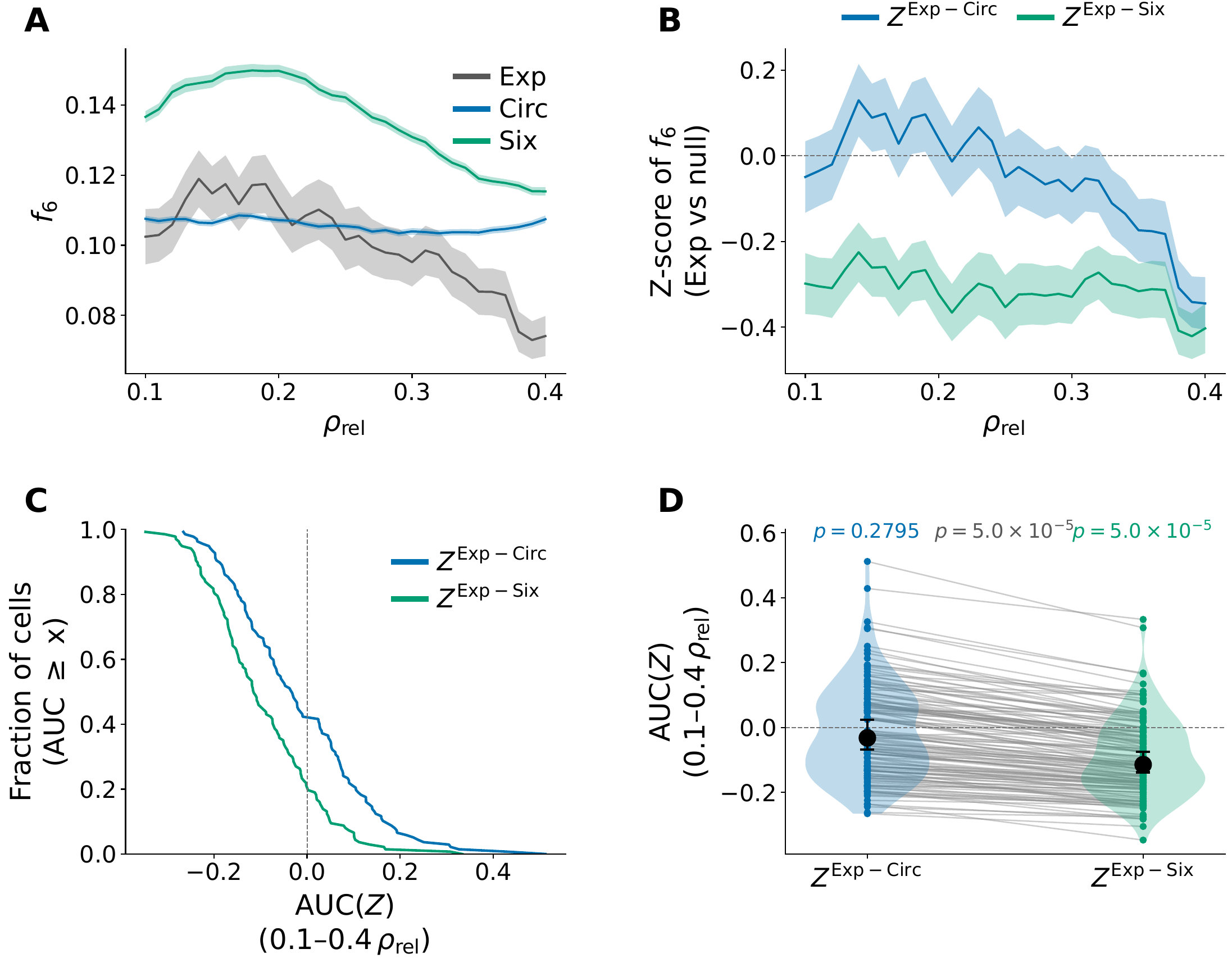}
		\caption[Sixfold modulation relative to matched reference models]{\textbf{Local sixfold modulation is consistent with the matched circular reference and lower than the matched sixfold control.}\par\smallskip
			\textbf{(A)} Population mean $\pm$ SEM of sixfold harmonic power fraction $f_6(\rho_{\mathrm{rel}})$ for recorded cells and matched simulations. Simulation realizations were first averaged within each cell and population summaries were then computed across cells.
			\textbf{(B)} Population mean $\pm$ SEM of cell-matched standardized deviations (Z-scores) of $f_6$ relative to each reference condition (SEM across cells). Positive values indicate a higher $f_6$ in the experimental field than expected from the corresponding matched reference.
			\textbf{(C)} Survival function of per-cell window summary statistics $AUC(Z)$ within the predefined local-field window ($0.1 \leq \rho_{\mathrm{rel}} \leq 0.4$); the dashed line marks $AUC(Z)=0$.
			\textbf{(D)} Paired per-cell $AUC(Z)$ values for experiment (Exp) vs. circular (Circ) prior and experiment vs. sixfold (Six) prior (dots: cells; lines: paired comparisons). Violin plots summarize the distributions. Black points mark the median; black error bars show percentile-bootstrap $95$\% confidence intervals of the median. Colored $p$-values: two-sided sign-flip permutation tests of whether the population mean $AUC(Z)$ differs from 0. Gray $p$-value: paired test of $AUC(Z^{\mathrm{Exp-Six}})-AUC(Z^{\mathrm{Exp-Circ}})$.
		}
		\label{fig:fig3}	
	\end{figure}

	We next quantified these differences across the complete matched population. The sixfold harmonic power fraction $f_6(\rho_{\mathrm{rel}})$ was used as the	primary measure because experimental fields exhibited broadband elevation of harmonic power relative to the matched circular reference, making absolute $P_6$ insufficient to test selective sixfold structure. The fractional metric isolates the relative contribution of the $k=6$ component and tests for sixfold specificity, rather than overall angular variability. Population-averaged radial profiles are shown in \cref{fig:fig3}A.

	To assess deviations relative to matched references on a per-cell basis, we	computed null-standardized Z-scores. Population mean $\pm$ SEM Z-score curves are shown in \cref{fig:fig3}B. Relative to the circular reference, the population-mean standardized deviation fluctuated around zero across much of the inner and intermediate part of the local-field window and became negative	toward its outer boundary. Relative to the sixfold-prior reference, the population-mean standardized deviation was negative throughout the analyzed window. These radius-resolved profiles are descriptive and do not account for the substantial cell-to-cell variation in the matched reference distributions.
	
	Separately, the negative experimental--circular deviation toward the outer portion of the local-field window was accentuated in a post hoc subgroup of fields with larger fitted widths. Because the radius-wise curves are descriptive, this feature was not used for primary inference. Its dependence on field width, harmonic composition, radial-window choice, and matched-model assumptions was examined in supplementary sensitivity analyses (Supplementary \cref{sec:supp-radius-excursion,tab:supp-window-robustness}).

	To summarize the effects at the cell level while avoiding radius-wise multiple testing, we computed the area under the Z-score curve ($AUC$) within the predefined local-field window (\cref{fig:fig3}C-D). Relative to the circular reference, the mean cell-wise $AUC(Z)$ did not differ significantly from zero (mean $=-0.0138$, median $=-0.0319$, two-sided sign-flip test $p=0.279$; $N=137$), indicating no reliable elevation of relative sixfold modulation above matched circular-field expectations.
	
	In contrast, relative to the sixfold-prior reference ($\beta=0.2$), $AUC(Z)$ values were significantly negative (mean $=-0.0936$, median $=-0.1143$, $p=5.0\times10^{-5}$), indicating that experimental sixfold 	modulation was weaker than that produced by simulations with the imposed local sixfold structure. Direct comparison between the two reference conditions ($AUC(Z^{\mathrm{Exp-Six}})-AUC(Z^{\mathrm{Exp-Circ}})$) likewise yielded a significant negative effect (mean $=-0.0798$, median $=-0.0710$,	$p=5.0\times10^{-5}$).
	
	Survival curves of per-cell $AUC(Z)$ values (\cref{fig:fig3}C) showed that, although some cells exhibited relatively large positive standardized deviations, the distribution was not shifted positively relative to the circular reference and was substantially more negative relative to the sixfold-prior reference. However, these population-level comparisons do not exclude heterogeneous effects or a small subset of cells with stronger local modulation. Consistently, $58/137$ cells had positive and $79/137$ negative $AUC(Z)$ values (two-sided exact sign test, $p=0.087$; Supplementary \cref{sec:supp-sign-test}). Relative to the sixfold-prior reference, only $28/137$ cells were positive and $109/137$ were negative.
	
	As a complementary non-compositional measure of sixfold fit strength, analyses using $R_6^2$ from the sixfold harmonic regression as the angular modulation metric yielded the same qualitative result: experimental fields did not differ significantly from the circular reference ($p=0.427$) and remained below the imposed-sixfold reference ($p=5.0\times10^{-5}$; Supplementary \cref{fig:supp-R2}).
	
	\subsection{Compositional sensitivity and harmonic leave-one-out}\label{sec:loo-methods}
	
	By construction, $f_6$ is a compositional statistic: a change in $f_6=P_6/\sum_{k=2}^{11}P_k$ can arise either from a change in $P_6$ itself or from changes in the non-sixfold harmonics contributing to the denominator. This limits the interpretation of the precise numerical value of an $f_6$ contrast and motivated a harmonic leave-one-out sensitivity analysis in which each non-sixfold denominator component was omitted in turn while the numerator $P_6$ was left unchanged.
	
	The primary conclusion was stable across all omissions (\cref{tab:harmonic-loo}). Every comparison with the circular reference remained non-significant ($0.142\leq p_{\mathrm{Circ}}\leq0.695$), whereas experimental fields remained below the imposed-sixfold reference for every omission ($p_{\mathrm{Six}}=5.0\times10^{-5}$ throughout). The paired contrast between the two reference comparisons also remained at the permutation-resolution limit throughout.
	
	Omitting $P_2$ produced the largest directional change, from $-0.0138$ to $0.0073$ ($p_{\mathrm{Circ}}=0.564$). Thus, the experimental–circular mean depended partly on the contribution of $P_2$ to the denominator, while the statistical conclusion remained unchanged.

	\begin{table}[htbp]
		\centering
		\small
		\caption{\textbf{Harmonic leave-one-out sensitivity of the compositional sixfold power fraction.}
			For each row, the indicated non-sixfold harmonic was omitted from the denominator of
			$f_6=P_6/\sum_{k=2}^{11}P_k$, while the numerator $P_6$ was unchanged.
			Mean Exp--Circ and Exp--Six values denote the population mean cell-wise $AUC(Z)$
			relative to the matched circular and sixfold reference distributions, respectively.
			The paired effect is the population mean of
			$AUC(Z^{\mathrm{Exp-Six}})-AUC(Z^{\mathrm{Exp-Circ}})$.
			$p$-values are two-sided sign-flip permutation tests of the corresponding population means
			(20,000 permutations). The leave-one-out variants are dependent sensitivity analyses and
			are not interpreted as independent hypothesis tests.}
		\resizebox{\textwidth}{!}{\begin{tabular}{lrrrrrr}
				\toprule
				Omitted harmonic &
				Mean Exp--Circ &
				$p_{\mathrm{Circ}}$ &
				Mean Exp--Six &
				$p_{\mathrm{Six}}$ &
				Mean paired effect &
				$p_{\mathrm{pair}}$ \\
				\midrule
				none     & -0.0138 & 0.279 & -0.0936 & $5.0\times10^{-5}$ & -0.0798 & $5.0\times10^{-5}$ \\
				$P_2$    &  0.0073 & 0.564 & -0.0755 & $5.0\times10^{-5}$ & -0.0828 & $5.0\times10^{-5}$ \\
				$P_3$    & -0.0052 & 0.695 & -0.0859 & $5.0\times10^{-5}$ & -0.0806 & $5.0\times10^{-5}$ \\
				$P_4$    & -0.0143 & 0.274 & -0.0932 & $5.0\times10^{-5}$ & -0.0789 & $5.0\times10^{-5}$ \\
				$P_5$    & -0.0159 & 0.196 & -0.0948 & $5.0\times10^{-5}$ & -0.0789 & $5.0\times10^{-5}$ \\
				$P_7$    & -0.0155 & 0.227 & -0.0946 & $5.0\times10^{-5}$ & -0.0791 & $5.0\times10^{-5}$ \\
				$P_8$    & -0.0179 & 0.146 & -0.0968 & $5.0\times10^{-5}$ & -0.0789 & $5.0\times10^{-5}$ \\
				$P_9$    & -0.0172 & 0.173 & -0.0958 & $5.0\times10^{-5}$ & -0.0786 & $5.0\times10^{-5}$ \\
				$P_{10}$ & -0.0181 & 0.151 & -0.0965 & $5.0\times10^{-5}$ & -0.0784 & $5.0\times10^{-5}$ \\
				$P_{11}$ & -0.0182 & 0.142 & -0.0969 & $5.0\times10^{-5}$ & -0.0787 & $5.0\times10^{-5}$ \\
				\bottomrule
			\end{tabular}}
		\label{tab:harmonic-loo}
	\end{table}
	
	\subsection{Phase alignment of sixfold modulation}
	
	A local sixfold component with a common orientation across fields should also yield a consistent fitted $k=6$ phase. We therefore extracted the sixfold phase $\phi(\rho)$ and quantified the population-level phase coherence using the $P_6$-weighted vector strength $V(\rho)$. Simulations in which the imposed sixfold prior had a fixed phase showed strong phase coherence throughout the local-field window, demonstrating recovery of the imposed phase. Experimental coherence was substantially weaker than in	this positive control, while the circular reference remained low (Supplementary \cref{fig:supp-pop-vec}). Because this analysis was descriptive, it was used as a phase-recovery diagnostic rather than as an additional population hypothesis test.

	\subsection{Robustness analyses}\label{sec:robusness-methods}
	
	The primary inference was based on the predefined local-field window
	($0.1 \leq \rho_{\mathrm{rel}} \leq 0.4$), selected to avoid poorly sampled near-center radii and to minimize the overlap with adjacent lattice fields. We assessed whether the result depended on
	the radial bounds, the amount of background activity included in the matched
	simulations, the rate-map smoothing parameters, or the strength of the imposed
	local sixfold modulation.
	
	\subsubsection{Window bounds}
	
	Varying the radial window did not reveal a positive experimental--circular $f_6$ shift (Supplementary \cref{tab:supp-window-robustness}). Windows	confined to the inner and intermediate field remained consistent with the	circular reference, whereas windows extending to or isolating the outer field showed significant negative deviations. In particular, the $0.1\lambda$--$0.5\lambda$ and $0.3\lambda$--$0.4\lambda$ windows yielded negative experimental--circular shifts. Experimental fields remained below the imposed-sixfold reference in every tested window, and the paired difference between the two reference comparisons remained significant throughout.
	
	\subsubsection{Uniform background noise in simulations}
	
	To assess sensitivity to background activity in the matched simulations, we
	varied the fraction of uniformly distributed spikes between $0$ and $0.2$
	(Supplementary \cref{tab:supp-noise-robustness}). The experimental--circular
	comparison remained non-significant at all three tested noise levels, whereas
	experimental fields remained significantly below the matched sixfold
	reference. The separation from the sixfold reference weakened modestly as the
	background fraction increased, but remained significant at all three
	tested levels.
	
	\subsubsection{Smoothing bandwidth}
	
	We repeated the full matched-reference analysis for spike/occupancy smoothing-width pairs $(\sigma_s, \sigma_o) = (3,2)$, $(3,3)$, $(4,3)$, the canonical $(5,4)$, and $(6,4)$, with all widths in cm (Supplementary \cref{tab:supp-bandwidth-robustness}).
	For $(3,3)$, $(4,3)$, $(5,4)$, and $(6,4)$, the primary experimental--circular $f_6$ comparison remained non-significant, while the experimental population remained below the imposed-sixfold reference.
	At the least-smoothed $(3,2)$ setting, $f_6$ and $R_6^2$ showed a negative deviation from the circular reference, whereas $f_6^{\neg2}$ remained circular-consistent. Thus, no tested smoothing setting revealed a positive local sixfold excess, and the main conclusion was not specific to the canonical
	$(5,4)$ pair.
	
	\subsubsection{Detectability calibration
		(\texorpdfstring{$\beta$}{beta} sweep)}
	
	To quantify sensitivity to imposed local sixfold structure, we repeated the matched-reference comparison across $\beta=0.1$, $0.2$, and $0.4$ (Supplementary \cref{fig:supp-beta-sweep}, \cref{tab:supp-table-beta}). 
	The experimental--circular comparison was unchanged across the sweep, whereas separation from the sixfold-prior reference increased monotonically with $\beta$. For $f_6$, the mean experimental--sixfold $AUC(Z)$ was $-0.0395$ at $\beta=0.1$ ($p=9.0\times10^{-4}$), $-0.0936$ at $\beta=0.2$ ($p=5.0\times10^{-5}$), and $-0.2414$ at $\beta=0.4$ ($p=5.0\times10^{-5}$). Although the latter comparisons reached the	permutation-resolution limit for $20{,}000$ permutations, the effect sizes and distributions continued to separate with increasing $\beta$. At an additional lower-strength calibration of $\beta=0.05$, the two reference comparisons were already measurably separated, but experimental $f_6$ was not significantly different from this weak sixfold reference itself. Thus, the analysis showed graded sensitivity to the strength of imposed local sixfold modulation; these discrete calibration points do not define a hard detection threshold. 
	
	\subsection{Temporal firing modes do not reveal sixfold organization}
	\label{results-temporal-modes}
	
	Burst firing may represent a distinct functional mode of grid-cell activity and could, in principle, express angular structure that is diluted in the complete spike train. Of the $140$ cells with successfully defined all-spike field geometry, $108$ contained at least $282$ independent burst-onset events within the local analysis support and entered the primary burst analysis. For burst events, $f_6$ did not exceed the matched circular reference (mean cell-wise $AUC(Z)=-0.0355$, sign-flip $p=0.012$) and remained below the reference with weak sixfold modulation (mean $=-0.0550$, $p=1.0\times10^{-4}$; Supplementary \cref{fig:temporal-modes}A). The burst-event $f_6$ also did not differ detectably from the paired, equal-count, radially matched tonic samples	($p=0.058$; Supplementary \cref{fig:temporal-modes}B); the complementary $R_6^2$ comparison yielded the same conclusion ($p=0.072$). The negative circular-reference departure weakened and was no longer significant under the stricter sampling criterion, whereas sixfold modulation in burst events remained below the imposed-sixfold reference; restriction to the canonical all-spike population gave the same overall conclusion (Supplementary \cref{sec:supp-temporal-modes}). Thus, restricting the analysis to burst events did not reveal enhanced local sixfold organization.
	
	We next asked whether local sixfold selectivity differed among independently defined temporal grid-cell phenotypes. Applying the temporal classification method of \textcite{gardnerToroidalTopologyPopulation2022} to the day-1 recordings assigned the $137$ cells in the	canonical analysis population to bursty ($n=41$), theta-modulated ($n=75$), and non-bursty ($n=21$) classes (\cref{fig:temporal-modes}C). All-spike local sixfold selectivity did not differ detectably among the three classes ($F=0.517$, permutation $p=0.596$, $\eta^2=0.0077$; \cref{fig:temporal-modes}D). The complementary $R_6^2$ analysis gave the same result ($F=0.365$, $p=0.696$, $\eta^2=0.0054$). Thus, neither transient burst firing nor temporal firing phenotype accounted for selective local sixfold organization.
	
	\subsection{Comparison with the model of Burak and Fiete}
	
	Finally, we asked whether the globally hexagonal state generated by a continuous-attractor model (CAN) also produces selective sixfold structure within individual firing fields. In the canonical Burak--Fiete (BF) CAN, all $137$ observation-matched readouts yielded valid local-field geometry. The median recovered grid scale was $\SI{46.06}{\centi\meter}$ (IQR $45.85$--$46.39$~cm), whereas the median relative field width was $\sigma/\lambda=0.109$ (IQR $0.106$--$0.112$), substantially narrower than in the experimental population. Representative rate maps are shown in Supplementary \cref{fig:supp-CAN}A. 
	
	Canonical BF did not show selective local sixfold enhancement (Supplementary \cref{fig:supp-CAN}B--D). Relative to each readout's matched circular reference, the primary sixfold power fraction was shifted downward (mean $AUC(Z)=-0.136$, $p=5.0\times10^{-5}$), and the deviation from the imposed-sixfold reference was larger (mean $AUC(Z)=-0.180$, $p=5.0\times10^{-5}$). This normalized deficit did not reflect weak absolute sixfold power. Instead, canonical BF exhibited a broad elevation of absolute harmonic power relative to its circular reference, including substantially elevated $P_6$. Within this elevated spectrum, however, the relative composition was dominated by still stronger low-order components, particularly $P_2$ (Supplementary \cref{fig:supp-CAN}C). Omitting $P_2$ from the $f_6$ denominator ($f_6^{\neg2}$) substantially attenuated the BF--circular difference, although a small negative shift remained (mean $AUC(Z)=-0.0224$, $p=0.026$; Supplementary \cref{fig:supp-CAN}D).
	
	We repeated the analysis in the width-matched condition over the first $30$~min. Its local geometry was much closer to the experimental regime (median $\lambda=\SI{53.92}{\centi\meter}$, IQR $53.08$--$55.31$~cm; median $\sigma/\lambda=0.187$, IQR $0.172$--$0.201$). The primary $f_6$ comparison remained below the circular reference (mean $AUC(Z)=-0.0336$, $p=0.00545$) and below the imposed-sixfold reference (mean $AUC(Z)	=-0.0532$, $p=5.0\times10^{-5}$). After omitting $P_2$, however, the width-matched model was consistent with its circular reference (mean $AUC(Z)=0.0062$, $p=0.622$). Thus, matching field width and grid scale more closely to the experimental regime did not reveal a robust selective local sixfold enhancement. Additional model-specific diagnostics are reported in Supplementary \cref{sec:supp-CAN-diagnostics}.

	\section{Discussion}
	
	In this study, we tested whether individual grid-cell firing fields exhibit an intrinsic local sixfold angular modulation related to the global hexagonal lattice. 
After correction for global lattice distortion, experimental fields showed no reliable elevation of sixfold modulation relative to per-cell circular references matched for spike count, occupancy, grid scale, field width, central-field position, and geometric transformations. In contrast, simulations with imposed local sixfold modulation produced clear, detectable deviations.
Together, these controls demonstrate that the analysis responds to imposed sixfold structure: separation from the experimental population increased monotonically with $\beta$, and fixed-phase sixfold simulations produced strong phase coherence. The calibration does not, however, define a sharp detection threshold, and the lowest-strength condition shows that sufficiently weak local sixfold modulation cannot be reliably distinguished from the experimental data. The same qualitative conclusion was obtained using the complementary $R_6^2$ metric, while the qualitative ACV diagnostics recovered both the imposed local sixfold pattern and the expected global lattice periodicity at larger radii.
	
	The radius-resolved Exp--Circ mean $Z$ profile was not perfectly flat, with	experimental $f_6$ falling below the circular reference toward the outer part of the predefined local-field window. Post hoc diagnostics showed that this deviation was accentuated among broader fields and was sensitive to harmonic composition and to increasing contributions from neighboring fields in the idealized matched lattice (Supplementary
	\cref{sec:supp-radius-excursion}). Because these effects were localized toward the outer field and did not alter the predefined $0.1\lambda$--$0.4\lambda$ cell-wise inference, we interpret the outer excursion as a radius-dependent model and field-shape sensitivity rather than evidence for a separate local sixfold effect.
	
	The primary statistic $f_6$ is compositional rather than an absolute measure of $P_6$. Its numerical value can therefore in principle change when non-sixfold components of the angular spectrum change. The harmonic leave-one-out analysis addressed this dependence directly: the experimental--circular comparison remained non-significant after omission of every individual non-sixfold denominator harmonic, whereas separation from the imposed-sixfold reference persisted throughout. Omitting $P_2$ produced the largest numerical change and shifted the already small experimental--circular effect toward zero, showing that twofold power contributes to the value of the normalized metric without revealing evidence for a separate sixfold mismatch.
	
	Importantly, the absence of detectable selective sixfold modulation does not imply that individual firing fields are angularly featureless or perfectly radially symmetric. Rather, other local angular anisotropies may be present, and sixfold modulations weaker than the sensitivity of the present matched-reference analysis cannot be excluded (\cref{fig:fig2}C). Our conclusion is therefore specific: after correction for global elliptic deformation, we find no robust evidence for an additional local sixfold modulation of individual firing fields at the spatial scale tested here.
	
	Global sixfold periodicity was robustly observed at larger radii, consistent with the well-established lattice structure in the rodent medial entorhinal cortex \parencite{sargolini2006conjunctive, hafting2005microstructure, stensola2012entorhinal}. Our findings therefore suggest a dissociation between global hexagonal lattice organization and selective sixfold angular structure within individual fields.
	
	In modeling work, individual grid fields are often treated as approximately radially symmetric units arranged in a hexagonal lattice \parencite{low2018probing, vollan2025left}. The BF simulations provide a mechanistic example of the distinction between global and local symmetry. The canonical network formed a globally periodic grid representation, yet its local angular spectrum was dominated by strong low-order structure rather than a selective sixfold component. The 30-min width-matched condition showed a substantially less extreme low-order spectral imbalance, but still did not reveal selective local sixfold enhancement. The similarity between experiment and BF therefore lies in the dissociation between global hexagonal organization and local sixfold selectivity, not in an identity of their local harmonic spectra.
	The non-sixfold components themselves were not the target of the present study and their origin remains unresolved. Likewise, the BF comparison concerns this specific model implementation and these parameter regimes rather than continuous-attractor models as a class. Model-specific geometry, long-duration behavior, and possible mechanisms absent from the present BF implementation are considered in Supplementary \cref{sec:supp-CAN-diagnostics}.
	
	Because burst spikes in grid cells have been reported to yield higher spatial information, coherence, and grid scores than an equal number of tonic spikes \parencite{bant2020topography}, temporal firing structure provides a plausible route by which local angular organization could be obscured in the all-spike analysis. The burst-event analysis had substantially lower effective sampling than the all-spike analysis. Its primary comparison showed a negative deviation from the circular reference, but this departure weakened under the stricter sampling criterion, while direct, radius-matched comparisons between burst and tonic samples did not show enhanced sixfold selectivity during burst firing. We therefore interpret the temporal-event analyses as providing no support for a hidden burst-specific enhancement, rather than as evidence for robust suppression of sixfold structure during bursts.

	Moreover, a population-average result could also conceal enhanced sixfold modulation in only a subset of cells. This possibility is biologically plausible because grid cells show distinct temporal firing phenotypes \parencite{gardnerToroidalTopologyPopulation2022}, and bursty subclasses have been reported to contribute disproportionately to theta-sweep representations \parencite{vollan2025left}. We therefore compared the canonical all-spike sixfold statistic among bursty, theta-modulated, and non-bursty cells. No detectable difference between the three classes was observed. However, the smaller bursty and particularly non-bursty groups limit the sensitivity to subtle phenotype-specific effects. Together, the temporal analyses provide no evidence that local sixfold organization occurs during burst episodes or exists in one of these temporal phenotypes.
	
	Our analyses were restricted to central fields in a large open-field environment. Extremely weak modulations below the sensitivity of the present calibration cannot be excluded, but would likely play a rather limited biological role. The extensive sensitivity analyses nevertheless support the stability of the primary inference across the analytical choices examined here. The present analysis is further based on cells from a single recording session, grid-cell module, and animal. The cell-level population statistics therefore quantify consistency within this recorded module and should not be interpreted as animal-level replication or as establishing generality across animals or grid-cell modules.
	
	The absence of robust local sixfold modulation also constrains a possible microscopic contribution to hexadirectional fMRI signals. Such signals are conventionally defined by sixfold modulation with movement, gaze, or inferred trajectory direction relative to an estimated grid orientation \parencite{doeller2010evidence, constantinescu2016organizing, nau2018hexadirectional}. Local field geometry can influence direction-dependent population coding as an animal traverses a field: sixfold anisotropy could produce direction-dependent variation in firing-rate gradients and consequently in population sensitivity. A fixed local field shape could therefore contribute to such a modulation at the population level and thereby to the signal measured in fMRI analyses. Our results consequently argue against a direct account in which macroscopic hexadirectionality simply reflects six-lobed individual firing fields, while leaving intact mechanisms based on directionally conjunctive coding, population organization, trajectory statistics, adaptation, or nonlinear aggregation. Consistent with this distinction, quantitative population modeling has shown that macroscopic hexadirectional modulation can arise from grid-axis-aligned head-direction tuning and, under some conditions, from firing-rate adaptation \parencite{binkhalid2024quantitative}. Neither mechanism is formulated as a direct inheritance of sixfold angular structure from the shape of individual firing fields. More broadly, the interpretation of reported hexadirectional fMRI effects is currently under reassessment \parencite{mei2026neural}.
	
	Grid cells are often conceptualized as generating periodic tilings composed of repeating units \parencite{rowland2016ten}. While hexagonal lattice geometry is well established, less attention has been paid to the angular structure within individual fields. Our results do not support the stronger expectation that global hexagonal lattice organization is inherited as a fixed sixfold angular modulation of individual firing fields. We conclude that global lattice periodicity does not imply intrinsic local sixfold structure at the level of a single grid-cell firing field.
	
	\subsection*{Data availability} 
	This study reanalyzes publicly available electrophysiological and behavioral
	data from \textcite{gardnerToroidalTopologyPopulation2022}. The dataset is
	available on Figshare at
	\url{https://doi.org/10.6084/m9.figshare.16764508.v6}.
	No new experimental data were generated in this study.
	
	\subsection*{Code availability}
	All custom analyses and simulations were implemented in Python.
	The corresponding analysis code is available from the author upon reasonable request.
	
	\subsection*{AI declaration} 
	The author used generative AI tools to assist with the development and debugging of analysis code, literature searches and synthesis, and the drafting and revision of portions of the manuscript. All AI-assisted code, analyses, citations, and text were critically reviewed and verified by the author, who takes full responsibility for the analyses, interpretations, and content of the manuscript.
	
	\subsection*{Acknowledgments} I thank Andreas Herz for suggesting the original research question, for supervision, scientific discussions, and stimulating feedback throughout this project, and Martin Stemmler for helpful discussions and comments on the manuscript.
	
	\subsection*{Competing interests}
	The author declares no competing interests.

	\FloatBarrier
	\clearpage 
	
	\printbibliography
	
	\FloatBarrier
	\clearpage 
	
\section{Supplementary information} 

\renewcommand{\thefigure}{S\arabic{figure}}
	\setcounter{figure}{0}
	
\renewcommand{\thetable}{S\arabic{table}}
	\setcounter{table}{0}
	
	\renewcommand{\theequation}{S\arabic{equation}}
	\setcounter{equation}{0}

\renewcommand{\thesubsection}{S\arabic{subsection}}
	\renewcommand{\thesubsubsection}{S\arabic{subsection}.\arabic{subsubsection}}
	\setcounter{subsection}{0}
	\setcounter{subsubsection}{0}
	
	\subsection{Robustness of rate-map smoothing bandwidths}\label{sec:supp-ratemap}
	
\begin{figure}[htbp]
		\centering
		\includegraphics[width=0.9\textwidth]{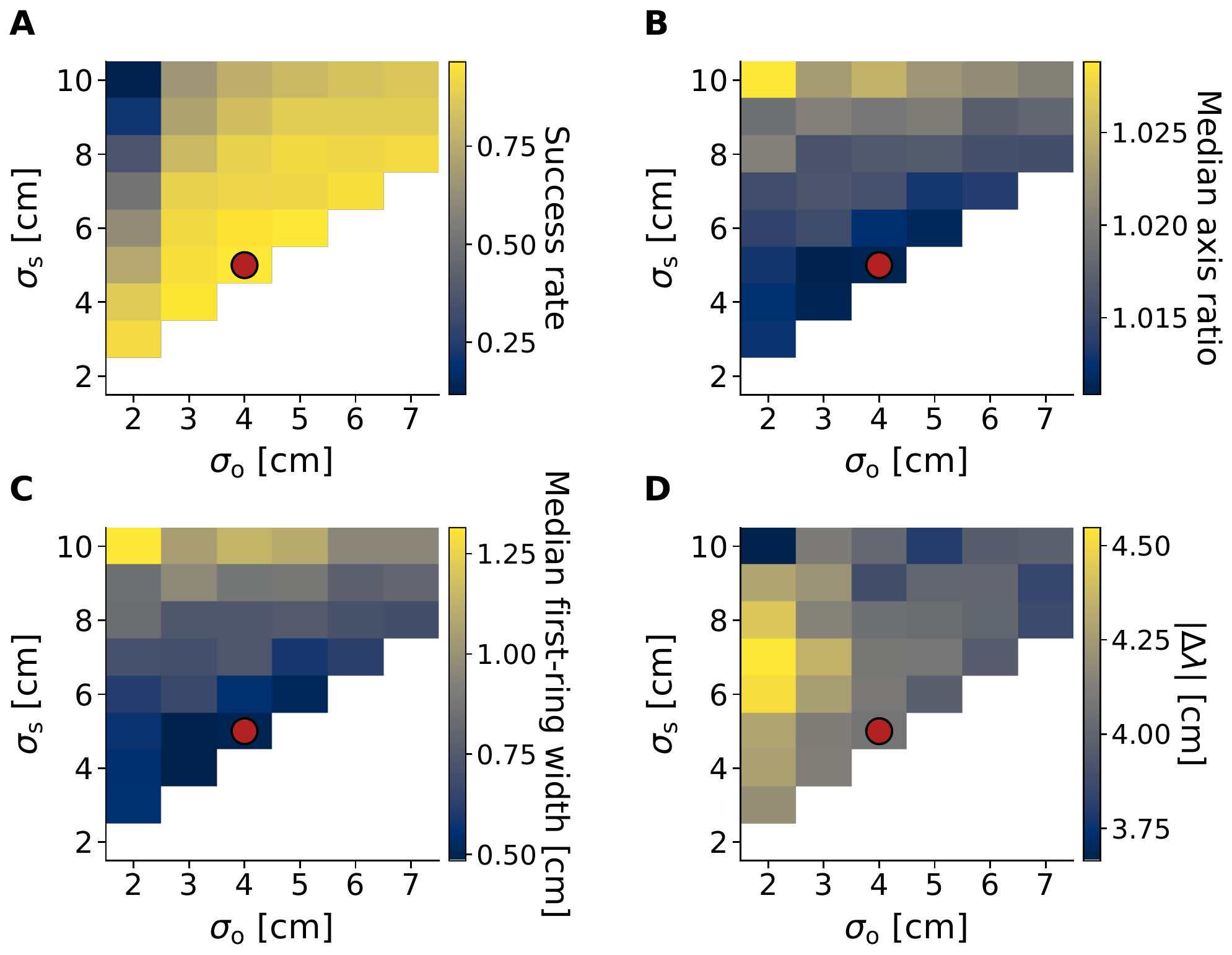}
		\caption[Selection of smoothing bandwidths for 
		rate-map construction.]{\textbf{Evaluation of Gaussian smoothing bandwidths for rate-map construction.}\par\smallskip
			\textbf{(A)} Fraction of eligible cells yielding valid grid-scale estimates across candidate bandwidths for spike-map ($\sigma_s$) and occupancy-map ($\sigma_o$) smoothing.
			\textbf{(B)} Median post-destretch autocorrelogram (ACF) axis ratio across cells.
			\textbf{(C)} Median first-ring ACF peak-distance spread (maximum minus minimum distance of the six first-order peaks).
			\textbf{(D)} Median absolute change in grid scale after affine normalization ($|\Delta\lambda|$).
Feasible pairs satisfied $\sigma_s > \sigma_o$, a success rate $\geq 0.6$, a median post-destretch axis ratio $\leq 1.1$, and axis-ratio IQR $\leq0.1$. Near-optimal candidates were defined as bandwidth pairs within $15$\% of the best feasible first-ring width. The marker denotes the canonical bandwidth pair used in the main analysis ($\sigma_s=\SI{5}{\centi\meter}$, $\sigma_o=\SI{4}{\centi\meter}$). This pair lay within a connected stable near-optimal region. Alternative bandwidths spanning lower and higher smoothing were evaluated in full matched-reference sensitivity analyses.
		}
		\label{fig:bandwidth_sweep}
	\end{figure}

	To evaluate the canonical Gaussian smoothing bandwidths for spike-density and occupancy maps and their local robustness, a systematic grid search was performed (Supplementary \cref{fig:bandwidth_sweep}) over spike-map smoothing $\sigma_s\in[2,10]$~cm and occupancy-map smoothing $\sigma_o\in[2,7]$~cm, both in $\SI{1}{\centi\meter}$ increments. 
		
	For each parameter pair, spatial firing rate maps were constructed as described in Methods and subjected to grid-scale estimation and affine normalization using the Pearson spatial autocorrelation described in \cref{methods-grid-scale-est}. Cells containing fewer than $10{,}000$ spikes were excluded before the sweep, leaving $146$ eligible cells. Parameters were evaluated according to predefined reliability criteria:
	
	\begin{enumerate}
		\item Success rate: At least 60\% of eligible cells yielded valid grid-scale estimates under the distortion-reduced pipeline.
		\item Post-destretch isotropy: Median fitted first-ring ACF ellipse axis ratio had to be $\leq 1.1$, with interquartile range $\leq 0.1$.
		\item Design constraint: $\sigma_s > \sigma_o$, motivated by the sparser sampling of spike events relative to positional occupancy.
	\end{enumerate}
	Within the feasible region defined by these criteria, peak consistency was quantified as the width of the first-ring distance distribution, defined as the difference between the maximum and minimum distances of the six first-order ACF peaks from the central peak. Parameter pairs were considered near-optimal when their median first-ring width was within 15\% of the minimum value among feasible pairs.
	
	To ensure stability, local robustness under $\pm \SI{1}{\centi\meter}$ perturbations in both bandwidth dimensions was additionally required. Specifically, candidate parameter pairs were retained only if at least two neighboring bandwidth pairs also satisfied the near-optimal criterion and exhibited $\leq15\%$ relative deviation in first-ring width.
	
	Under the strict design constraint $\sigma_s>\sigma_o$, the corrected sweep contained seven locally stable near-optimal bandwidth pairs. The canonical $(5,4)$~cm pair was one of these candidates and yielded
	valid geometry estimates for $141/146$ eligible cells ($96.6$\%). At this pair, the median post-destretch ACF axis ratio was $1.011$ (IQR $0.0098$), and the median first-ring peak-distance spread was $\SI{0.496}{\centi\meter}$. The minimum feasible first-ring spread was $\SI{0.485}{\centi\meter}$. The pre-existing $(5,4)$~cm setting was therefore retained as the canonical analysis pair and assessed lower and higher smoothing choices separately in full matched-reference sensitivity analyses. The least-smoothed locally stable candidate was $(3,2)$~cm.	
	
	\subsection{Affine ACF normalization: stability and metric effects}\label{sec:supp-ellipse}
	
	To reduce the global elliptical anisotropy in the spatial autocorrelogram (ACF), an affine normalization (“destretching”) estimated from the six first-order ACF peaks was applied. Because the downstream analyses quantify local, center-aligned angular modulation in annuli, the evaluation tested whether the affine normalization (i) is stable, (ii) reduces ACF anisotropy out-of-sample, and (iii) alters local angular metrics in a manner consistent with removal of global distortion rather than introduction of spurious structure.
	
	\noindent
	For each retained experimental grid cell, matched control simulations were generated using the same matching procedure as in the main analysis, with two simulations per cell for each local-modulation condition ($\beta=0$ (\enquote{circular}) and $\beta=0.2$ (\enquote{sixfold}) in Eq.~\ref{eq:I-hex}). Simulated data were processed using the rate-map smoothing parameters ($\sigma_s=\SI{5}{\centi\meter}$, $\sigma_o=\SI{4}{\centi\meter}$) and the Pearson ACF pipeline described in Methods.
	
\begin{figure}[tbp]
		\centering
		\includegraphics[width=0.9\textwidth]{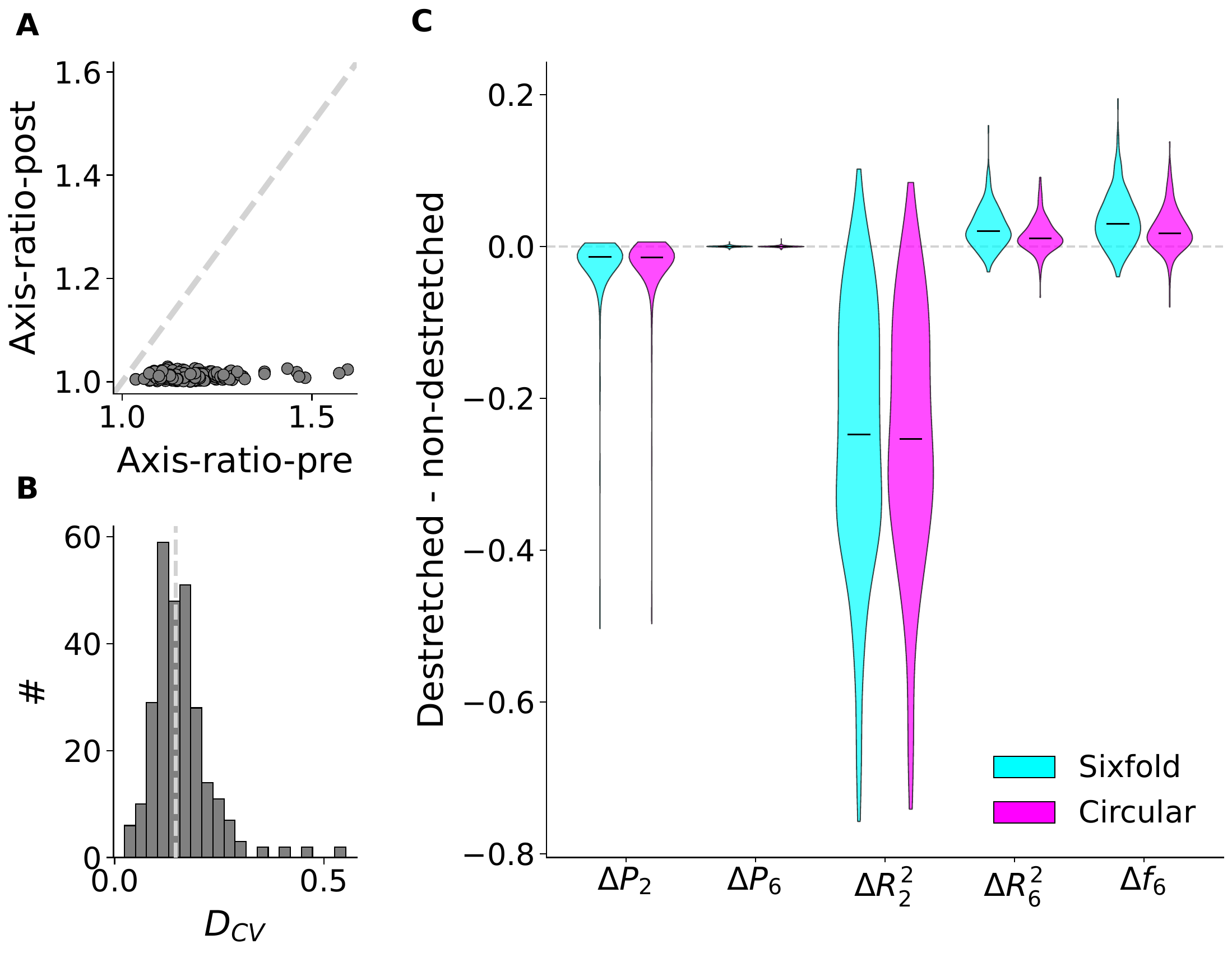}
		\caption[Effects of affine normalization on lattice anisotropy and harmonic metrics.]{\textbf{Affine normalization reduces global lattice anisotropy and selectively suppresses twofold angular structure.}\par\smallskip
			\textbf{(A)} First-ring ACF axis ratio before and after affine normalization in simulated grid fields ($N=274$). Most points fall below the identity line (dashed line), indicating reduced anisotropy.
			\textbf{(B)} Cross-validated reduction in anisotropy after split-half estimation of the affine transform. The transform estimated from one half of the spikes was applied to the complementary half. $D_{\mathrm{CV}}$ denotes the mean reduction across the two cross-applications. Positive values indicate that the estimated transform reduces anisotropy in independent spike samples; the dashed line marks the median.
			\textbf{(C)} Cell-wise changes in inner-range ($\rho \leq 0.4\lambda$) harmonic metrics induced by affine normalization. Violins show the distributions; black horizontal bars mark the medians. Shown are changes (destretched minus original) in twofold power ($\Delta P_2$), sixfold power ($\Delta P_6$), the respective $R^2$ values, and the sixfold power fraction ($\Delta f_6$). Across both circular and sixfold simulations, the normalization reduces twofold structure while absolute sixfold power ($P_6$) remains essentially unchanged. The modest increases in $R_6^2$ and $f_6$ are therefore consistent primarily with removal of competing twofold structure rather than generation of additional sixfold power.
		}
		\label{fig:supp-ellipse}
	\end{figure}

	To assess whether affine normalization reduced known global anisotropy, the $\beta=0.2$ matched controls were first examined. Among simulations yielding valid geometry estimates ($N=274$), affine normalization reduced the first-ring ACF axis ratio in all simulations. The median fitted axis ratio decreased from $1.162$ before normalization to $1.010$ after normalization, corresponding to a median reduction of $0.150$. Before normalization, the estimated axis ratio closely tracked the injected global distortion (Pearson $r=0.982$). After normalization, residual anisotropy retained a weaker association with the injected distortion ($r=0.209$), indicating substantial removal of the imposed global stretch (\cref{fig:supp-ellipse}A).
	
	To test whether the estimated transform generalized beyond the spike sample from which it was obtained, spikes were randomly divided into two halves. Only spike events were split; both halves used the same behavioral occupancy trajectory. The affine transform estimated from one half was applied to the complementary half, and the procedure was repeated in the opposite direction. We used the cross-validated reduction
	$$
	D_{\mathrm{CV}}
	=
	\frac{1}{2}
	\left[
	(r_A-r_{A\leftarrow B})
	+
	(r_B-r_{B\leftarrow A})
	\right],
	$$
	where $r_A$ and $r_B$ denote the raw first-ring axis ratios in the two halves and $r_{A\leftarrow B}$ and $r_{B\leftarrow A}$ the corresponding ratios after applying the transform estimated from the other half. Positive $D_{\mathrm{CV}}$ therefore indicates out-of-sample reduction of anisotropy. Cross-validated reduction was positive in all $274$ simulations, with median $D_{\mathrm{CV}}=0.146$. The median residual deviation of the cross-validated post-normalization axis ratio from circularity was $0.015$ (IQR $[0.011,\,0.020]$; \cref{fig:supp-ellipse}B).
	
	Because the primary analysis is based on center-aligned annular angular profiles, we next quantified the effect of affine normalization on the harmonic metrics used in the main analysis. Destretched and non-destretched analyses were applied to the same matched simulations, allowing paired estimation of the change induced by the affine transform. In both the $\beta=0$ and $\beta=0.2$ controls, normalization strongly reduced twofold harmonic power: median $\Delta P_2=-0.0143$ for $\beta=0$ and $\Delta P_2=-0.0135$ for $\beta=0.2$. In contrast, absolute sixfold power was essentially unchanged (median $\Delta P_6=-2.9\times10^{-5}$ and $8.35\times10^{-5}$, respectively; \cref{fig:supp-ellipse}C).
	
	The corresponding twofold fit strength also decreased substantially (median $\Delta R_2^2=-0.254$ for $\beta=0$ and $-0.248$ for $\beta=0.2$), whereas sixfold fit strength increased only modestly (median $\Delta R_6^2=0.0107$ and $0.0203$, respectively). The sixfold power fraction $f_6$ increased modestly in both conditions (median $\Delta f_6=0.0173$ for $\beta=0$ and $0.0300$ for $\beta=0.2$). Because $f_6=P_6/\sum_{k=2}^{11}P_k$, this increase occurred despite essentially unchanged absolute $P_6$ and is therefore consistent primarily with reduction of competing twofold power in the denominator rather than generation of additional sixfold power.
	
	Although affine normalization reduced twofold harmonic power in the large majority of simulations, a minority showed changes in the opposite direction. The population-level effect was nevertheless strongly biased toward reduced global anisotropy and twofold suppression. Absolute sixfold power, by contrast, showed no systematic directional shift.
	
	Because the preprocessing-induced harmonic shifts were similar in the $\beta=0$ and $\beta=0.2$ controls, local angular metrics were interpreted relative to matched control distributions processed through the identical affine-normalization pipeline. In particular, suppression of $P_2$ can increase relative measures such as $f_6$ without increasing absolute $P_6$; this preprocessing effect is therefore represented in the matched null distributions. Under this null-referenced framework, affine normalization reduces anisotropy-driven twofold structure without systematically increasing absolute sixfold harmonic power.
	
	The affine transform was estimated from the global ACF. Thus, it targeted the lattice-wide affine component of anisotropy rather than local field shape; individual firing fields were not constrained to be circular, and residual local twofold structure could remain after normalization.

	\subsection{Self-organized grid clustering (SOGC)}\label{sec:supp-SOGC}
	
	To identify grid-field centers in a robust and spatially consistent manner, a self-organized grid clustering (SOGC) algorithm was employed. SOGC is a continuous, mode-seeking procedure related to mean-shift clustering, augmented by a weak isotropic repulsive interaction between candidate nodes. The attraction component localizes nodes to firing-density modes, whereas the repulsive component provides finite-range spatial regularization at the scale of a firing field.
	
	Let $x_i \in \mathbb{R}^2$ denote spike locations. To reduce bias caused by nonuniform spatial sampling, each spike was weighted by the inverse smoothed occupancy $w_i \geq 0$ evaluated at its location in the destretched coordinate system; these weights were normalized by their median. Candidate field centers are represented by node positions $c_k \in \mathbb{R}^2$, initialized on a dense regular grid covering the arena.
Nodes are updated iteratively under the influence of two forces:
	\begin{itemize}
		\item \textbf{Attraction to spike density} \\
		Each candidate center is attracted toward regions of high local spike density via a kernel-weighted centroid update, analogous to the mean-shift vector:
		\begin{equation}
			\Delta_k^\mathrm{att} = \frac{\sum_i w_i K_\sigma(\|x_i - c_k\|)x_i}{\sum_i w_i K_\sigma(\|x_i - c_k\|)} - c_k, \quad K_\sigma(r) = e^{-\frac{r^2}{2\sigma^2}}	
		\end{equation}
This term causes candidate centers to ascend the local spike-density landscape and converge toward firing field maxima.
		
		\item \textbf{Repulsion between nearby centers} \\
		A weak repulsive interaction was introduced between candidate centers:
		\begin{equation}
			\Delta_k^\mathrm{rep} = \sum_{l \neq k} G_\rho(\|c_k - c_l\|) \frac{c_k - c_l}{\|c_k - c_l \|+\epsilon}, \quad G_\rho(r) = e^{-\frac{r^2}{2\rho^2}}
		\end{equation}
where $\rho$ controls the spatial scale of repulsion and $\epsilon$ avoids numerical singularities. Because the displacement vector $c_k - c_l$ points away from the neighboring node, this term produces an outward displacement when added to the update. The interaction discourages local crowding and acts as a weak spatial regularizer at the field scale. It is isotropic and does not itself specify a lattice orientation or lattice phase.
		
		\item \textbf{Update rule and convergence} \\
		Candidate centers are updated according to:
		\begin{equation}
			c_k \leftarrow c_k + \eta_\mathrm{att} \Delta_k^\mathrm{att} + \eta_\mathrm{rep} \Delta_k^\mathrm{rep}
		\end{equation}
where $\eta_\mathrm{att}$ and $\eta_\mathrm{rep}$ control the relative update strengths. Both interactions are truncated beyond finite cut-off distances for computational efficiency. After each update, convergence was assessed from the maximum node displacement; if convergence had not been reached, nodes separated by less than the merging radius $r_\mathrm{merge}$ were merged before the next iteration. Iteration terminated when the displacement criterion was met or the maximum number of iterations was reached.
		
		When the repulsion term is omitted ($\eta_\mathrm{rep}=0$), the mode-seeking component corresponds to occupancy-weighted, multi-seed mean-shift-like clustering followed by merging of nearby nodes.
		
		\item \textbf{Occupancy correction} \\
		Occupancy correction affects only the attraction term by modulating the weights $w_i$, thereby reducing the influence of spatially nonuniform sampling on the attraction step. Repulsion is treated as a structural prior and remains independent of occupancy.
		
	\end{itemize}
	For the analyses reported here, candidate nodes were initialized on a regular grid with spacing $0.25\lambda$. The attraction kernel used $\sigma=0.2\lambda$, an attraction cutoff of $0.6\lambda$, and $\eta_\mathrm{att}=1$. The weaker repulsive interaction used $\rho=0.6\lambda$, a cutoff of $1.2\lambda$, and $\eta_\mathrm{rep}=0.1$. Nodes separated by less than $0.2\lambda$ were merged. Iteration was limited to 50 updates, with convergence defined as a maximum node displacement below $10^{-4}\lambda$.
	
	All SOGC operations were performed in the destretched coordinate system, and detected centers were subsequently transformed back to the original spatial coordinates.
	
	The interaction range and strength were evaluated using synthetic grid-like spike data with known field-center locations. Parameter variants were compared based on field-center recovery rather than on the experimental sixfold-modulation statistics, which were not used as a selection criterion. These simulations were used to verify accurate center recovery and to identify a weak regularization regime rather than to establish a unique optimal parameterization. Because the synthetic benchmarks assumed ideal triangular lattice geometry, they do not establish optimality for the full range of local lattice irregularities present in experimental grid cells.
	
	\subsection{Lattice-constrained filtering of candidate centers}\label{sec:supp-lattice-fit}
	
	SOGC may occasionally identify spurious candidate centers arising from noise fluctuations or partial fields near arena boundaries. To suppress such detections and assess consistency with the global grid structure, a lattice-constrained filtering step was applied based on the estimated grid scale.
	
	Using the grid spacing $\lambda$ obtained from the spatial autocorrelogram, an anchor-free sixfold lattice was fitted to the set of candidate centers. The spacing was held fixed at the ACF-derived value throughout this fit; only lattice orientation and translation were optimized. A hexagonal lattice is fully specified by a spacing $\lambda$, an orientation $\theta$, and a translation (phase) $\alpha \in [0, 1]^2$ in lattice coordinates. The lattice basis vectors were defined as
	\begin{equation}
		B = [b_1, b_2], \quad 
		b_1 = \lambda 
		\begin{pmatrix}
			\mathrm{cos} (\theta) \\
			\mathrm{sin} (\theta)
		\end{pmatrix},
		\quad
		b_2 = \lambda 
		\begin{pmatrix}
			\mathrm{cos}(\theta + \pi/3) \\
			\mathrm{sin}(\theta + \pi/3)
		\end{pmatrix}.
	\end{equation}
	Candidate centers $X \in \mathbb{R}^{2 \times N}$ were transformed into lattice coordinates $U = B^{-1}X$. For ideal lattice points, the coordinates satisfy
	\begin{equation}
		u_i = n_i + \alpha, \quad n_i \in \mathbb{Z}^2,
	\end{equation}
	up to small noise. The lattice translation $\alpha$ was estimated robustly using a two-stage procedure on the torus $[0, 1)^2$: a coarse histogram-based mode estimate followed by a weighted circular mean that down-weighted outliers based on lattice residuals.
	
	For each candidate center, the Euclidean distance to the nearest fitted lattice site was computed. Centers whose distance from the nearest fitted lattice site exceeded $0.3\lambda$ were discarded. The fitted lattice therefore served as a consistency filter for candidate-center selection. To preserve biological variability, retained centers were not displaced toward their nearest lattice sites; subsequent annulus-based analyses were centered on the empirically detected SOGC locations. In particular, only the most central eligible field per cell was retained for downstream analyses, so the fitted lattice was not used to redefine the geometry of the analyzed field.

	A reliable estimation of the lattice translation and orientation requires multiple candidate centers per cell. The lattice-fitting procedure therefore assumes that a majority of detected centers correspond to true firing fields, while allowing for a minority of outliers arising from noise or boundary truncation. To increase the robustness of the algorithm, lattice phase estimation was performed using a two-stage procedure that combines a coarse, histogram-based estimate on the torus with a subsequent weighted circular mean that down-weights centers with large lattice residuals. This approach reduces the sensitivity to spurious detections while preserving an accurate alignment when true lattice structure is present.

	\subsection{Angle-resolved firing-rate estimation within grid fields}\label{sec:supp-annulus-rate-estimation}
	
	To quantify the angular structure within grid fields without imposing hard spatial boundaries, firing activity was analyzed in polar coordinates relative to each detected field center.
	
	\begin{itemize}
		\item \textbf{Polar binning} \\
		Spike counts and occupancies were accumulated in fine-grained polar bins with a resolution of $\ang{1}$. Radial distances were expressed in units of the destretched grid scale $\lambda$. Expressing radius relative to $\lambda$ assumes that field width scales
		approximately with grid scale. Residual between-cell variation in relative field width was represented in the matched simulations through the cell-specific fitted field width. Population inference then integrated $Z$ across the predefined radial window rather than testing individual radii separately. Occupancy maps were normalized to unit mass prior to analysis, such that angular structure reflects relative spatial modulation rather than absolute firing rate.

		\item \textbf{Radial smoothing} \\
		Radial smoothing was performed using a symmetric sliding rectangular window with a fixed relative-radius width of $0.11 \lambda$. The window size was constrained to be odd in bin units to maintain centered annuli. Annuli were evaluated over a predefined radial range (for the main analysis: $0.1\lambda$--$0.4\lambda$), stepped in increments of $0.01 \lambda$.
		
		\item \textbf{Angular smoothing} \\
		Angular smoothing was performed using a circular Hann window (default width $\ang{29}$). Convolution was implemented with circular wrap-around to avoid edge artifacts at $\ang{0}/\ang{360}$. The chosen width balances angular resolution and sampling noise while preserving periodic structure at angular scales larger than the window width, including sixfold (\ang{60}) modulation.
		
		\item \textbf{Occupancy correction and regularization} \\
		Angular firing rates were computed as:
		$$
		r = \frac{S+\alpha}{O+\beta},
		$$
		where $S$ and $O$ denote smoothed spike and occupancy mass, respectively. To stabilize estimates in sparsely sampled bins, small pseudocounts $\alpha$ and $\beta$ were included after smoothing. For the reported analyses, adaptive regularization was enabled. $\beta$ was set as a fixed fraction of the median non-zero occupancy ($\beta = 10^{-3} \mathrm{median}(O_+)$), and $\alpha$ was scaled to the median firing rate in well-sampled bins ($\alpha = r_0 \beta$). Under typical sampling densities, pseudocounts were negligible relative to empirical occupancy mass and served only to regularize low-occupancy regions.
		
		\item \textbf{Coordinate transformations} \\
		All polar binning and smoothing were performed in the distortion-reduced coordinate system described in
		\cref{methods-grid-scale-est}. Spike coordinates, occupancy maps, and field centers were transformed prior to angular analysis to perform the analysis in the distortion-reduced coordinate system.
		
	\end{itemize}
	As expected for a finite-width Hann kernel, angular smoothing acts as a low-pass filter. For the canonical $29$-bin kernel, the power transfer was approximately $0.969$ at $k=2$, $0.752$ at $k=6$, and $0.370$ at $k=11$. A pure-harmonic diagnostic showed negligible cross-frequency leakage, confirming that smoothing attenuates higher harmonics without generating power at a different harmonic order. This frequency response is therefore retained in, and calibrated by, the matched reference distributions used for inference.
	
	\subsection{Sign-based population sensitivity analysis}\label{sec:supp-sign-test}
	
	The primary population analysis used a sign-flip permutation test of the mean cell-wise $AUC(Z)$. Because the test statistic is the mean, it retains information about effect magnitude: a smaller subset of cells with large deviations can therefore contribute substantially to the population statistic. This choice was appropriate for the primary biological question, which allowed either a broadly distributed weak modulation or stronger modulation in only a subset of cells.
	
	As a complementary sensitivity analysis, we also asked the distinct question of whether cell-wise deviations showed a consistent direction irrespective of their magnitude. For each reference condition, with $N$ cells, and $A_i$ denoting the $AUC(Z)$ of the corresponding null-standardized curve over $0.1\lambda$--$0.4\lambda$, we counted:
	\begin{equation}
		B = \sum_i {\mathbf{1}(A_i > 0)}
	\end{equation} 
	and evaluated $B$ using an exact two-sided binomial test under $B \sim \mathrm{Binomial}(N, 0.5)$. Under this null, each cell is equally likely to yield a positive or negative $A_i$, such that the expected fraction of positive values is $0.5$; finite-sample departures from this proportion are described by the binomial distribution. Unlike the mean sign-flip statistic, this sign test assigns the same weight to a small and a large positive deviation and therefore tests directional consistency across cells rather than the mean magnitude-weighted effect.
	
	For the primary $f_6$ comparison with the circular reference, $58/137$ cells had positive and $79/137$ negative $AUC(Z)$ values (two-sided exact sign test, $p=0.087$), consistent with the absence of a population-wide directional shift. Relative to the imposed-sixfold reference, only $28/137$ cells were positive and $109/137$ were negative ($p=1.74\times10^{-12}$), providing complementary evidence for a consistent directional separation from the sixfold control.

	\subsection{Exploratory analysis of radius-dependent deviations}
	\label{sec:supp-radius-excursion}
	
	The negative experimental--circular deviation toward the outer boundary of the local-field window was examined post hoc because it was visually apparent in the radius-resolved profile (\cref{fig:fig3}B), although radius-wise values were not used for primary inference. Cells were first stratified by the
	experimental field-width estimate. Among cells with	$\sigma/\lambda\leq0.20$ ($N=95$), the experimental--circular deviation over $0.3\lambda$--$0.4\lambda$ was weaker and non-significant
	(mean $AUC(Z)=-0.0103$, $p=0.089$), whereas cells with $\sigma/\lambda>0.20$ ($N=42$) showed a larger negative deviation (mean $=-0.0317$, $p=0.0012$). The difference between the broad- and narrow-field groups in the raw experimental--circular $f_6$ contrast was significant in this binary post hoc comparison ($p=0.028$), although fitted field width was not significantly associated with the contrast as a continuous variable.
	Importantly, over the predefined $0.1\lambda$--$0.4\lambda$ analysis window, both width strata remained consistent with the circular reference and below the imposed-sixfold reference. Thus, the width-stratified effect was localized toward the outer field and did not alter the primary full-window inference.
	
	The same broad-field regime showed the modest compression in recovered field width visible in Supplementary \cref{fig:supp-model-matching}. Because each matched simulation assigns the selected cell's field-width estimate to all fields of the regular lattice, we tested whether overlap among neighboring modeled fields was itself sufficient to produce this compression. In an idealized diagnostic, we set $\lambda=1$, removed spike sampling, occupancy weighting, background	activity, affine transformations, and ACF estimation, and compared an isolated Gaussian field with identical Gaussian fields tiled on a regular hexagonal	lattice. The resulting radial profiles were fitted over $0.1\lambda$--$0.5\lambda$ using the same Gaussian-plus-baseline width model as in the experimental analysis. Isolated Gaussian fields recovered the generating width essentially exactly. In contrast, tiled fields showed increasing negative recovery bias as the generating width increased: approximately $-2.9\%$ at $\sigma/\lambda=0.20$, $-5.8\%$ at $0.22$, and $-9.6\%$ at $0.24$. This range encompasses the approximately $-7.3\%$ mean signed recovery error in absolute field width observed among the experimental cells with $\sigma/\lambda>0.20$. Thus, overlap from neighboring modeled fields is sufficient to produce recovery compression at the broad end of the width range without invoking a failure of the width-fitting algorithm for isolated fields.
		
	Neighboring-field overlap also affected the angular structure of the matched references. Within $0.3\lambda$--$0.4\lambda$, outer $f_6$ in the circular reference increased with the matched experimental field width	(Spearman $\rho=0.274$, $p=0.0012$), as did sixfold regression strength $R_6^2$ ($\rho=0.275$, $p=0.0012$). The broader-field circular-reference subset likewise had slightly larger outer $f_6$ than the narrower subset ($p=0.029$). This behavior is expected because increasing $\sigma/\lambda$ increases the contribution of the six regularly positioned neighboring modeled fields to outer annuli. Experimental neighboring fields need not share the selected central field's width or regular Gaussian shape,	and can additionally differ in width, amplitude, position, and directional extension. Accordingly, this diagnostic identifies a contribution from the matched model to the outer mismatch but does not assign a unique mechanism to the experimental component.
	
	\noindent
	The outer-radius shift was also sensitive to harmonic composition. When $P_2$ was excluded from the $f_6$ denominator, the experimental--circular deviation over $0.3\lambda$--$0.4\lambda$ was no longer significant in the full	population ($p=0.112$), in the narrower-field subset ($p=0.553$), or in the	broader-field subset ($p=0.063$). $R_6^2$ retained a negative outer-radius deviation. Thus, competing low-order structure contributes substantially to the visible $f_6$ decline, but its compositional denominator does not account for the entire outer-radius difference.
	
	The small systematic grid-scale recovery offset was also examined as a possible	source of the outer-radius deviation. Across cells, the mean recovered simulation grid scale was approximately $1.3\%$ larger than the matched experimental value. Re-expressing the stored circular-reference radial profiles approximately on the scale defined by the experimental input $\lambda$ had a negligible effect on the  $0.3\lambda$--$0.4\lambda$ comparison: the absolute change in mean $AUC(Z)$ was $<1.1\times10^{-4}$ in the full population and $<4\times10^{-4}$ in the broader-field subgroup. The small grid-scale recovery bias therefore cannot account for the outer-radius deviation.
	
	Together, these diagnostics indicate that the outer-radius feature occurs where contributions from neighboring fields become increasingly important. For the matched model, propagating a single cell-specific width to identical untruncated fields on a regular lattice produces both predictable compression of recovered broad-field widths and a modest increase in outer sixfold specificity. Real fields and their neighbors are more heterogeneous, providing additional possible sources of radius-dependent angular structure. We therefore regard the outer excursion as a localized sensitivity to neighboring-field structure and idealized model assumptions rather than as a separate inferential result. Crucially, neither the narrower- nor the broader-field subset showed a positive experimental--circular shift over the predefined local-field window, and experimental fields remained below	the imposed-sixfold reference in both subsets.
	
	\subsection{Monte Carlo convergence of matched reference distributions}
	
	To assess convergence of the cell-specific reference calibration, the primary analysis was repeated on the fixed final population ($N=137$) using the first $25$, $50$, and $75$ successful matched replicates and compared with the complete retained reference bank. For the experimental--circular comparison, mean $AUC(Z)$ values were $-0.00749$, $-0.01247$, $-0.01271$, and $-0.01376$, with corresponding sign-flip $p$-values of $0.560$, $0.320$, $0.316$, and $0.279$, respectively. Cell-wise effects correlated with the complete-bank estimates at $r=0.977$, $0.992$, and $0.998$ using 25, 50, and 75 successful replicates, respectively, indicating stable population inference at the final simulation depth.

	\subsection{Burst-event and temporal-phenotype analyses}
	\label{sec:supp-temporal-modes}

	\subsubsection{Burst-event sampling geometry and matched references}
	
	The burst analysis retained the same $31$ reported radial centers from $0.1\lambda$ to $0.4\lambda$ as the canonical analysis. The annulus estimator is constructed from $0.01\lambda$ radial bins, and the nominal $0.10\lambda$ rectangular radial smoothing window is represented by an odd number of bins, yielding $11$ bins ($0.11\lambda$ total width, $0.055\lambda$ half-width; Section~\ref{sec:supp-annulus-rate-estimation}). Consequently, as is the case in the canonical analysis, events contributing to the complete set of reported radial estimates span $0.045\lambda$--$0.455\lambda$ around the frozen field center. 
	
	The $0.05\lambda$ radial strata introduced for the burst analysis are separate from both the $0.01\lambda$ estimator bins and the overlapping radial analysis windows. They were non-overlapping conditioning strata with edges $0.045$, $0.05$, $0.10$, $0.15$, $\ldots$, $0.45$, and $0.455\lambda$.	For each cell, the empirical burst-onset count was determined independently in	each of these strata.
	
	The existing matched-model parameter bank supplied the cell-specific grid
	scale, field width, transform, selected center, occupancy map, and spatial support. The all-spike simulated event realizations themselves were not reused. Instead, new burst-specific circular and sixfold reference event sets were drawn from the corresponding occupancy-weighted spatial model, restricted to the frozen local support. Each reference realization contained exactly the empirical burst count of that cell in every radial conditioning stratum; polar angle was not conditioned. Reference events were then passed directly to the frozen annulus estimator, without re-estimation of an autocorrelogram, grid scale, distortion transform, or field center. As in the canonical analysis, $10$\% of the generating mixture consisted of a uniform component defined over the physical arena. However, this uniform component was restricted and renormalized separately within each radial stratum. Consequently, the realized fraction of retained events originating from this uniform component was not constrained to equal $10$\% within an individual reference realization.
	
	The production reference bank contained exactly $100$ circular and $100$ sixfold realizations for every one of the $108$ primary cells ($10{,}800$ realizations per condition). Because geometric recovery was not	rerun, the $\geq75$ successful-replicate criterion of the canonical matched simulation analysis was not used. Instead, complete $100$-replicate coverage was required for each cell and condition.

	\subsubsection{Burst--vs--tonic matching and sampling sensitivities}
	
	For radial matching, let $B_j$ and $T_j$ denote the available numbers of burst onsets and tonic spikes, respectively, in radial stratum $j$. Each paired realization sampled
	$
	M_j=\min(B_j,T_j)
	$
	observations without replacement from each firing mode. Burst and tonic samples therefore had exactly the same event count within every radial stratum, and hence the same total sample size and binned radial distribution, while their local polar angles remained unconstrained. $25$ such paired realizations were generated per cell. The burst-minus-tonic $AUC(f_6)$ difference was calculated separately for every realization and then averaged within each cell before the population sign-flip test; the $25$ resamples therefore did not increase the inferential sample size.
	
	The primary sampling criterion retained $108$ cells. As a stricter sensitivity analysis, at least $50$ burst events within every sliding radial analysis window were required, retaining $90$ cells. This criterion concerned the total event support of each radial window rather than individual radius--angle bins. A separate direct-comparability analysis restricted the burst population to cells also present in the canonical $137$-cell all-spike population, retaining $105$ cells. The experimental--circular $f_6$ comparison was negative (mean $AUC(Z)=-0.0355$, $p=0.012$), and burst-event $f_6$ remained below the imposed-sixfold reference (mean $=-0.0550$,	$p=1.0\times10^{-4}$). Under the stricter sampling criterion, the negative circular-reference departure weakened and was no longer significant (mean $=-0.0275$, $p=0.082$), whereas the comparison with the imposed-sixfold	reference remained negative ($p=0.00125$). Restriction to the canonical	all-spike population gave mean $AUC(Z)=-0.0305$ relative to circular ($p=0.0286$) and $-0.0501$ relative to the sixfold reference ($p=2.5\times10^{-4}$). The complementary $R_6^2$ analysis showed the same sampling dependence: $p=0.015$ relative to circular in the primary population, $p=0.114$ under the stricter sampling criterion, and $p=0.0388$ in the canonical-population subset.
	
	\subsubsection{Temporal-ACF phenotype analysis}
	
	Temporal grid-cell phenotypes were defined by applying the classification procedure accompanying \textcite{gardnerToroidalTopologyPopulation2022}. Temporal autocorrelograms were computed over lags up to $\SI{200}{\milli\second}$ using $\SI{1}{\milli\second}$ bins, normalized by the first lag bin with that bin subsequently set to zero, and Gaussian-smoothed
	with a standard deviation of four bins. Pairwise cosine distances were used to construct an $80$-nearest-neighbor graph, and clustering was performed using the ToMATo procedure used in the released analysis code. The resulting cluster identities were matched to the bursty (B), theta-modulated (T), and non-bursty (N) reference phenotypes.
	
	As an implementation check, the pooled reference context used by the released classification analysis was first reconstructed: R1--R3 from rat R day 2 together with Q1, Q2, and S1 ($N=847$). This yielded $524$ bursty, $229$ theta-modulated, and $94$ non-bursty cells, compared with the reported pooled counts of $523$, $229$, and $95$, respectively. The one-cell discrepancy was unchanged across GUDHI versions 3.3.0, 3.4.0, 3.4.1, 3.4.1.post1, and 3.5.0, and no clustering parameter was adjusted to force agreement.
	
	For application to the recordings analyzed here, the three R day-2 modules in this pooled context were replaced by their day-1 counterparts while Q1, Q2,	and S1 were retained. The resulting three day-1 cluster identities were arbitrary labels; they were assigned one-to-one to the bursty, theta-modulated, and non-bursty phenotypes by minimizing cosine distance between the day-1 mean temporal autocorrelograms and the corresponding reconstructed reference prototypes.

	Across all $166$ R1 day-1 cells, the resulting classes contained $54$ bursty, $88$ theta-modulated, and $24$ non-bursty cells. All $137$ cells in the canonical sixfold-analysis population could be classified, yielding $41$, $75$, and $21$ cells, respectively. Between-class differences in the primary all-spike $AUC(Z)$ statistic were evaluated using a one-way permutation ANOVA ($F=0.517$, $p=0.596$, $\eta^2=0.0077$). The complementary $R_6^2$ endpoint likewise showed no detectable class difference ($F=0.365$, $p=0.696$, $\eta^2=0.0054$).

	\subsection{Burak--Fiete comparison: additional diagnostics and scope}
	\label{sec:supp-CAN-diagnostics}
	
	The canonical and width-matched BF conditions differ substantially in relative field width, which affects how much of the selected central field contributes across the common $0.1\lambda$--$0.4\lambda$ analysis window. For canonical BF, the median $\sigma/\lambda=0.109$ implies that $0.3\lambda\approx2.75\sigma$ and $0.4\lambda\approx3.67\sigma$; the central field therefore contributes little firing activity toward the outer part of the window. In the width-matched condition ($\sigma/\lambda=0.187$), $0.4\lambda\approx2.14\sigma$, so the selected field remains appreciable over a larger fraction of the same relative-radius interval. This difference is also reflected in the matched imposed-sixfold references (Supplementary \cref{fig:supp-CAN}B): the sixfold-reference $f_6$ decreases strongly with radius in canonical BF but remains more stable after width matching. The imposed modulation parameter specifies the local generating field, not a radius-independent recovered value of $f_6$ after full lattice tiling, finite event sampling, occupancy correction, and harmonic normalization. The width-matched condition was therefore included to test whether the primary conclusion depended on the unusually narrow fields of the released canonical parameterization rather than to optimize angular symmetry.
	
	Observation depth also differed substantially by design between the two BF conditions. The canonical full-trajectory observations contained a median of $28{,}827$ events per readout (IQR $19{,}256$--$39{,}099$) over $8475.99$~s of valid trajectory support, corresponding to a median observation rate of $3.40$~Hz (IQR $2.27$--$4.61$~Hz). In contrast, the 30-min width-matched observations contained a median of $4{,}968$ events (IQR $3{,}205$--$7{,}175$) over $1800$~s, corresponding to $2.76$~Hz (IQR $1.78$--$3.99$~Hz). These event counts were inherited from the corresponding experimental spike counts over the same temporal supports and therefore describe the finite observation model, not intrinsic BF firing rates.
	
	The height of the imposed-sixfold reference curves should therefore not be compared directly between the two BF conditions as though the same imposed modulation strength were expected to yield the same recovered $f_6$ value. The same local sixfold prior is observed after condition-specific finite-event sampling and lattice tiling. The substantially lower event count of the width-matched condition increases the contribution of sampling variability across harmonic orders, while its broader fields increase overlap between neighboring lattice fields across the analyzed radial window. In particular, lattice-derived sixfold structure contributes to both the circular and imposed-sixfold references as neighboring fields become more influential, reducing the incremental fractional prominence of the imposed local modulation. These contributions were not decomposed separately. Importantly, the imposed-sixfold reference nevertheless remained clearly distinguishable from the width-matched BF observations (mean $AUC(Z)=-0.0532, p=5.0\times{10^{-5}}$), showing that the analysis retained sensitivity to the imposed modulation despite the smaller visual $k=6$ peak.
	
	When the width-matched parameterization was extended beyond the predefined 30-min sensitivity window under the same uniform post-initialization input condition, its spatial registration deteriorated substantially. Median correlation between the dynamic and ideally integrated spatial maps decreased from $0.808$ at 30~min to $0.437$ at 60~min and $0.149$ at~90 min, with a value of $0.298$ at 120~min; median ordinary gridness likewise decreased from $0.848$ to $0.286$, $0.370$, and $0.125$, respectively. Because these measures were obtained from cumulative maps, their later non-monotonicity does not imply recovery of accurate path integration. The width-matched condition was therefore used for local-symmetry inference only over the first 30~min. We treat the longer-duration deterioration as a model-specific limitation of this parameter regime rather than as evidence for a general field-width--stability trade-off in continuous-attractor networks. Environmental boundaries can provide error-correcting information for grid representations \parencite{hardcastleEnvironmentalBoundaries2015}, and structured interactions between grid modules can reduce deterministic attractor drift \parencite{sahniInteractionsContinuousAttractors2024}. More elaborate attractor-based architectures also include mechanisms absent from the present single-module BF implementation \parencite{jiSystemsModelAlternating2025}. Whether such mechanisms alter local field microstructure was not tested here.
	
	\FloatBarrier
	\clearpage 
	
	\pdfbookmark[2]{Supplementary Figures}{supp-figures}
	\subsection*{Supplementary Figures}

	\begin{figure}[tbp]
		\centering
		\includegraphics[width=0.9\textwidth]{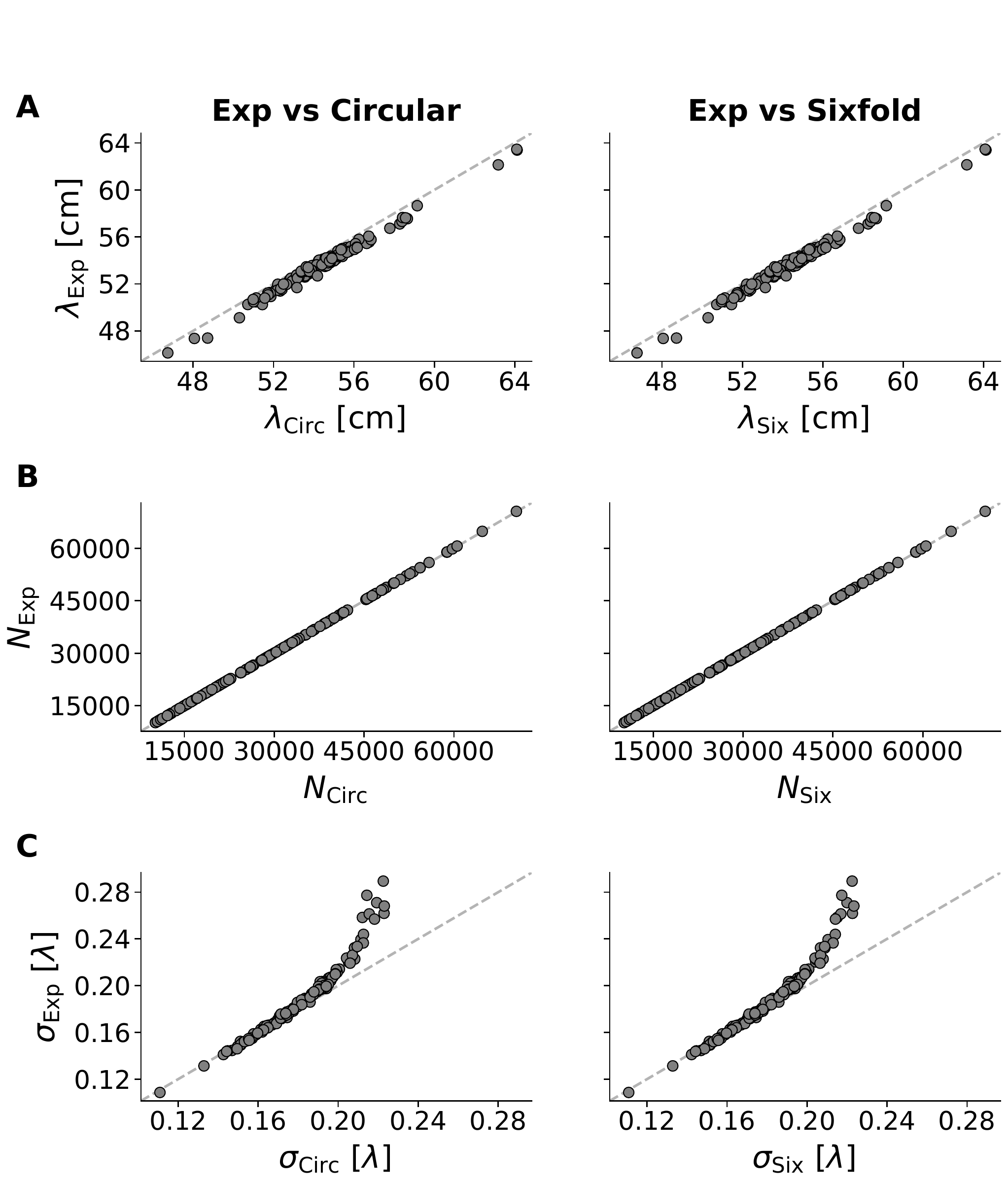}
		\caption[Cell-wise parameter matching for null simulations.]{\textbf{Matched simulations closely recover per-cell geometric and sampling parameters.}\par\smallskip
			\textbf{(A)} Recovered grid scale ($\lambda$) in circular (left) and sixfold (right) simulations versus experimental estimates used as simulation input parameters. Points represent cells (simulation values averaged across replicates). Dashed line indicates unity. Across the final $N=137$ cells, median absolute grid-scale recovery errors were $\SI{0.697}{\centi\meter}$ and $\SI{0.707}{\centi\meter}$ for the circular and sixfold-prior references, respectively.
			\textbf{(B)} Total spike count per cell in simulations versus experimental data. The requested spike count was matched before inverse transformation and arena cropping; recovered spike counts remained close to the experimental values after cropping.
			\textbf{(C)} Recovered local-field width ($\sigma$), estimated from the radial envelope fit over $0.1\lambda$--$0.5\lambda$. Recovered values closely matched experimental estimates overall
			(median absolute relative recovery error: $1.77\%$ and $1.85\%$ for circular and sixfold simulations, respectively), with some compression at the broad end of the observed width range. 
		}
		\label{fig:supp-model-matching}
	\end{figure}

	\begin{figure}[tbp]
		\centering
		\includegraphics[width=0.9\textwidth]{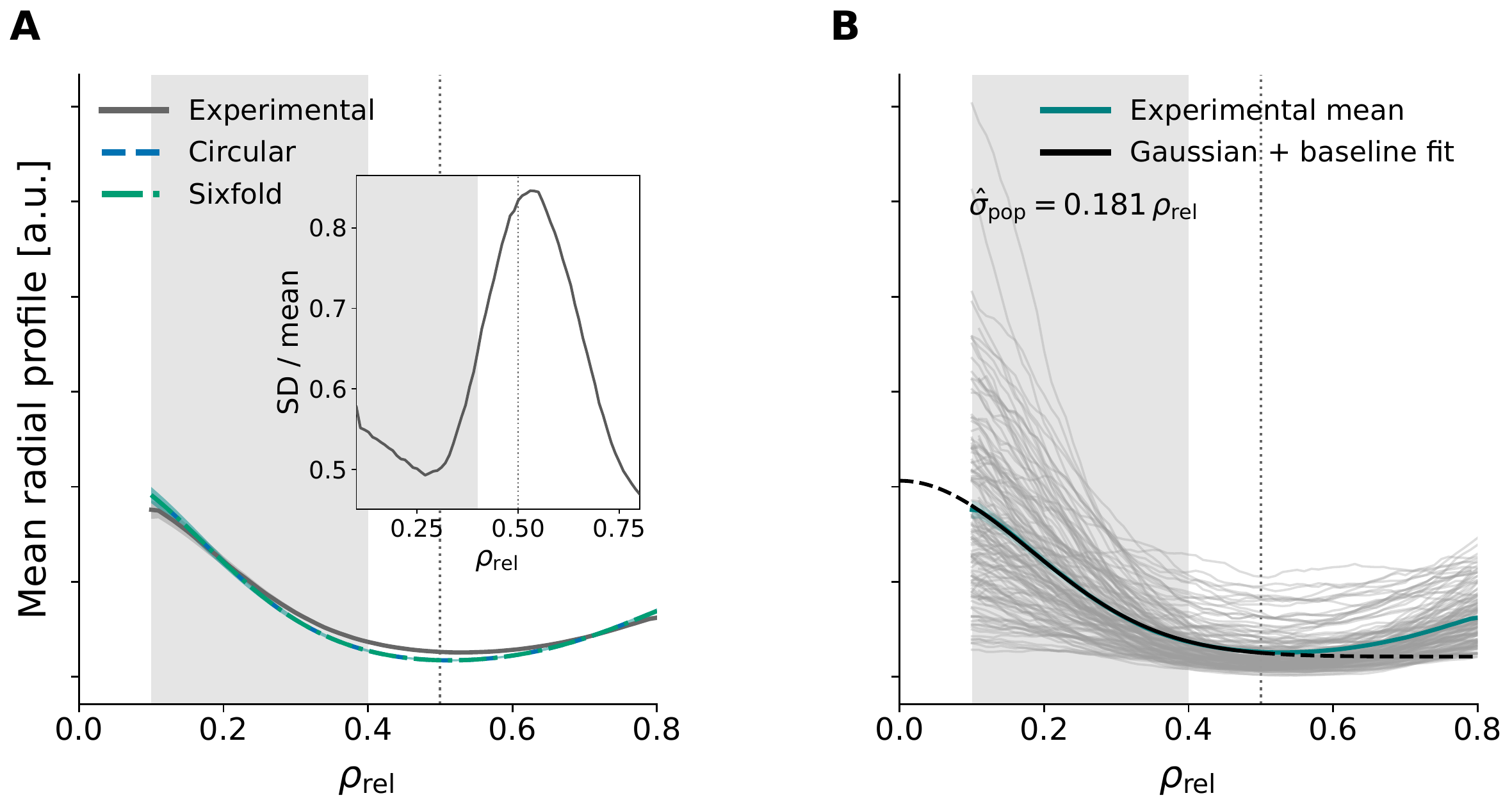}
		\caption[Population radial envelope and Gaussian-envelope fit.]{\textbf{Population radial envelope and Gaussian-envelope fit.}\par\smallskip
			\textbf{(A)} Population mean ($\pm$ SEM across cells, $N=137$) of the annulus-averaged firing rate $\langle r(\rho,\theta)\rangle_\theta$ for experimental cells and matched circular and sixfold-prior simulations. Radii are expressed relative to grid scale ($\rho_{\mathrm{rel}}=\rho/\lambda$). Dotted line: upper bound of the field-width fit window. Shaded region indicates the predefined local-field window ($0.1\lambda$--$0.4\lambda$). The population mean reaches its minimum around $0.5\lambda$ and rises at larger radii, consistent with increasing contributions from neighboring lattice fields. Inset: across-cell SD of the experimental radial profile normalized by the population mean, showing increasing relative across-cell variability outside the local-field window.
			\textbf{(B)} Experimental data: faint gray lines are individual cells; teal curve and shading show the population mean $\pm$ SEM. Black curve is the Gaussian-plus-baseline fit for the population mean; solid in the fit window ($0.1\lambda$--$0.5\lambda$), dashed where extrapolated. The fitted width $\hat{\sigma}_{\mathrm{pop}}$ is shown.
		}
		\label{fig:supp-gauss-field_width}
	\end{figure}
	
	\begin{figure}[tbp]
		\centering
		\includegraphics[width=0.9\textwidth]{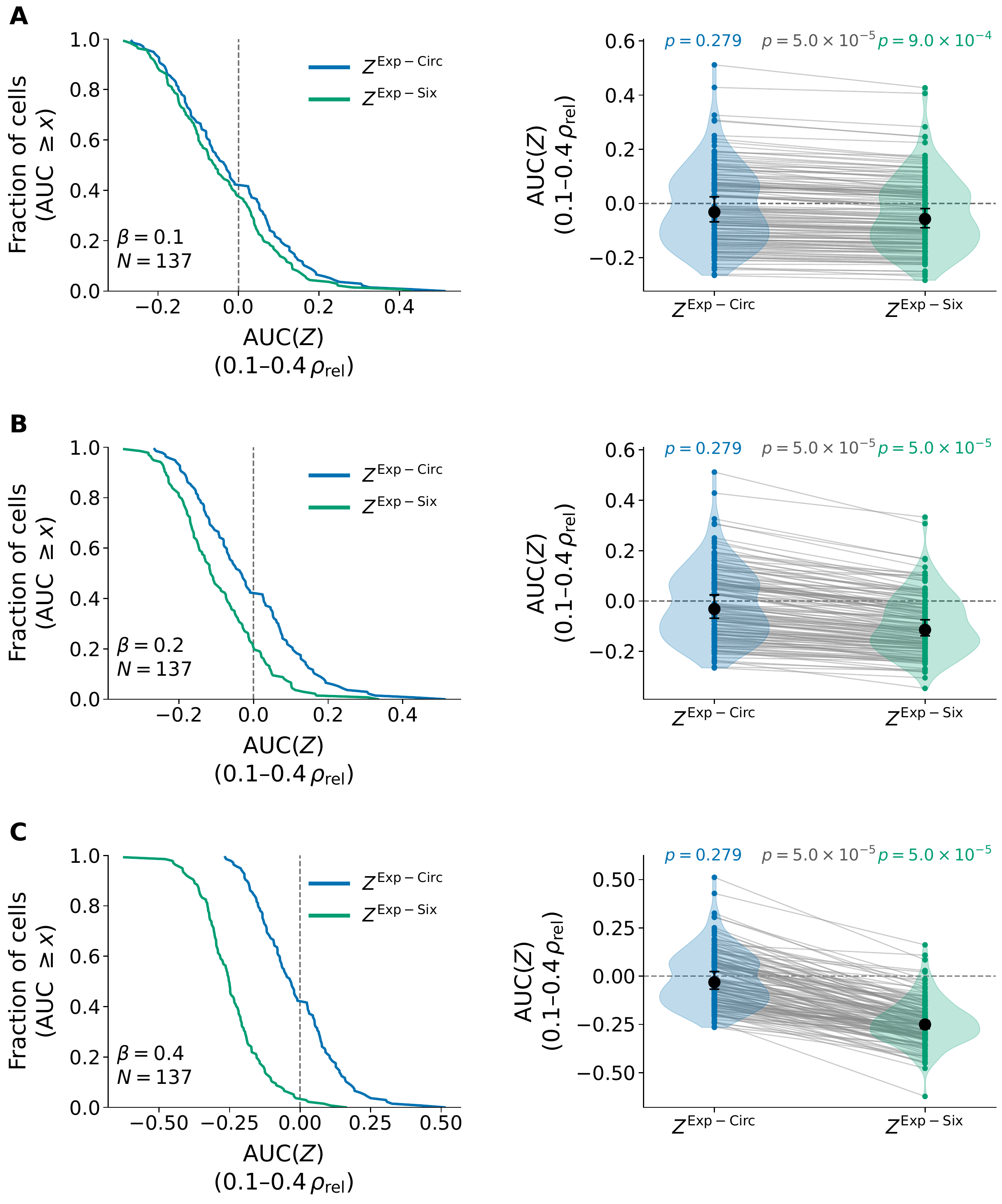}
		\caption[Detectability of imposed local sixfold modulation (beta sweep).]{\textbf{Detectability of imposed local sixfold modulation under matched sampling conditions.}\par\smallskip
			Matched simulations were generated with increasing sixfold-prior strength (\textbf{A}, $\beta=0.1$; \textbf{B}, $\beta=0.2$; \textbf{C}, $\beta=0.4$). Per-cell summaries were evaluated within the predefined local-field window ($0.1\lambda$--$0.4\lambda$).
			Left column: survival curves of per-cell window summaries $AUC(Z)$.
			Right column: paired per-cell $AUC(Z)$ values (dots: cells; lines: paired comparisons; violins: distributions; black markers: median with bootstrap 95\% CI). Blue and green $p$-values indicate two-sided population sign-flip tests
			of the mean $AUC(Z)$ against zero for Exp--Circ and Exp--Six, respectively;
			the gray $p$-value indicates the paired contrast
			$AUC(Z^{\mathrm{Exp-Six}})-AUC(Z^{\mathrm{Exp-Circ}})$.
Increasing $\beta$ produces graded increases in detectable sixfold deviation relative to matched circular controls. Experimental data remain near the circular null and fall below sixfold-prior expectations for larger $\beta$, demonstrating graded sensitivity to the imposed local sixfold structure under matched sampling conditions.
		}
		\label{fig:supp-beta-sweep}
	\end{figure}

	\begin{figure}[tbp]
		\centering
		\includegraphics[width=0.9\textwidth]{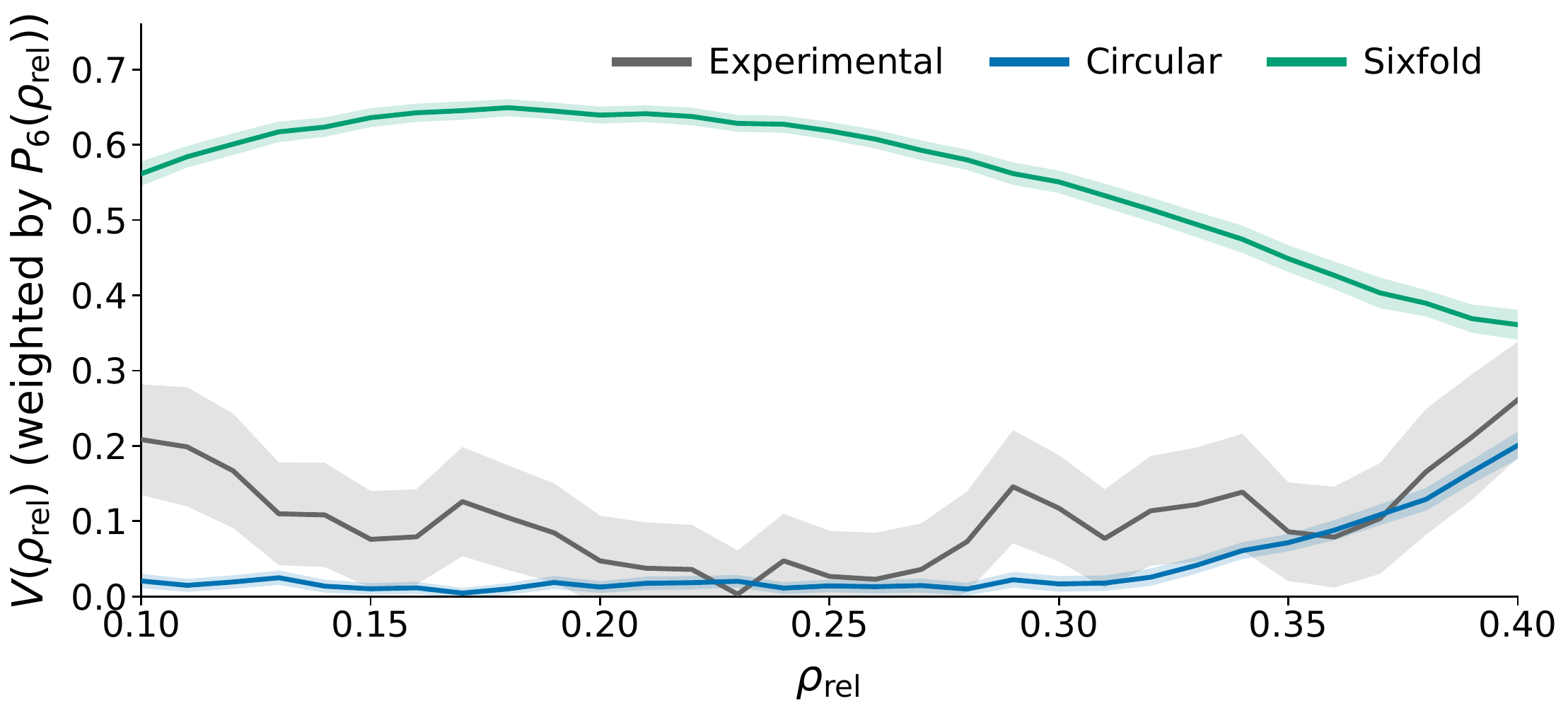}
		\caption[Population alignment of sixfold phase across radii.]{\textbf{Population alignment of sixfold phase across relative radii.}\par\smallskip
			Vector strength $V(\rho_{\mathrm{rel}})$ of the sixfold phase ($k=6$) across cells, weighted by sixfold power $P_6(\rho_{\mathrm{rel}})$. Shading indicates bootstrap-estimated standard error of $V$ obtained by resampling cells. Within the local-field window ($0.1\lambda$--$0.4 \lambda$), the experimental alignment is low and close to the circular null, whereas sixfold-prior simulations show substantially stronger phase alignment.
		}
		\label{fig:supp-pop-vec}
	\end{figure}
	
	\begin{figure}[tbp]
		\centering
		\includegraphics[width=0.9\textwidth]{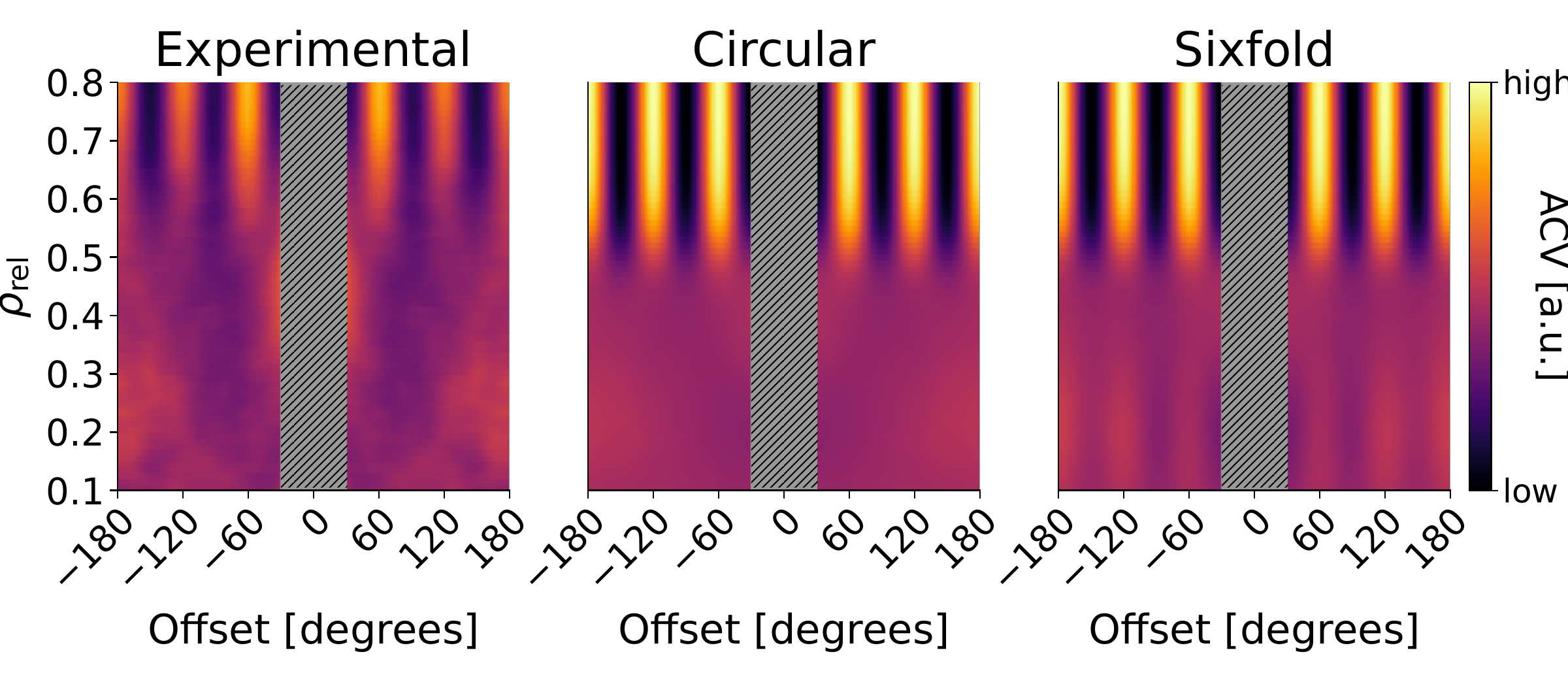}
		\caption[Extended angular autocovariance maps across radii.]{\textbf{\boldmath Extended angular autocovariance (ACV) maps across $0.1\lambda$--$0.8\lambda$.}
			\par\smallskip
			Population-averaged ACV maps for experimental data and matched circular and sixfold-prior simulations ($\beta=0.2$). ACVs are computed from normalized, mean-centered angular profiles as in \cref{fig:fig2} and displayed using the same color scale. At larger radii, all conditions exhibit pronounced sixfold periodicity consistent with the global lattice organization. In simulations, the imposed local sixfold structure becomes apparent at smaller relative radii. The gray hatched region denotes the excluded $\pm\ang{30}$ neighborhood around zero offset.
		}
		\label{fig:supp-acv-full}
	\end{figure}

	\begin{figure}[tbp]
		\centering
		\includegraphics[width=0.9\textwidth]{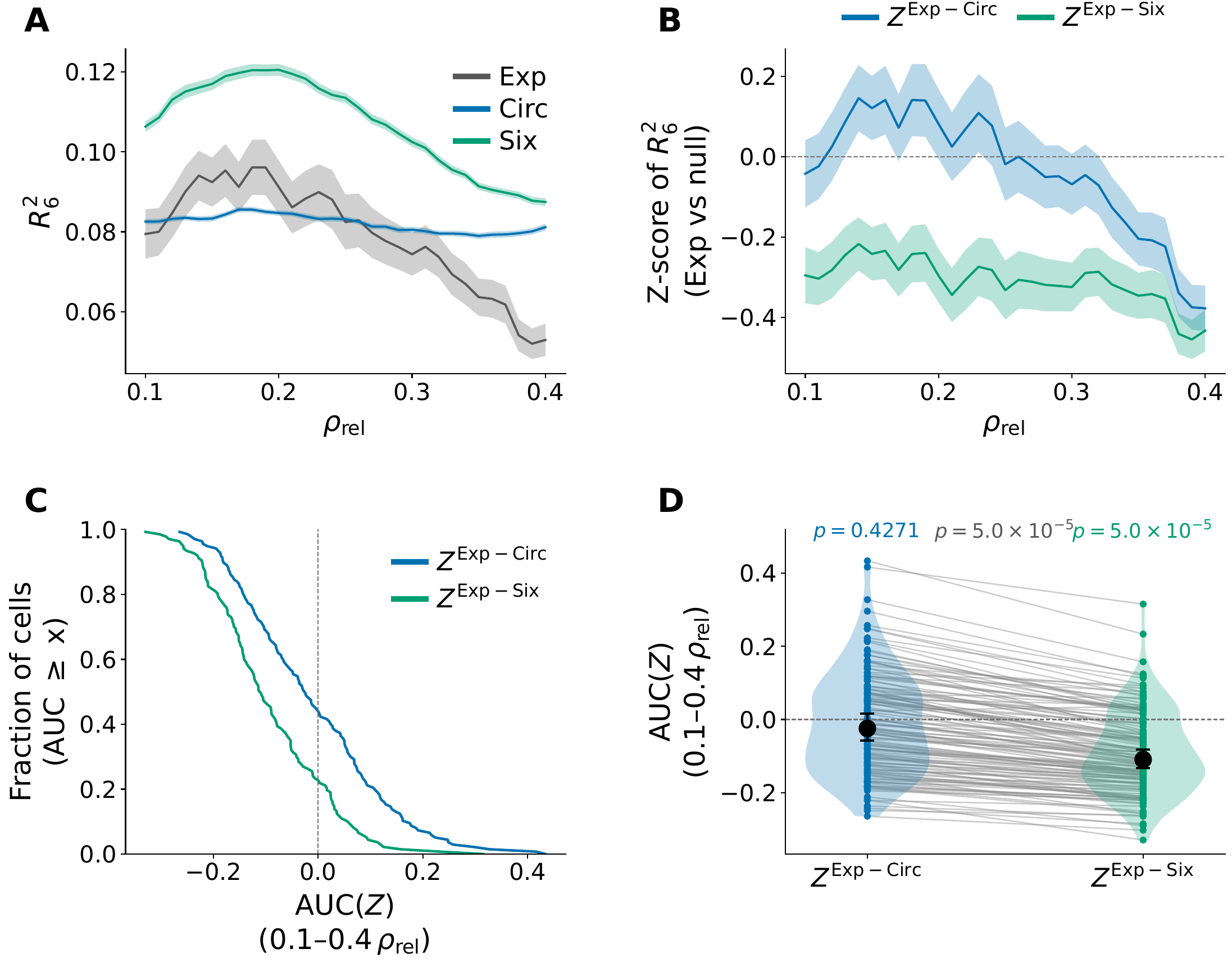}
		\caption[Sixfold modulation quantified using $R_6^2$.]{\textbf{\boldmath Sixfold modulation quantified using harmonic regression $R_6^2$ yields results consistent with $f_6$.}\par\smallskip
			\textbf{(A)} Population mean $\pm$ SEM of sixfold $R_6^2(\rho_{\mathrm{rel}})$ for experimental cells and matched simulations. For each cell, model replicates were first averaged within the cell; population
			statistics were computed across cells.
			\textbf{(B)} Population mean $\pm$ SEM of cell-matched standardized deviations (Z-scores).
			\textbf{(C)} Survival curves of per-cell window summaries $AUC(Z)$ within $0.1\lambda$--$0.4\lambda$.
			\textbf{(D)}
			Paired per-cell $AUC(Z)$ values for Exp--Circ and Exp--Six
			(dots: cells; lines: paired comparisons). Violin plots summarize
			the distributions. Black points mark the median; black error bars
			show percentile-bootstrap 95\% confidence intervals of the median.
			Colored $p$-values indicate two-sided sign-flip permutation tests
			of whether the population mean $AUC(Z)$ differs from 0.
			The gray $p$-value indicates the paired test of
			$AUC(Z^{\mathrm{Exp-Six}})-AUC(Z^{\mathrm{Exp-Circ}})$.
		}
		\label{fig:supp-R2}
	\end{figure}

	\begin{figure}[tbp]
		\centering
		\includegraphics[width=0.9\textwidth]{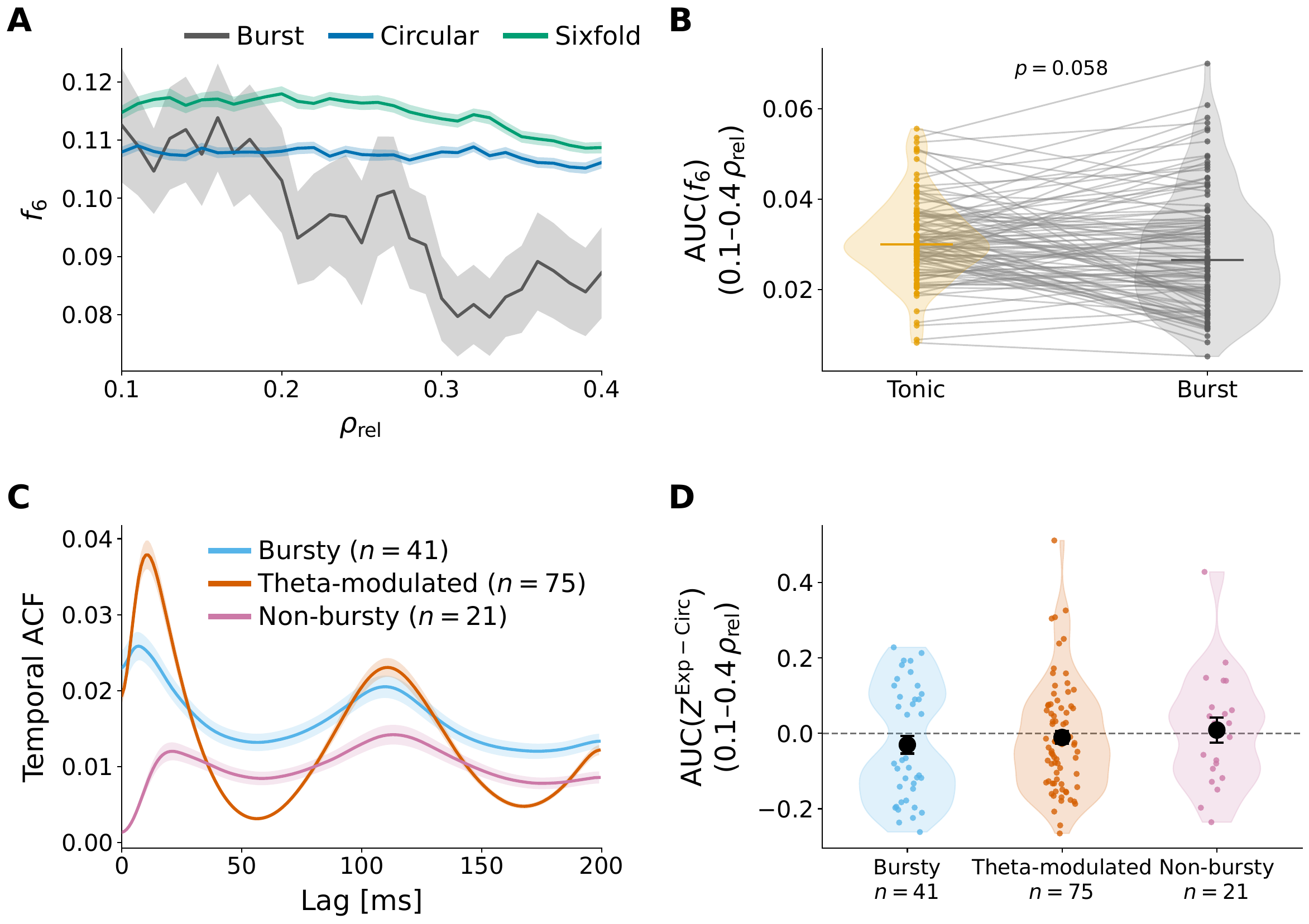}
		\caption[Temporal firing modes do not reveal enhanced local sixfold organization.]{\textbf{Temporal firing modes do not reveal enhanced local sixfold organization.}
			\par\smallskip
			\textbf{(A)}
			Population mean $\pm$ SEM of sixfold power fraction $f_6$ for independent burst-onset events ($N=108$ cells), together with cell-matched circular and weak sixfold-modulation reference populations. Reference realizations were conditioned on each cell's observed radial distribution of burst events before application of the frozen annulus estimator.
			\textbf{(B)}
			Direct within-cell comparison of burst and tonic firing. Within each
			$0.05\lambda$ radial matching stratum, burst-onset events and tonic spikes were
			sampled to identical counts while polar angle remained unconstrained.
			Twenty-five paired realizations were generated per cell; displayed values are
			the within-cell mean $AUC(f_6)$ over the predefined
			$0.1\lambda$--$0.4\lambda$ interval (dots), with lines connecting the burst and
			tonic estimates from the same cell. The indicated $p$-value is the two-sided
			cell-wise sign-flip test of the burst-minus-tonic difference. Violins show the across-cell distributions; colored horizontal bars mark their medians.
			\textbf{(C)}
			Mean $\pm$ SEM temporal autocorrelograms for bursty ($n=41$),
			theta-modulated ($n=75$), and non-bursty ($n=21$) phenotypes defined using the \textcite{gardnerToroidalTopologyPopulation2022} method
			within the canonical $137$-cell population.
			\textbf{(D)}
			Canonical all-spike sixfold selectivity, expressed as cell-wise
			$AUC(Z^{\mathrm{Exp-Circ}})$, across the same temporal phenotypes.
			Dots denote cells, violins show the class distributions, and black markers
			denote mean $\pm$ SEM. The omnibus one-way permutation ANOVA found no detectable
			between-class difference ($F=0.517$, $p=0.596$, $\eta^2=0.0077$).
		}
		\label{fig:temporal-modes}
	\end{figure}

	\begin{figure}[tbp]
		\centering
		\includegraphics[width=0.9\textwidth]{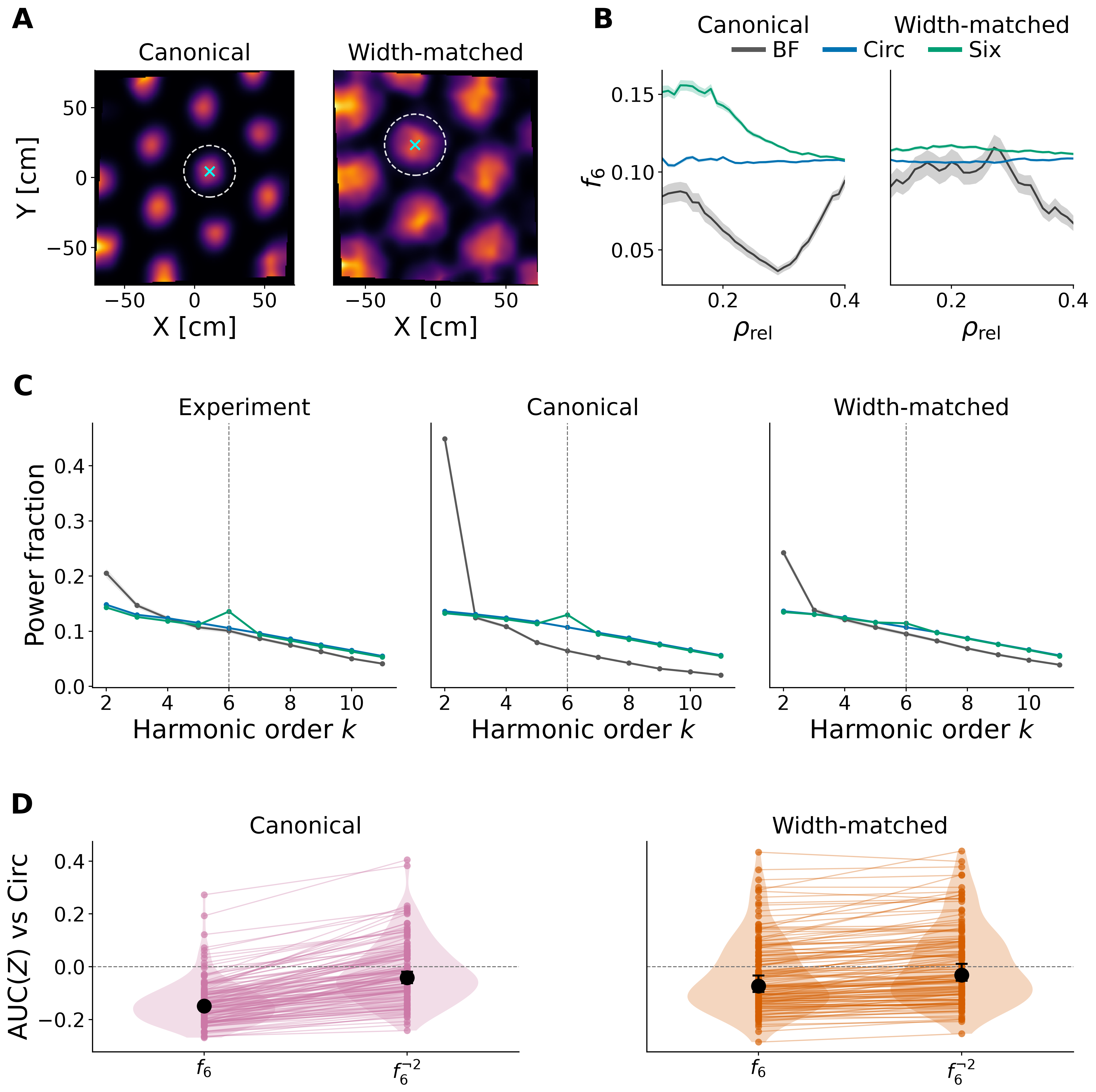}
		\caption[\textbf{Local angular structure in canonical and width-matched
			Burak--Fiete simulations.}]{\textbf{Local angular structure in canonical and width-matched Burak--Fiete
				simulations.} \par\smallskip
			\textbf{(A)} Representative distortion-reduced rate maps from the canonical
			full-trajectory condition and the width-matched condition (first $30$~min).
			The same deterministic readout identity is shown in both conditions; dashed
			circles indicate $0.4\lambda$ around the selected field center.
			\textbf{(B)} Population mean $\pm$ SEM of the primary sixfold power fraction
			$f_6$ for BF readouts and their own matched circular (Circ) and imposed-sixfold
			(Six) references.
			\textbf{(C)} Mean within-readout harmonic power fractions
			$P_k/\sum_{j=2}^{11}P_j$, averaged across the predefined local-field window,
			for experiment, canonical BF, and width-matched BF together with their
			corresponding matched references. The matched references provide the
			condition-specific spectral baseline; imposed-sixfold references show a
			selective $k=6$ peak, whereas canonical BF is strongly weighted toward low
			harmonic order. Because these fractions are compositional, a smaller fraction
			at another $k$ does not imply lower absolute harmonic power.
			\textbf{(D)} Effect of omitting $P_2$ from the $f_6$ denominator on
			per-readout $AUC(Z)$ relative to the corresponding matched circular
			reference, shown separately for the canonical and width-matched BF
			conditions. $f_6$ denotes the predefined sixfold power fraction, whereas
			$f_6^{\neg2}$ retains the same $P_6$ numerator with $P_2$ omitted from
			the denominator. Lines connect statistics
			from the same readout; violins summarize the distributions and black markers show the median and percentile-bootstrap $95$\% confidence interval. Corresponding population-level tests are reported in the text. Within each BF condition, readouts are spatial phases from a single deterministic network realization and are not independent network replicates.
		}	
		\label{fig:supp-CAN}
	\end{figure}
	
	\begin{figure}[tbp]
		\centering
		\includegraphics[width=0.9\textwidth]{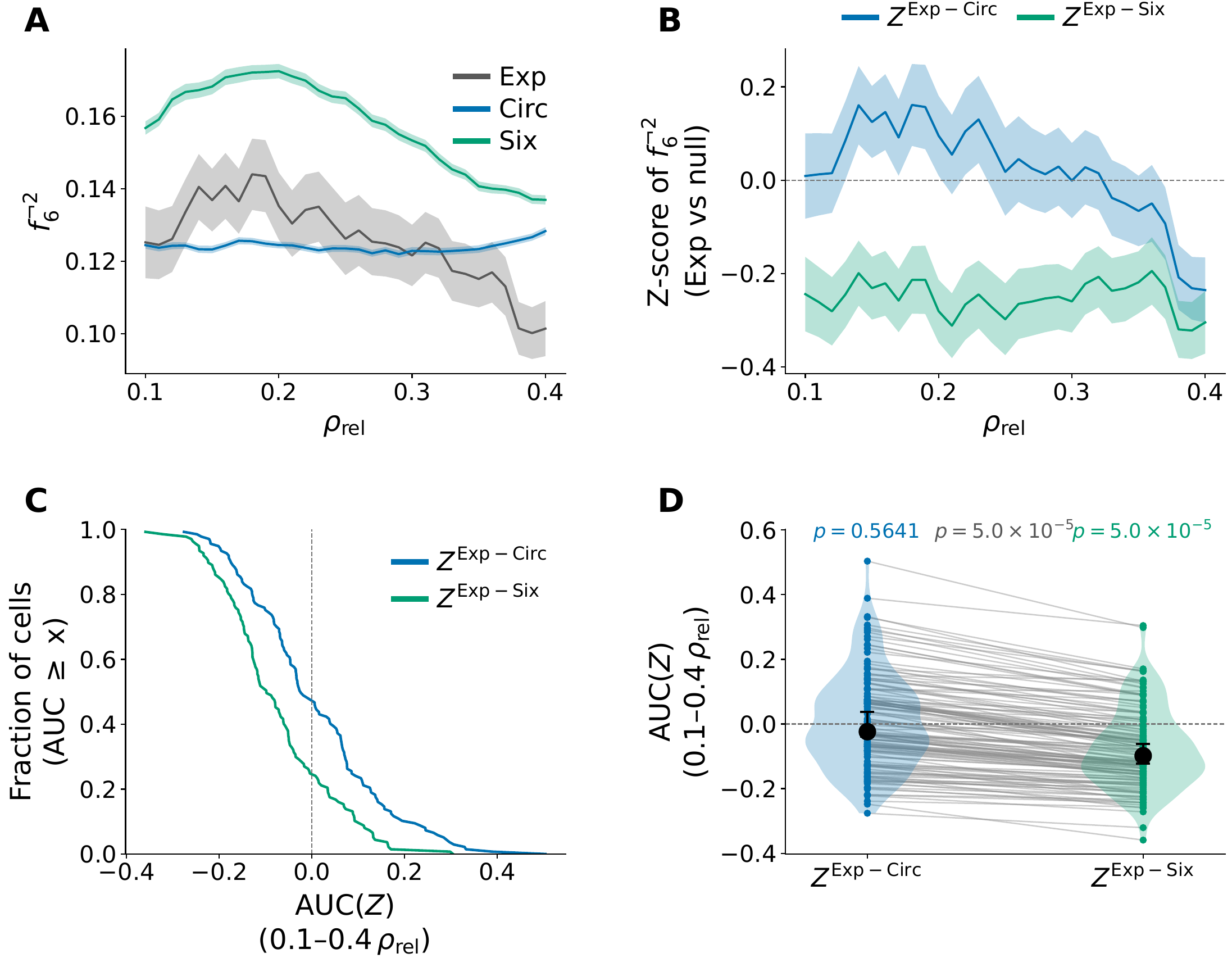}
		\caption{\LOOFigureCaption{2}}
		\label{fig:supp-loo-k2}
	\end{figure}
	
	\begin{figure}[tbp]
		\centering
		\includegraphics[width=0.9\textwidth]{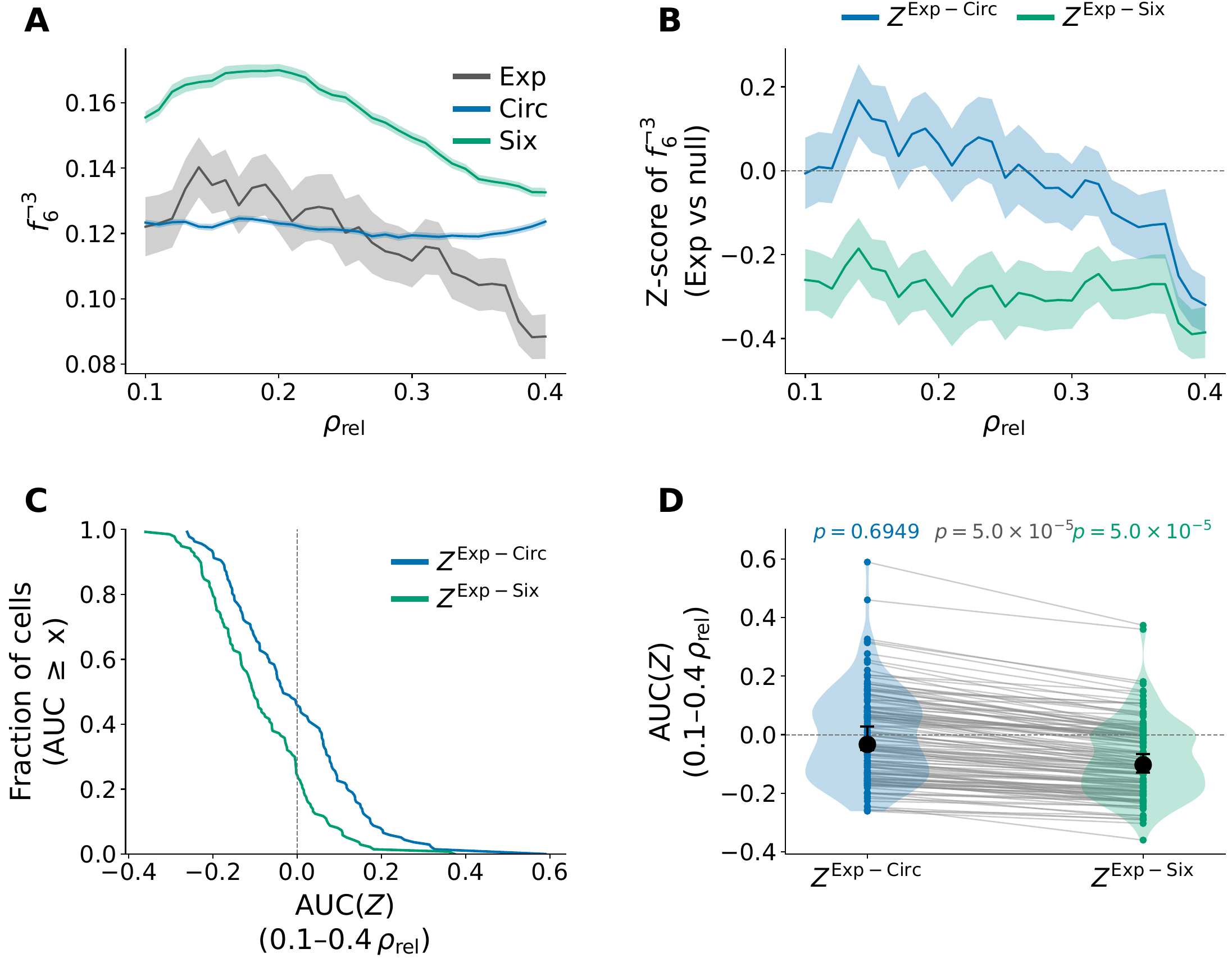}
		\caption{\LOOFigureCaption{3}}
		\label{fig:supp-loo-k3}
	\end{figure}
	
	\begin{figure}[tbp]
		\centering
		\includegraphics[width=0.9\textwidth]{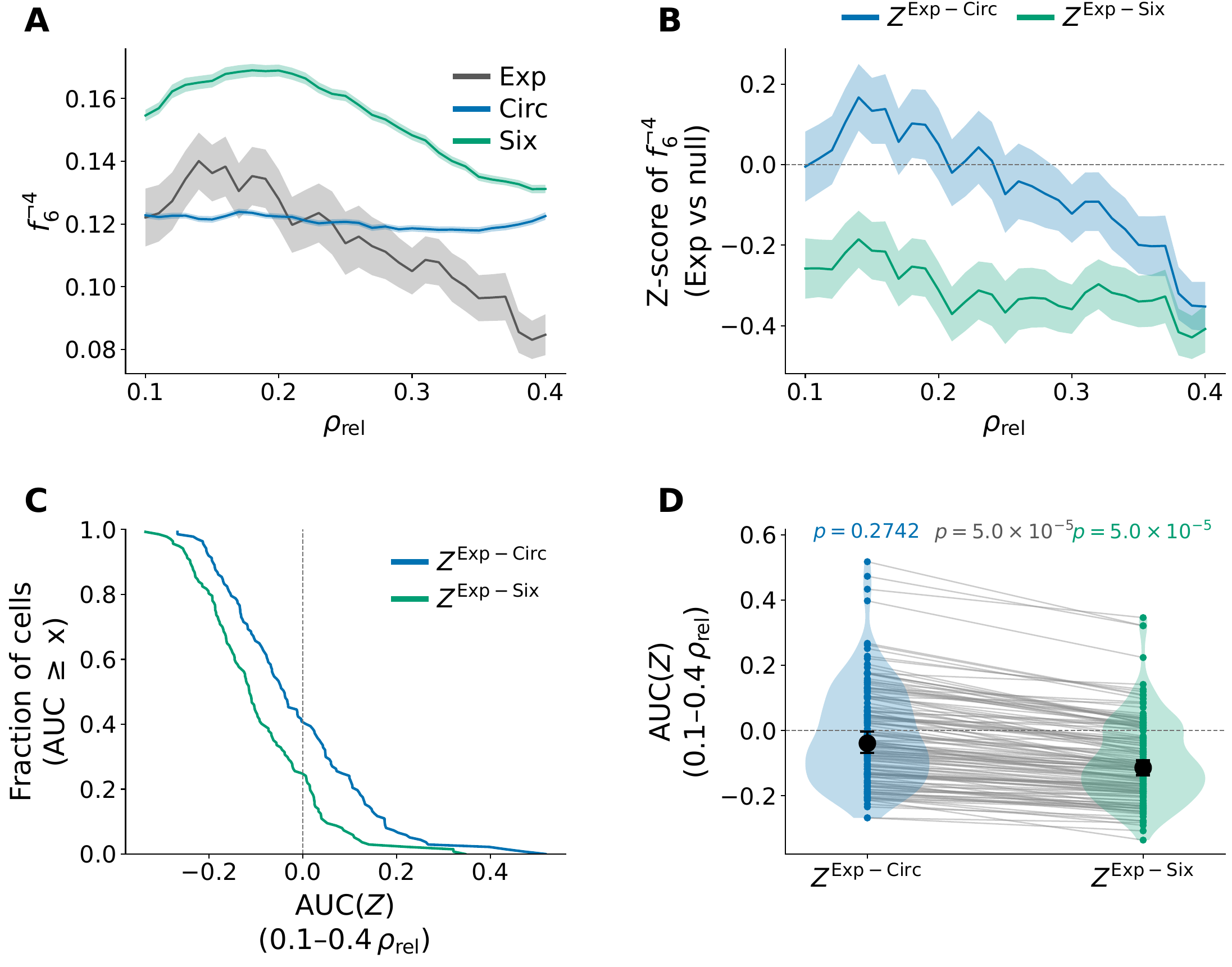}
		\caption{\LOOFigureCaption{4}}
		\label{fig:supp-loo-k4}
	\end{figure}
	
	\begin{figure}[tbp]
		\centering
		\includegraphics[width=0.9\textwidth]{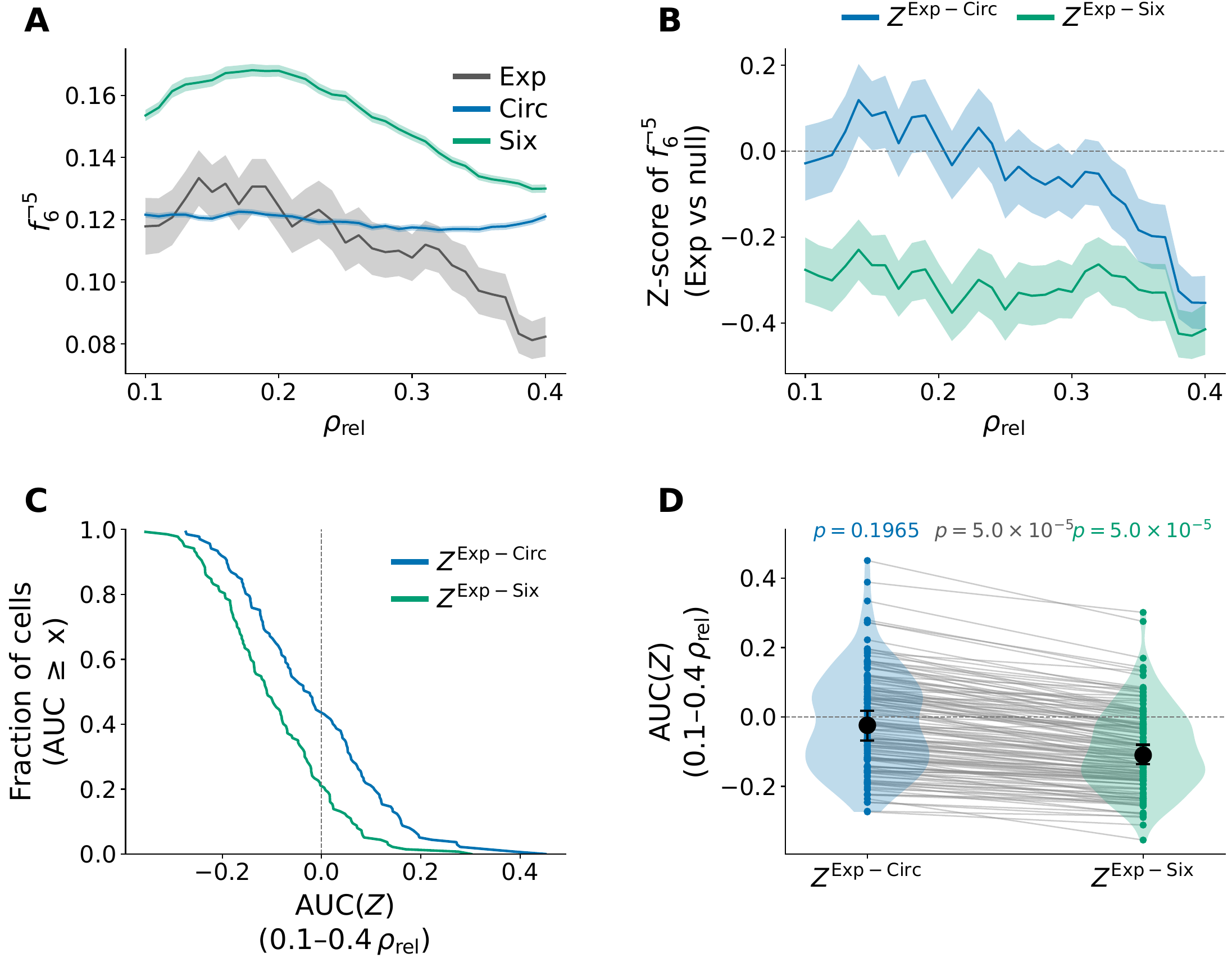}
		\caption{\LOOFigureCaption{5}}
		\label{fig:supp-loo-k5}
	\end{figure}
	
	\begin{figure}[tbp]
		\centering
		\includegraphics[width=0.9\textwidth]{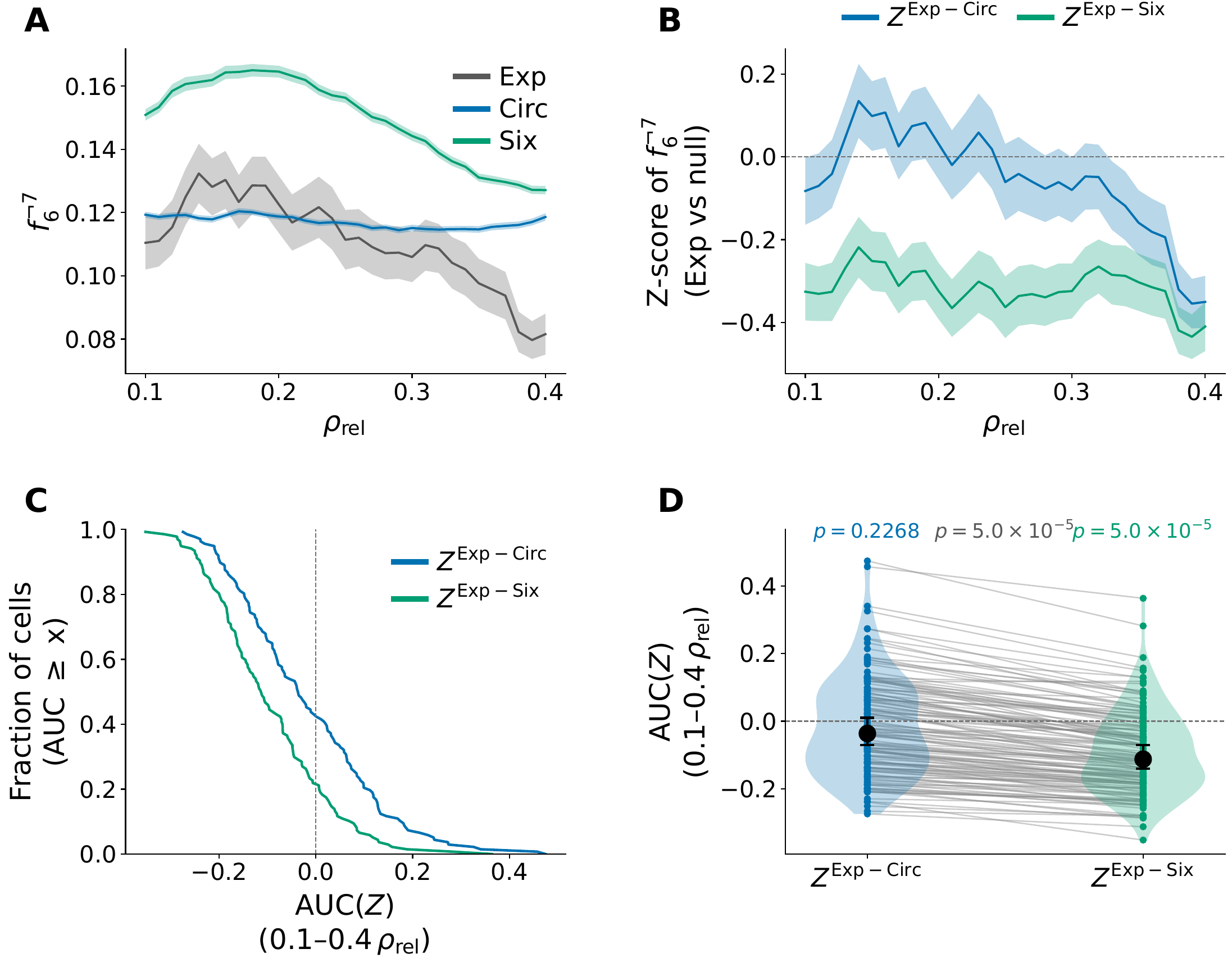}
		\caption{\LOOFigureCaption{7}}
		\label{fig:supp-loo-k7}
	\end{figure}
	
	\begin{figure}[tbp]
		\centering
		\includegraphics[width=0.9\textwidth]{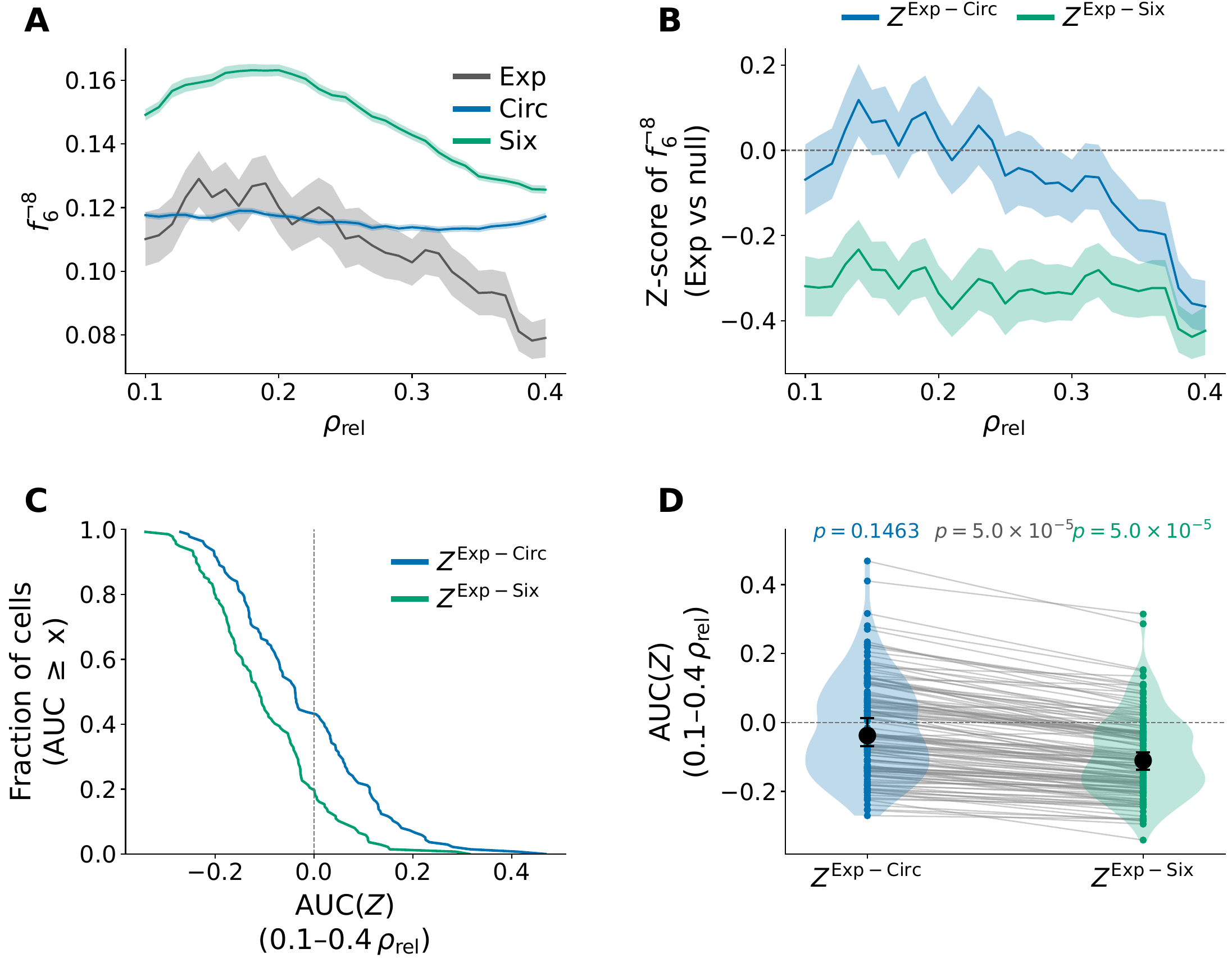}
		\caption{\LOOFigureCaption{8}}
		\label{fig:supp-loo-k8}
	\end{figure}
	
	\begin{figure}[tbp]
		\centering
		\includegraphics[width=0.9\textwidth]{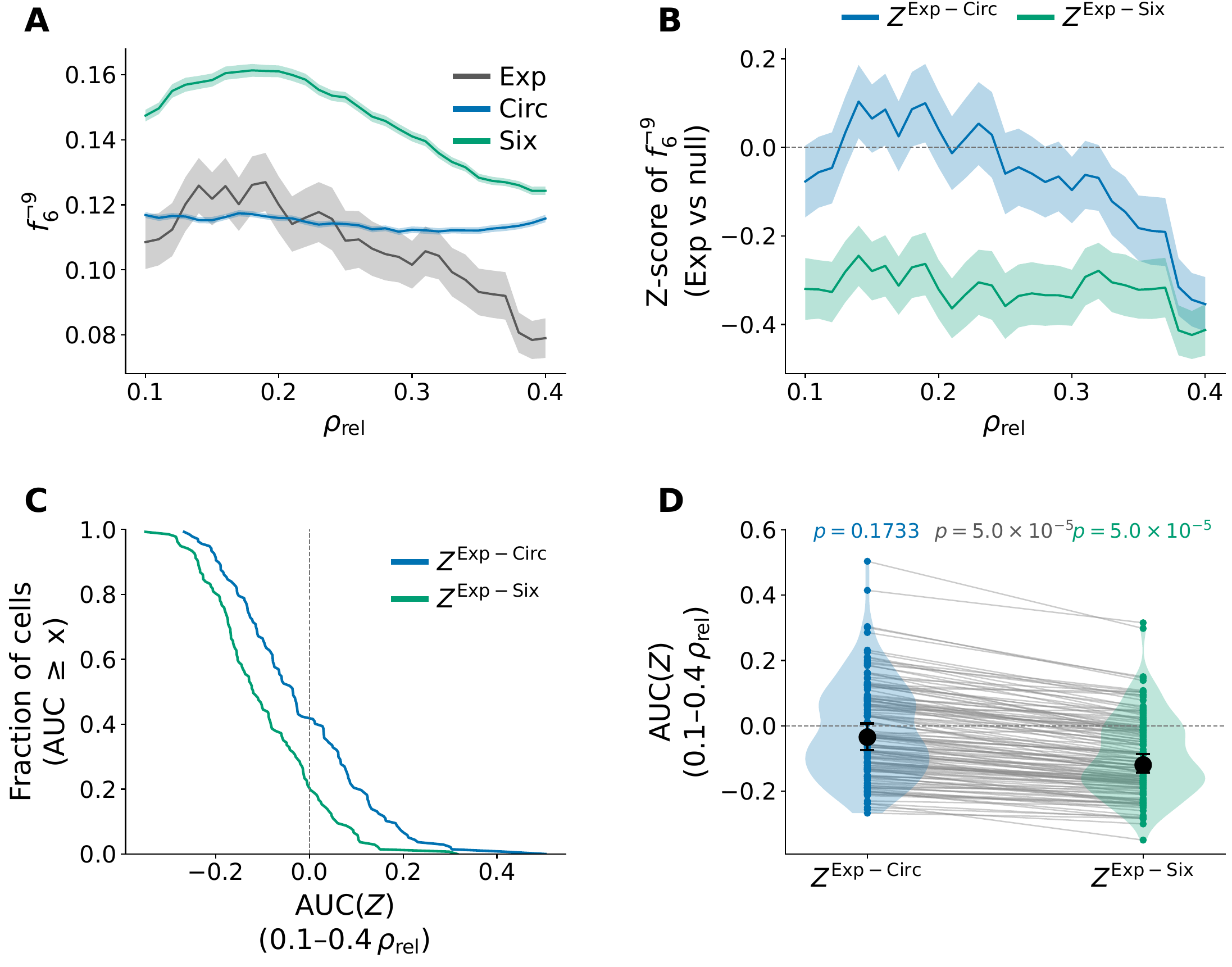}
		\caption{\LOOFigureCaption{9}}
		\label{fig:supp-loo-k9}
	\end{figure}
	
	\begin{figure}[tbp]
		\centering
		\includegraphics[width=0.9\textwidth]{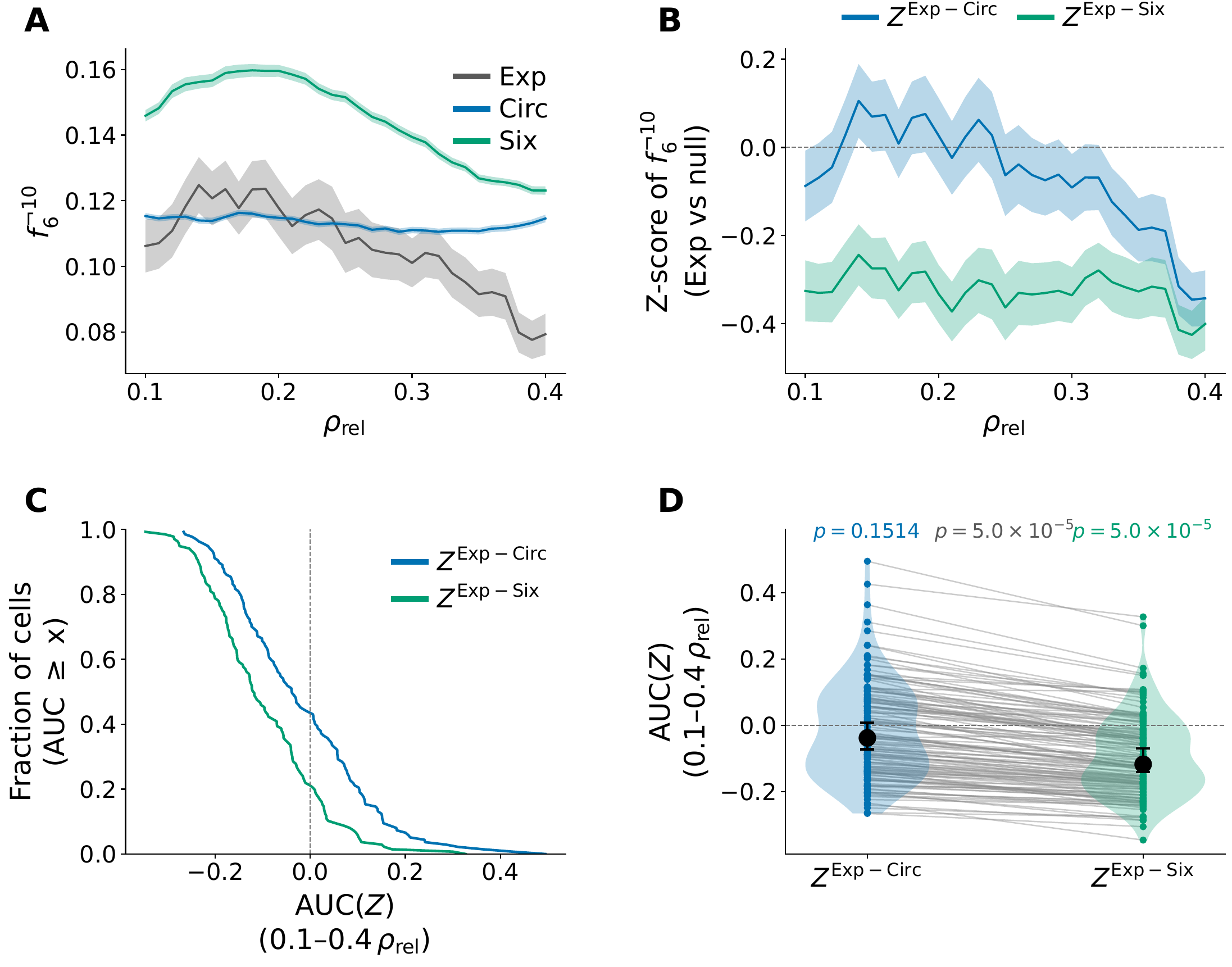}
		\caption{\LOOFigureCaption{10}}
		\label{fig:supp-loo-k10}
	\end{figure}
	
	\begin{figure}[tbp]
		\centering
		\includegraphics[width=0.9\textwidth]{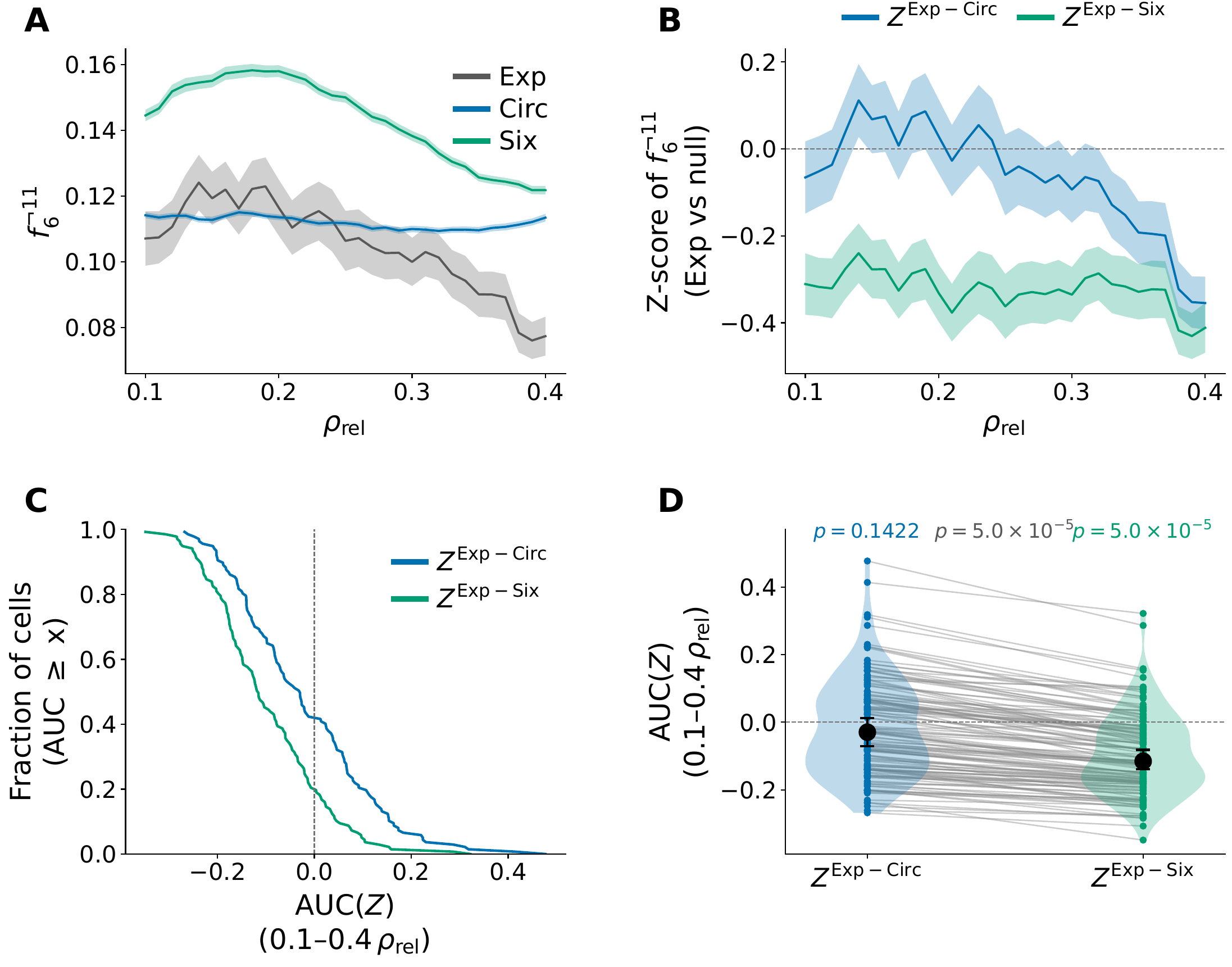}
		\caption{\LOOFigureCaption{11}}
		\label{fig:supp-loo-k11}
	\end{figure}
	
	\FloatBarrier
	\clearpage 
	
	\pdfbookmark[2]{Supplementary Tables}{supp-tables}
	\subsection*{Supplementary Tables}
	
	\captionsetup[table]{skip=4pt}
	
	\setlength{\floatsep}{6pt plus 2pt minus 2pt}
	\setlength{\textfloatsep}{8pt plus 2pt minus 2pt}
	
	\setcounter{topnumber}{10}
	\setcounter{bottomnumber}{10}
	\setcounter{totalnumber}{10}
	
	\renewcommand{\topfraction}{0.95}
	\renewcommand{\bottomfraction}{0.95}
	\renewcommand{\textfraction}{0.05}
	\renewcommand{\floatpagefraction}{0.8}
	
	\begin{table}[!htbp]
		\centering
		\caption{\textbf{Robustness of the primary $f_6$ inference to radial-window choice.}
			Values are population means of the cell-wise $AUC(Z)$ summaries. $p$-values are two-sided sign-flip permutation tests of the corresponding mean (20,000 permutations). The predefined main-analysis window is highlighted. Since $AUC(Z)$ is an integral, its numerical magnitude scales with window width.}
		\resizebox{\textwidth}{!}{\begin{tabular}{lrrrrrrr}
				\toprule
				Window ($\rho_{\mathrm{rel}}$) & $N$ &
				Mean Exp--Circ & $p_{\mathrm{Circ}}$ &
				Mean Exp--Six & $p_{\mathrm{Six}}$ &
				Mean paired effect & $p_{\mathrm{pair}}$ \\
				\midrule
				0.1--0.20 & 137 &  0.0053 & 0.407 & -0.0278 & $5.0\times10^{-5}$ & -0.0331 & $5.0\times10^{-5}$ \\
				0.1--0.25 & 137 &  0.0064 & 0.460 & -0.0442 & $5.0\times10^{-5}$ & -0.0506 & $5.0\times10^{-5}$ \\
				0.1--0.30 & 137 &  0.0038 & 0.718 & -0.0606 & $5.0\times10^{-5}$ & -0.0644 & $5.0\times10^{-5}$ \\
				\textbf{0.1--0.40} & \textbf{137} & \textbf{-0.0138} & \textbf{0.279} & \textbf{-0.0936} & \textbf{$5.0\times10^{-5}$} & \textbf{-0.0798} & \textbf{$5.0\times10^{-5}$} \\
				0.1--0.50 & 137 & -0.0665 & $5.0\times10^{-5}$ & -0.1466 & $5.0\times10^{-5}$ & -0.0801 & $5.0\times10^{-5}$ \\
				0.3--0.40 & 137 & -0.0168 & $1.0\times10^{-3}$ & -0.0299 & $5.0\times10^{-5}$ & -0.0131 & $5.0\times10^{-5}$ \\
\bottomrule
		\end{tabular}}
\label{tab:supp-window-robustness}
\end{table}

	\begin{table}[!htbp]
		\centering
		\caption{\textbf{Robustness of the primary $f_6$ inference to simulated uniform background fraction.}
			The radial window was fixed at $0.1\lambda$--$0.4\lambda$. Values are mean cell-wise $AUC(Z)$ summaries; $p$-values are two-sided sign-flip permutation tests of the corresponding mean.}
		\resizebox{\textwidth}{!}{\begin{tabular}{lrrrrrrr}
				\toprule
				Noise fraction & $N$ &
				Mean Exp--Circ & $p_{\mathrm{Circ}}$ &
				Mean Exp--Six & $p_{\mathrm{Six}}$ &
				Mean paired effect & $p_{\mathrm{pair}}$ \\
				\midrule
				0.0 & 137 & -0.0157 & 0.221 & -0.1090 & $5.0\times10^{-5}$ & -0.0933 & $5.0\times10^{-5}$ \\
				0.1 (canonical) & 137 & -0.0138 & 0.279 & -0.0936 & $5.0\times10^{-5}$ & -0.0798 & $5.0\times10^{-5}$ \\
				0.2 & 137 & -0.0071 & 0.582 & -0.0743 & $5.0\times10^{-5}$ & -0.0672 & $5.0\times10^{-5}$ \\
				\bottomrule
		\end{tabular}}
		\label{tab:supp-noise-robustness}
	\end{table}

	\begin{table}[!htbp]
		\centering
		\caption{\textbf{Robustness of the primary $f_6$ inference to rate-map smoothing.}
			Spike and occupancy smoothing widths are shown as $(\sigma_s,\sigma_o)$ in cm. The radial window was fixed at $0.1\lambda$--$0.4\lambda$.}
		\resizebox{\textwidth}{!}{\begin{tabular}{lrrrrrrr}
				\toprule
				Bandwidth (cm) & $N$ &
				Mean Exp--Circ & $p_{\mathrm{Circ}}$ &
				Mean Exp--Six & $p_{\mathrm{Six}}$ &
				Mean paired effect & $p_{\mathrm{pair}}$ \\
				\midrule
				(3,2) & 129 & -0.0401 & 0.00190 & -0.1083 & $5.0\times10^{-5}$ & -0.0682 & $5.0\times10^{-5}$ \\
				(3,3) & 138 & -0.0184 & 0.163 & -0.0938 & $5.0\times10^{-5}$ & -0.0755 & $5.0\times10^{-5}$ \\
				(4,3) & 137 & -0.0221 & 0.0805 & -0.0982 & $5.0\times10^{-5}$ & -0.0761 & $5.0\times10^{-5}$ \\
				(5,4) canonical & 137 & -0.0138 & 0.279 & -0.0936 & $5.0\times10^{-5}$ & -0.0798 & $5.0\times10^{-5}$ \\
				(6,4) & 135 & -0.0173 & 0.181 & -0.0980 & $5.0\times10^{-5}$ & -0.0806 & $5.0\times10^{-5}$ \\
				\bottomrule
		\end{tabular}}
		\label{tab:supp-bandwidth-robustness}
	\end{table}

	\begin{table}[!htbp]
		\centering
		\caption{\textbf{Detectability calibration across imposed sixfold modulation strengths.}
			The radial window was fixed at $0.1\lambda$--$0.4\lambda$.
			The $\beta=0.2$ row is the canonical sixfold reference used in the main analysis.
			The $\beta=0.05$ condition provides an additional lower-strength calibration.}
		\resizebox{\textwidth}{!}{\begin{tabular}{lrrrrrrr}
				\toprule
				$\beta$ & $N$ &
				Mean Exp--Circ & $p_{\mathrm{Circ}}$ &
				Mean Exp--Six & $p_{\mathrm{Six}}$ &
				Mean paired effect & $p_{\mathrm{pair}}$ \\
				\midrule
				0.05 & 137 & -0.0138 & 0.279 & -0.0215 & 0.0767 & -0.00775 & $5.0\times10^{-5}$ \\
				0.10 & 137 & -0.0138 & 0.279 & -0.0395 & $9.0\times10^{-4}$ & -0.0257 & $5.0\times10^{-5}$ \\
				0.20 & 137 & -0.0138 & 0.279 & -0.0936 & $5.0\times10^{-5}$ & -0.0798 & $5.0\times10^{-5}$ \\
				0.40 & 137 & -0.0138 & 0.279 & -0.2414 & $5.0\times10^{-5}$ & -0.2277 & $5.0\times10^{-5}$ \\
				\bottomrule
		\end{tabular}}
		\label{tab:supp-table-beta}
	\end{table}

\end{document}